\documentclass[journal,onecolumn]{IEEEtran}
\usepackage{longtable}
\usepackage{subfigure}
\usepackage{bm}
\usepackage{times}
\usepackage{epsf,epsfig}
\usepackage{float}
\usepackage{placeins}
\usepackage{psfrag}
\usepackage{amsmath}
\usepackage{bm}
\usepackage{amssymb}
\usepackage{graphicx}
\usepackage{epstopdf}
\usepackage{multirow}
\usepackage{enumerate}
\usepackage{color}
\usepackage{mathrsfs}
\usepackage{setspace}
\usepackage{diagbox}
\usepackage{enumitem}

\makeatletter
\newif\if@restonecol
\makeatother

\usepackage[linesnumbered,ruled,vlined]{algorithm2e}
\usepackage{algpseudocode}

\ifCLASSINFOpdf
\else
\fi
\begin{document}
%

\title{Protograph-Based Low-Density Parity-Check Hadamard Codes }




%
\author{
Peng W.  Zhang, 
        Francis C.M. Lau,~\IEEEmembership{Fellow,~IEEE,} 
        and~Chiu-W. Sham,~\IEEEmembership{Senior Member,~IEEE}
\thanks{P.W. Zhang and 
        F.C.M. Lau are with the Future Wireless Networks and IoT Focus Area, 
         Department
of Electronic and Information Engineering, The Hong Kong Polytechnic University, Hong Kong (e-mail: pengwei.zhang@connect.polyu.hk and francis-cm.lau@polyu.edu.hk).}
\thanks{C.-W. Sham is with the Department of Computer Science,  The University of Auckland,
New Zealand (e-mail: b.sham@auckland.ac.nz).}
\thanks{This paper was presented in part at WCNC 2020 \cite{ZhangPW2020}.}}


\maketitle

\begin{abstract}
In this paper, we propose a new method to design low-density parity-check Hadamard (LDPC-Hadamard) codes --- a type of ultimate-Shannon-limit approaching channel  codes.
The technique is based on applying Hadamard constraints to the check nodes  in a generalized protograph-based LDPC code, followed by lifting the generalized protograph. We name the codes formed  \textit{protograph-based LDPC Hadamard} (PLDPC-Hadamard) codes.
We also propose a modified Protograph Extrinsic Information Transfer (PEXIT)  algorithm for analyzing and optimizing PLDPC-Hadamard code designs.
The proposed algorithm further allows the analysis of PLDPC-Hadamard codes
with degree-$1$ and/or punctured nodes.
We find codes with decoding thresholds ranging from $-1.53$~dB to  $-1.42$~dB.
At a BER of $10^{-5}$, the gaps of our codes to the ultimate-Shannon-limit
range from $0.40$~dB (for rate = $0.0494$) to $0.16$~dB (for rate = $0.003$).
Moreover, the error performance of our codes is comparable to that of the traditional
 LDPC-Hadamard codes.
Finally, the BER performances of our codes after puncturing are simulated and compared.

\end{abstract}

\begin{IEEEkeywords}
Protograph LDPC code, PLDPC-Hadamard code, PEXIT algorithm, ultimate Shannon limit.
\end{IEEEkeywords}

\IEEEpeerreviewmaketitle

\section{Introduction}
In 1943, Claude Shannon derived the channel capacity theorem \cite{Shannon1948}.
Based on the theorem, the maximum rate that information can be sent through a channel without errors can be evaluated.
In the 50 years that followed, huge efforts were spent in designing error correction codes that would allow communication systems to work close to the channel capacity.
Despite the efforts made, the gaps to the capacity remained large. It was not until 1993 when a major breakthrough took place.

In 1993, Berrou \textit{et al}. invented the turbo codes and demonstrated that with a code rate of $0.5$, the proposed turbo code and decoder could work within $0.7$ dB from the capacity limit at a bit error rate (BER) of $10^{-5}$ \cite{Berrou1993}, \cite{Berrou1996}.
Subsequently, a lot of research effort has been spent on investigating coding schemes that are capacity-approaching.
Besides turbo codes, other most well-known ones are low-density parity-check codes (proposed by Gallager in 1960s \cite{Gallager1963} and rediscovered by MacKay and Neal in 1990s \cite{MacKay1995}) and polar codes (proposed by Arikan in 2009 \cite{Arikan2009}).
These capacity-approaching codes have since been used in many wireless communication systems (e.g., 3G/4G/5G, Wifi, satellite communications) and optical communication systems.

As the first type of capacity-approaching codes, turbo codes use a concatenation of two or more convolutional component encoders (with one or more interleavers) to encode information.
Correspondingly, two or more serial Bahl-Cocke-Jelinek-Raviv (BCJR) decoders are iteratively exchanging message information in the decoding processing.
Using a similar iterative decoding mechanism, LDPC codes whose parity-check matrix possesses a sparse structure use the belief-propagation (BP) algorithm to  implement the iterative decoding in a parallel manner.
As for the third type of capacity-approaching codes, polar codes are constructed by multiple recursive concatenation of a short kernel code and its decoder applies the successive cancellation (SC) mechanism.
The progresses of the aforementioned three types of capacity-approaching codes  over the past decades can be
found in the survey papers \cite{Shao2019, Fang2015} and the references therein.

LDPC codes can be represented by a matrix containing a low density of ``1''s and also by its corresponding Tanner graph \cite{Tanner1981}.
In the Tanner graph, there are two sets of nodes, namely variables nodes and check nodes, sparsely connected by links. Messages are updated and passed iteratively along the links during the decoding process.
Density evolution (DE) \cite{Richardson2001} is a kind of analytical method that tracks the probability density function (PDF) of the messages after each iteration.
 It not only can predict the convergence of the decoder, but also can be used for optimizing LDPC code designs.
Through the DE method, irregular LDPC codes with optimal variable-node/check-node degree distributions have been found to possess performance thresholds close to Shannon limits under an additive white Gaussian noise (AWGN) channel \cite{Richardson2001b}.
The extrinsic information transfer (EXIT) chart is another common technique
 employed to analyze and optimize LDPC codes  \cite{Brink2001}, \cite{Brink2004}.
An optimal LDPC code design is found when the EXIT curves of
the variable nodes and check nodes are ``matched'' with the smallest bit-energy-to-noise-power-spectral-density ratio ($E_b/N_0$).

For an LDPC code with given degree distributions and code length, the progressive-edge-growth (PEG) method \cite{Hu2005} is commonly used to connect the variable nodes and check nodes with an aim to maximizing the girth (shortest cycle) of the code.
The method is simple and the code can achieve good error performance. However, the code has a quadratic encoding complexity with its length because it is un-structured.
The hardware implementation of the encoder/decoder also consumes a lot of resources and has high routing complexity.

Subsequently, structured quasi-cyclic (QC) LDPC codes are proposed  \cite{Fossorier2004}.
QC-LDPC codes have a linear encoding complexity and allow
parallel processing in the hardware implementation.
Other structured codes, such as the repeat-accumulate (RA) codes and their variants, can be formed by the repeat codes and the accumulators \cite{Divsalar1998}, \cite{Jin2000}.
They also belong to a subclass of LDPC codes that have a fast encoder structure and good error performance.
Structured LDPC codes can also be constructed by protographs \cite{Thorpe2003}.
By expanding the protomatrix (corresponding protograph) with a small size, we can obtain a QC matrix (corresponding lifted graph) that possesses the same properties as the protomatrix.
The codes corresponding to the lifted graphs are called protograph-based LDPC (PLDPC) codes.
The traditional EXIT chart cannot be used to analyze protographs where
degree-$1$ or punctured variable nodes exist.
Subsequently, the protograph EXIT (PEXIT) chart method is developed \cite{Liva2007}   for analyzing and designing PLDPC codes, and
 well-designed PLDPC codes are found to achieve performance close to the Shannon limit \cite{Fang2015}, \cite{Divsalar2005}.
{\color{black}
When coupled spatially, PLDPC codes can enhance their theoretical thresholds and decoding performance  \cite{Stinner2015}, \cite{Stinner2016}, \cite{Mitchell2015}.
In the case of block-fading channels, root-protograph LDPC codes are found to achieve near-outage-limit performance  \cite{Fang2015b}, \cite{Fang2019}.
}

In the Tanner graph of an LDPC code, the VNs are equivalent to
repeat codes while
and CNs correspond to single-parity-check (SPC) codes.
If other block codes, such as Hamming codes  and BCH codes, are used to replace the repeat codes and/or SPC codes, generalized LDPC (GLDPC) codes are obtained \cite{Lentmaier1999}, \cite{Boutros1999}, \cite{Min2012}.
With more powerful constraints on VNs or CNs, the GLDPC codes can not only have better error performance, but also speed up convergence and lower the error-floor.
Yet, the two component block codes should be ``match'' in well-designed GLDPC code.

In \cite{Abu-Surra2006}, \cite{Liva2008} and \cite{Surra2011}, doped-Tanner codes are formed by replacing  the SPC component codes
in the structured LDPC codes with Hamming codes and  recursive systematic convolutional codes.
Ensemble codeword weight enumerators are used to search GLDPC codes 
and Hamming codes are used to design medium-length GLDPC codes with performances approaching the channel capacity ($>0$ dB).
{\color{black}
In \cite{Sharon2006a} and \cite{Sharon2006b}, EXIT functions of block codes over binary symmetric channels have been derived and used for analyzing LDPC codes. The same author further demonstrates the use of linear programming algorithm to optimize a rate-$8/9$ GLDPC code from the perspective of degree distribution  \cite{Sharon2019}.
}

To achieve good performance (BER = $10^{-5}$) at very low $E_b/N_0$, say $<-1.15$ dB,  Hadamard codes have been proposed to replace the SPC codes, forming the low-rate LDPC-Hadamard codes \cite{Yue2005}, \cite{Yue2007}.
By adjusting the degree distribution of the VNs and using the EXIT chart technique,
 the EXIT curves of the Hadamard ``super CNs'' and VNs are matched and  excellent error performance at low $E_b/N_0$ is obtained.

In practice, different channels possess different capacities, depending on factors such as modulation scheme, signal-to-noise ratio and code rate.
However, the ``ultimate Shannon limit'' over an additive white Gaussian channel remains at $-1.59$ dB, i.e., $E_b / N_0 = -1.59$ dB.
It defines the $E_b / N_0$ value below which no digital communications can be error-free.
Scenarios where digital communications may need to work close to the ultimate Shannon limit include space communications, multiple access (e.g. code-division multiple-access \cite{Buehrer2006}
and interleave-division multiple-access \cite{LiPing2006}) with severe inter-user interferences, or
embedding low-rate information in a communication link.
The most notable channel codes with performance close to this limit are turbo-Hadamard codes \cite{Li2003}, \cite{Xu2018}, \cite{Jiang2018}, \cite{Jiang2019}, concatenated zigzag Hadamard codes \cite{Leung2006}, \cite{Jiang2020}, and  LDPC-Hadamard codes \cite{Yue2005}, \cite{Yue2007}.
Both turbo-Hadamard codes and concatenated zigzag Hadamard codes suffer from long decoding latency due to the forward/backward decoding algorithms \cite{Li2003}, \cite{Leung2006}.
The LDPC-Hadamard codes allow parallel processing and hence the decoding latency can be made shorter \cite{Yue2007}.
However, in optimizing the threshold of LDPC-Hadamard codes, only the degree distribution of the variable nodes has been found for a given order of the Hadamard code used.
Therefore, the  method used in optimizing LDPC-Hadamard codes has the following drawbacks.

\begin{enumerate}
\item
For the same variable-node degree distribution, many different code realizations with very diverse bit-error-rate performances can be obtained.
As a consequence, the degree distribution should be further refined to provide more information for code construction.

{\color{black}
\item
When the information length and hence code length are determined, the degrees of each individual variable node and each individual Hadamard check node can be computed. Then the connections between the variable nodes and the Hadamard check nodes in the Tanner graph are assigned randomly or semi-randomly, say by the PEG algorithm  \cite{Hu2005}. Since the two sets of nodes are connected randomly or semi-randomly, there is little structure to follow, making both encoding and decoding very complex to realize in practice. Take the LDPC-Hadamard code with code rate $R=0.05$ and Hadamard code order $r=4$ as an example. The degree distributions optimized by \cite{Yue2007}  indicate the ratio of the number of rows $m$ (corresponds to the number of Hadamard check nodes) and the number of columns $n$ (corresponds to the number of variable nodes) in the connection matrix equals $0.6338$, i.e., $m/n = 0.6338$. For an information length of $65,536$, it can be computed that $m = 113,426$ and $n = 178,962$. When the connections between the $178,962$ variable nodes and the $113,426$ Hadamard check nodes in the Tanner graph are semi-randomly connected by the PEG algorithm, the graph has little structure and is therefore not conducive to parallel encoding/decoding and reduces encoding/decoding efficiency. In the hardware implementation, the unstructured conventional LDPC-Hadamard code further results in high routing complexity and low throughput.
}
\item
The degree distribution analysis requires a minimum variable-node degree of $2$ because an EXIT curve cannot be produced for degree-$1$ variable nodes.
Moreover, LDPC-Hadamard codes with punctured variable nodes cannot be analyzed.
In short, LDPC-Hadamard codes with degree-$1$ and/or punctured variable nodes cannot be analyzed.
Yet in many application scenarios, LDPC codes with degree-$1$ and/or punctured variable nodes can provide better theoretical thresholds and hence error performances.
New techniques should therefore be proposed to analyze ultimate-Shannon-limit-approaching codes with degree-$1$ and/or punctured variable nodes.
\end{enumerate}
The concept in \cite{Yue2007} has been applied in designing other low-rate generalized LDPC codes \cite{Liu2018}.
However, the main criterion of those codes is to provide low latency communications and hence their performance is relatively far from the ultimate Shannon limit.

Compared with LDPC code designs based on degree distributions,
PLDPC codes not only can form structured QC-LDPC codes which are advantageous to encoding and decoding, but also can be analyzed using the PEXIT algorithm. Moreover, the PEXIT technique can accurately analyze the decoding performance of PLDPC codes with degree-$1$ and/or punctured VNs.
Generalized PLDPC codes using other block codes in VNs/CNs have been proposed and analyzed, but their decoding thresholds are mainly designed to be close to
Shannon capacities greater than $0$ dB.

In this paper, we propose a method to design LDPC-Hadamard codes
which possesses degree-$1$ and/or punctured VNs.
The technique is based on applying Hadamard constraints to the CNs  in a generalized PLDPC code, followed by lifting the generalized protograph. We name the codes formed   \textit{protograph-based LDPC Hadamard} (PLDPC-Hadamard) codes.
We also propose a modified PEXIT algorithm for analyzing and optimizing PLDPC-Hadamard code designs.
Codes with decoding thresholds ranging from $-1.53$~dB to  $-1.42$~dB have been found,
and simulation results show a bit error rate  of $10^{-5}$ can be achieved at $E_b/N_0 = -1.43$~dB.
Moreover, the BER performances of these codes after puncturing are simulated and compared.
The main contributions of the paper can be summarized as follows:


\begin{itemize}
\item[1)]
It is the first attempt to use protographs to design codes with performance close to the ultimate Shannon limit.
By appending additional degree-$1$ Hadamard variable nodes to the check nodes of a protograph, the SPC check nodes are converted into more powerful Hadamard constraints, forming the generalized protograph of PLDPC-Hadamard codes.
Moreover, the proposed PLDPC-Hadamard code inherits the advantages of PLDPC codes.
{\color{black}After using the copy-and-permute operations to lift the protograph, the matrix corresponding to the lifted graph is a structured QC matrix which is greatly beneficial to linear encoding, parallel decoding and hardware implementation.}
\end{itemize}

\begin{itemize}
\item[2)]
To analyze the decoding threshold of a PLDPC-Hadamard code, we propose a modified PEXIT method.
We replace the SPC mutual information (MI) updating with our proposed Hadamard MI updating  based on Monte Carlo simulations.
Different from the EXIT method used in optimizing the degree distribution of VNs in an LDPC-Hadamard code \cite{Yue2007}, our proposed PEXIT method
 searches and analyzes protomatrices corresponding to the generalized protograph of the PLDPC-Hadamard codes. The proposed method, moreover, is applicable to analyzing  PLDPC-Hadamard codes with degree-$1$ VNs and/or punctured VNs.
Using the analytical technique, we have found PLDPC-Hadamard codes with very low decoding thresholds ($<-1.40$ dB) under different code rates.
\end{itemize}

\begin{itemize}
\item[3)]
Extensive simulations are performed under an AWGN channel. For each case, $100$ frame
errors are collected before the simulation is terminated. Results show that the PLDPC-Hadamard codes can obtain comparable BER performance to the traditional LDPC-Hadamard codes \cite{Yue2007}.
At a BER of $10^{-5}$, the gaps to the ultimate Shannon limit are $0.40$ dB for the rate-$0.0494$ code, $0.35$ dB for the rate-$0.021$ code, $0.24$ dB for the rate-$0.008$ code and $0.16$ dB for the rate-$0.003$ code, respectively.
The FER performance and the average number of decoding iterations required are
also recorded.
\end{itemize}

\begin{itemize}
\item[4)]
Punctured PLDPC-Hadamard codes are studied.
Puncturing different VNs in the protograph of a PLDPC-Hadamard code can produce quite different BER/FER performance improvement/degradation compared with the unpunctured code.
Moreover, when the order of the Hadamard code $r=5$ (i.e., $r$ is odd),
puncturing the extra degree-$1$ Hadamard VNs provided by the non-systematic Hadamard encoding is found to degrade the error performance.
\end{itemize}

The remainder of the paper is organized as follows.
Section \ref{sect:background} reviews some background knowledge
on Hadamard codes (Sect. \ref{sect:Hadamard}),
LDPC codes (Sect. \ref{sect:LDPC&LDPCH}),
LDPC-Hadamard codes (Sect. \ref{sect:LDPC-Hadamard}),
and protograph-based LDPC codes (Sect. \ref{sect:PLDPC}).
Section \ref{sect:PLDPCH} introduces our  proposed
 PLDPC-Hadamard code, including its structure, encoding and decoding methods, and code rate.  In particular, the cases in which the order of the Hadamard code used is even or odd are fully described and analyzed.
A low-complexity PEXIT method for analyzing PLDPC-Hadamard codes is proposed
and an optimization algorithm is provided.
Section~\ref{sect:sim_re} presents the protomatrices
of the PLDPC-Hadamard codes found by the proposed methods,
their decoding thresholds and simulation results.
The error performance of these codes after puncturing are further evaluated.
Section \ref{sect:conclusion} provides some final remarks.

\section{Background} \label{sect:background}
\subsection{Hadamard Codes} \label{sect:Hadamard}

A Hadamard code with an order $r$ is a class of linear block codes.
We consider a $q \times q$ positive Hadamard matrix $+\bm{H}_q = \{+\bm{h}_j, j=0,1, \dots, q-1\}$, which can be constructed recursively using
\begin{equation}
{+\bm{H}_q} = \left[ {\begin{array}{*{20}{c}}
{ + {\bm{H}_{q/2}}}&{ + {\bm{H}_{q/2}}}\\
{ + {\bm{H}_{q/2}}}&{ - {\bm{H}_{q/2}}}
\end{array}} \right]
\end{equation}
with $q = 2^r$ and $\pm\bm{H}_1 = [\pm1]$. Each column $+ \bm{h}_j$ is a Hadamard codeword and thus $\pm \bm{H}_q$ contains $2q= 2^{r+1}$ codewords $\pm \bm{h}_j$.
Note that Hadamard codewords can also be represented by mapping
 $+1$ in  $\pm \bm{h}_j$ to bit ``$0$'' and $-1$ to bit ``$1$''.


Considering an information sequence $\bm{u} \in \{0,1\}^{r+1}$ of length $r+1$ and
denoted by $\bm{u} = [u_0\ u_1\ \dots\ u_r]^T$, the Hadamard encoder encodes
 $\bm{u}$ into a codeword $\bm{c}^H$ of length $q$, i.e., $\bm{c}^H\in \{0,1\}^{2^r} = [c_0^H\ c_1^H\ \dots\ c_{2^r-1}^H]^T$,
 where $(\cdot)^T$ represents the transpose operation.
Assuming that the $+\bm{h}_j$ or  $- \bm{h}_j$ corresponding to $\bm{c}$, i.e., by mapping bit ``0'' in $\bm{c}$ to $+1$ and bit ``$1$'' to $-1$, is uniformly transmitted through an additive white Gaussian noise (AWGN) channel with mean $0$ and variance ${\sigma}_{ch}^2$, we denote the received signal by $\bm{y} = [y_0\ y_1\ \dots\ y_{2^r-1}]^T$.
Given $y_i$, the log-likelihood ratio (LLR) value $L(c^H_i \mid y_i)$ is computed by the ratio between the
conditional probabilities $\Pr( c_i^H = ``0"  \mid  y_i )$ and $\Pr( c_i^H = ``1"  \mid  y_i )$, i.e.,
\begin{equation}\label{eq:cpt_L}
L(c^H_i \mid y_i) = \ln \frac{ \Pr( c_i^H = ``0" \mid  y_i ) }{ \Pr( c_i^H = ``1" \mid  y_i ) } \;\;\;\; i = 0,1,...,2^r - 1.
\end{equation}
Applying the following Bayes' rule to \eqref{eq:cpt_L}
\begin{equation}
\Pr( c_i^H \mid  y_i ) = \frac{p( y_i \mid c_i^H  ) \cdot \Pr( c_i^H  ) }{p( y_i )},
\end{equation}
we obtain
\begin{equation}
L(c^H_i \mid y_i) = \ln \frac{ p( y_i \mid c_i^H = ``0" ) \cdot \Pr( c_i^H = ``0" ) }{ p( y_i \mid c_i^H = ``1" ) \cdot \Pr( c_i^H = ``1" ) }
\label{eq:cpt_L1}
\end{equation}
where $p( y_i \mid c_i^H = ``0" )$ and $ p( y_i \mid c_i^H = ``1"  )$  denote
the  channel output probability density functions (PDF) conditioned on the code bit $c_i^H = ``0" $ and $c_i^H = ``1" $, respectively, being transmitted;
$\Pr( c_i^H = ``0" ) $ and $\Pr( c_i^H = ``1" )$ denote the \textit{a priori} probabilities that $c_i^H = ``0" $ and $c_i^H = ``1" $ are transmitted, respectively;
and $p( y_i )$ denotes the PDF of received signal $y_i$.
We denote $L_{ch}^H(i)$ as the channel LLR value of the $i$-th bit, i.e.,
\begin{equation}\label{eq:cpt_ch}
L_{ch}^H(i) = \ln \frac{p(  y_i \mid c_i^H = ``0" )}{p( y_i \mid c_i^H = ``1" )}=\frac{2y_i}{{\sigma}_{ch}^2}, \;\;\;\; i = 0,1,...,2^r - 1;
\end{equation}
and denote ${L_{apr}^H}(i)$ as the  \textit{a priori} LLR of the $i$-th bit, i.e.,
\begin{equation}\label{eq:cpt_apr}
{L_{apr}^H}(i) = \ln \frac{\Pr(c_i^H = ``0")}{\Pr(c_i^H = ``1")}, \;\;\;\; i = 0,1,...,2^r - 1.
\end{equation}
Thus, \eqref{eq:cpt_L1}  is rewritten as
\begin{equation}\label{eq:cpt_L2}
L(c^H_i \mid y_i) = L_{ch}^H(i) + {L_{apr}^H}(i), \;\;\;\; i = 0,1,...,2^r - 1.
\end{equation}
We also define
\begin{eqnarray}
\bm{L}_{ch}^H &=& [{L}_{ch}^H(0) \; {L}_{ch}^H(1)  \; \cdots  \; {L}_{ch}^H(2^r-1)] ^T; \\
\bm{L}_{apr}^H &=& [{L}_{apr}^H(0) \; {L}_{apr}^H(1) \;  \cdots \;  {L}_{apr}^H(2^r-1)] ^T.
\end{eqnarray}

In  \cite{Li2003},  a symbol-by-symbol maximum \textit{a posteriori} probability (symbol-MAP) Hadamard decoder has been developed,
in which the \textit{a posteriori} LLR values of the code bits are computed based on the received vector $\bm{y}$. In the following, we show the steps to
derive the  \textit{a posteriori} LLR values.
\begin{enumerate}
\item
We re-write \eqref{eq:cpt_L} into
\begin{eqnarray}
&&\makebox[-2cm]{}\Pr( c_i^H = ``0" \mid  y_i ) =  \cr
&&\makebox[-0.5cm]{}  \frac{\exp(L(c^H_i \mid y_i)/2)  }{\exp(L(c^H_i \mid y_i)/2) + \exp(-L(c^H_i \mid y_i)/2)  };
\label{eq:ci=0|yi}\\
&&\makebox[-2cm]{} \Pr( c_i^H = ``1" \mid  y_i ) =  \cr
&& \makebox[-0.5cm]{}
\frac{\exp(-L(c^H_i \mid y_i)/2)  }{\exp(L(c^H_i \mid y_i)/2) + \exp(-L(c^H_i \mid y_i)/2)  }; \cr
&& \makebox[3cm]{} i = 0,1,...,2^r - 1.
\label{eq:ci=1|yi}
\end{eqnarray}
We also denote $\pm H[i, j]$ as the $i$-th bit in $\pm\bm{h}_j$.
Given $\bm{y}$ and applying \eqref{eq:ci=0|yi} and \eqref{eq:ci=1|yi},
the \textit{a posteriori} probabilities of the transmitted Hadamard codeword $\bm{c}^H$ being $+ \bm{h}_j$ or $- \bm{h}_j$ ($j = 0,1,\cdots,2^r-1$) are given by
\begin{eqnarray}\label{eq:cpt_app_yhj}
&&\Pr\left( \bm{c}^H = \pm \bm{h}_j  \mid  \bm{y} \right)  \cr
&&= \prod\limits_i{ \Pr \left( c^H_i = \pm H[i,j] \mid y_i \right) }\nonumber\\  
&&= \prod\limits_i{ \frac{ \exp \left( \pm H[i,j] \cdot L(c^H_i \mid y_i)/2 \right) }
{\exp(L(c^H_i \mid y_i)/2) + \exp(-L(c^H_i \mid y_i)/2)}  } \nonumber\\ 
&&= \kappa \cdot \gamma \left( \pm {\bm{h}_j} \right)
\end{eqnarray}
where
\begin{equation}
\kappa =\left[ \prod_i{ \left[ \exp \left(L(c^H_i \mid y_i)/2 \right) + \exp \left( -L(c^H_i \mid y_i)/2 \right) \right]} \right]^{-1}   \nonumber
\end{equation}
is independent of $\pm {\bm{h}_j}$;
\begin{equation}
\gamma \left( \pm {\bm{h}_j} \right) = \exp \left( {\frac{1}{2}\left\langle { \pm {\bm{h}_j}, \bm{L}_{ch}^H+\bm{L}_{apr}^H} \right\rangle } \right)
\label{eq:gamma}
\end{equation}
represents the \textit{a posteriori} ``information'' of the codeword $\pm\bm{h}_j$;
and $\left\langle  \cdot  \right\rangle $ denotes the inner-product operator.
\item Based on $\Pr\left( \bm{c}^H = \pm \bm{h}_j  \mid  \bm{y} \right)$,
the \textit{a posteriori} LLR of the $i$-th ($i = 0,1,...,2^r - 1$) code bit, which is denoted by $L_{app}^H(i)$, is
computed using
\begin{eqnarray}\label{eq:cpt_app}
L_{app}^H(i) &=&  \ln \frac{\Pr(c_i^H = ``0"\mid \bm{y} )}{\Pr(c_i^H = ``1"\mid \bm{y} )}\nonumber\\  
&=& \ln \frac{\sum\limits_{\pm H\left[ {i,j} \right] =  + 1} \Pr( \bm{c}^H = \pm\bm{h}_j \mid \bm{y} ) }
{\sum\limits_{\pm H\left[ {i,j} \right] =  - 1} \Pr( \bm{c}^H =  \pm\bm{h}_j \mid \bm{y} ) }\nonumber\\
&=& \ln \frac{{\sum\limits_{\pm H\left[ {i,j} \right] =  + 1} {\gamma \left( { \pm {\bm{h}_j}} \right)} }}
{{\sum\limits_{\pm H\left[ {i,j} \right] =  - 1} {\gamma \left( { \pm {\bm{h}_j}} \right)} }}.
\end{eqnarray}
\end{enumerate}
We  define
\begin{eqnarray}
\bm{L}_{app}^H &=& [{L}_{app}^H(0) \; {L}_{app}^H(1) \;  \cdots \;  {L}_{app}^H(2^r-1)] ^T.
\end{eqnarray}
Based on the butterfly-like structure of the Hadamard matrix, $\bm{L}_{app}^H$ can be computed using the fast Hadamard transform (FHT) and the dual FHT (DFHT)  \cite{Li2003,Jiang2018,Jiang2019}.
Hard decisions can then be made on $L_{app}^H(i)$ to estimate code bits.
In the case of iterative decoding, the Hadamard decoder subtracts the $L_{apr}^H(i)$ from $L_{app}^H(i)$ and feeds back ``new'' extrinsic information to other component decoders.


\subsection{LDPC Codes}\label{sect:LDPC&LDPCH}

An LDPC code with code length $N$, information length $k = N - M$ and code rate $R = k / N$ can be represented by a $M \times N$ parity-check matrix $\bm{H}_{M \times N}$ whose entries only include $0$ or $1$.
Moreover, the matrix $\bm{H}_{M \times N}$ needs to satisfy the following conditions.
\begin{enumerate}
\item The number of $``1"$s in the matrix should be much less than the number of elements $MN$, i.e., a low density of $``1"$s.
\item The codeword bits corresponding to the $``1"$s in each row of the matrix must take part in the same parity-check equation, i.e., each LDPC codeword $\bm{c}$ satisfies $\bm{c}\bm{H}_{M \times N}^T = \bm{0}$, where
$\bm{0}$ represents a zero vector of appropriate length.
\end{enumerate}

\begin{figure}[t]
\centerline{
\includegraphics[width=3.5in]{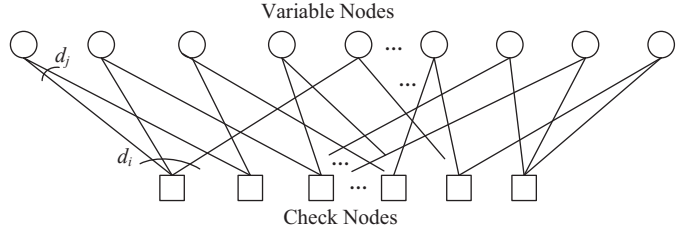}}
 \caption{Representation of an LDPC code by a Tanner graph.}
  \label{fig:Tanner_graph}
\end{figure}

The matrix $\bm{H}_{M \times N}$ can also be represented by a Tanner graph, as shown in Fig. \ref{fig:Tanner_graph}. 
The circles denote the variable nodes (VNs) corresponding to the columns of the matrix; the squares denote the check nodes (CNs) corresponding to the rows of the matrix; and the edges connecting the VNs and CNs correspond to the $``1"$s of the matrix.
Denote $\bm{\lambda} = \{\lambda_j\}$ and $\bm{\rho} = \{\rho_i\}$ as the fraction of degree-$d_j$ VNs and fraction of degree-$d_i$ CNs, respectively.
The degree distribution $(\bm{\lambda}, \bm{\rho})$ not only determines the $``1"$s distribution in $\bm{H}_{M \times N}$, but also can be used by the extrinsic information transfer (EXIT) chart technique to estimate the theoretical threshold of the LDPC code \cite{Brink2001}, \cite{Brink2004}.


\subsection{LDPC-Hadamard Codes}\label{sect:LDPC-Hadamard}
In the Tanner graph of an LDPC code, a VN with degree-$d_j$ emits $d_j ( d_j > 1 )$ edges connecting to $d_j$ different CNs and forms a $( d_j, 1 )$ repeat code;
whereas a CN with degree-$d_i$ emits $d_i ( d_i > 1 )$ edges connecting $d_i$ different VNs and forms a $( d_i, d_i - 1 )$ single-parity-check (SPC) code.
A generalized LDPC code is  obtained when  the repeat code and/or SPC code is/are replaced by other block codes.

In \cite{Yue2007},  the SPC codes of an LDPC code are replaced with Hadamard codes, forming
an LDPC-Hadamard code.
In particular,  Hadamard parity-check bits are added to the CNs in the Tanner graph such that the SPC constraints become the Hadamard constraints (see the Hadamard check node shown in Fig.~\ref{fig:PLDPCH_protograph}).
Also, with a fixed Hadamard order and a given code rate, the EXIT method is used to adjust the degree distribution of VNs. The aim is to find an optimal degree distribution of the VNs such that  the EXIT curves of the repeat codes (i.e., VNs) and Hadamard codes are matched
 under a low $E_b/N_0$.

%
%
%

\subsection{Protograph-based LDPC Codes}\label{sect:PLDPC}

When an LDPC code contains degree-$1$ VNs or punctured VNs, the traditional EXIT chart cannot evaluate its decoding performance.
However, for LDPC codes constructed based on protographs, their theoretical performance
can be estimated by the protograph EXIT (PEXIT) algorithm even if they contain degree-$1$ VNs or punctured VNs \cite{Liva2007}.


\begin{figure}[t]
\centerline{
\includegraphics[width=2.2in]{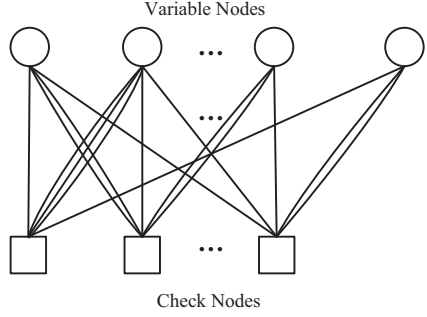}}
 \caption{ A protograph corresponding to the protomatrix in \eqref{eq:protomatrix}. }
  \label{fig:protograph}
\end{figure}

A protograph can be denoted by $G = (V, C, E)$ where $V$ is a set of VNs, $C$ is a set of CNs and $E$ is a set of edges \cite{Thorpe2003}. Fig.~\ref{fig:protograph} illustrates a protograph, and
the corresponding protomatrix (also called base matrix) is given by
\begin{equation}
{\bm{B}_{m \times n}}{\rm{ = }}\left[ {\begin{array}{*{20}{c}}
1&3& \cdots &0&1\\
2&1& \cdots &2&0\\
 \vdots & \vdots & \ddots & \vdots & \vdots \\
1&1& \cdots &1&2
\end{array}} \right]. 
\label{eq:protomatrix}
\end{equation}
The entries in $\bm{B}_{m \times n}=\{b_{i,j}: i=0,1,2\dots,m-1; j=0,1,2\dots,n-1\}$ are allowed to be larger than $1$ and they correspond to the multiple edges connecting the same pair of VN and CN in the protograph.
The parity-check matrix $\bm{H}_{M \times N}$ of a protograph-based LDPC (PLDPC) code can be constructed by expanding the protomatrix $\bm{B}_{m \times n}$ where $m \ll M$ and $n \ll N$.

To obtain a larger $\bm{H}_{M \times N}$, the following copy-and-permute operations
can be used to expand $\bm{B}_{m \times n}$.
\begin{enumerate}
\item Duplicate the protograph $z$ times.
\item Permute the edges which connect the same type of VNs and CNs among these duplicated protographs.
\end{enumerate}
This expansion process is also called lifting and the parameter $z$ is called the lifting factor.
The equivalent process in the ``matrix domain'' is to replace each  $b_{i,j}$ by
\begin{itemize}
\item a $z \times z$ zero matrix if $b_{i,j}=0$; or
\item a summation of $b_{i,j}$ non-overlapping $z \times z$
permutation matrices if $b_{i,j}\neq 0$.
\end{itemize}
As mentioned,  permutations  occur only among the edges connecting to the same type of nodes
and the lifted matrix $\bm{H}_{M \times N}$
(where $M=zm$ and $N=zn$)
keeps the same degree distribution and code rate as $\bm{B}_{m \times n}$.
The code represented by $\bm{H}_{M \times N}$  is called a PLDPC code.

To analyze the decoding performance of a PLDPC code,
the PEXIT algorithm is applied to $\bm{B}_{m \times n}$.
In the PEXIT method, the mutual information (MI) values on all types of edges are updated separately and iteratively  \cite{Liva2007}.
To illustrate the method, different types of MI are first defined as follows:
\begin{itemize}
\item $I_{ac}(i,j)$: \textit{a priori} MI from \textit{j}-th VN to \textit{i}-th CN in $\textbf{B}_{m \times n}$;
\item $I_{av}(i,j)$: \textit{a priori} MI from \textit{i}-th CN to \textit{j}-th VN in $\textbf{B}_{m \times n}$;
\item $I_{ev}(i,j)$: extrinsic MI from \textit{j}-th VN to \textit{i}-th CN in $\textbf{B}_{m \times n}$;
\item $I_{ec}(i,j)$: extrinsic MI from \textit{i}-th CN to \textit{j}-th VN in $\textbf{B}_{m \times n}$;
%
\item $I_{app}(j)$:  \textit{a posteriori} MI value of the  \textit{j}-th VN;
\item $I_{ch}$: MI from the channel.
\end{itemize}
Without going into the details,
the steps below show how to determine the threshold $(E_b/N_0)_{th}$.
\begin{enumerate}
\item Set all MI values to $0$.
\item Select a relatively large $E_b/N_0$.
\item \label{step:Eb/N0} Initialize $I_{ch}$ based on $E_b/N_0$ and the code rate.
\item \label{step:it_start} Compute $I_{ev}(i, j) \; \forall i,j$.
\item Set $I_{ac}(i, j)=I_{ev}(i, j)$.
\item Compute $I_{ec}(i, j) \; \forall i,j$.
\item \label{step:it_end} Set $I_{av}(i, j)=I_{ec}(i, j)$.
\item Repeat Steps \ref{step:it_start}) to \ref{step:it_end}) $I_{iter}$ times.
\item Compute $I_{app}(j)$.
\item If $I_{app}(j)=1 \; \forall j $, reduce $E_b/N_0$ and go to Step \ref{step:Eb/N0}); otherwise set the previous $E_b/N_0$ that achieves $I_{app}(j)=1 \; \forall j $ as the threshold $(E_b/N_0)_{th}$ and stop.
\end{enumerate}
The above analytical process can be regarded as the repeated computation and exchange between the \textit{a priori} MI matrices $\{I_{av}(i, j)\}/\{I_{ac}(i, j)\}$ and extrinsic MI matrices $\{I_{ev}(i, j)\}/\{I_{ec}(i, j)\}$. Moreover,  these matrices have the same size as $\bm{B}_{m \times n}$.
Note that the PEXIT algorithm can be used to analyze protographs with degree-$1$ VNs, i.e., columns in the protomatrix with weight $1$.
Protographs with punctured VNs will also be analyzed in a similar way, except that the code rate will be changed accordingly and the corresponding $I_{ch}$ will be initialized as $0$.

\section{Protograph-based LDPC-Hadamard Codes}\label{sect:PLDPCH}

\subsection{Code Structure}\label{sect:code_str}

\begin{figure}[t]
\centerline{
\includegraphics[width=3.0in]{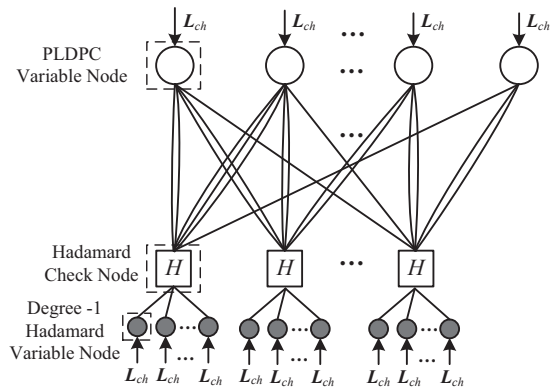}}
 \caption{ A protograph of PLDPC-Hadamard codes. }
  \label{fig:PLDPCH_protograph}
\end{figure}

We propose a variation of LDPC-Hadamard code called
 \emph{protograph-based LDPC Hadamard (PLDPC-Hadamard) code}.
Unlike LDPC-Hadamard codes which are constructed based on parity-check matrices containing
``0''s and ``1''s, PLDPC-Hadamard codes are designed based on protomatrices or protographs.
The base structure of a PLDPC-Hadamard code is shown in Fig.~\ref{fig:PLDPCH_protograph}.
 Referring to the figure, each blank circle denotes a protograph variable node (P-VN), each square with an ``H'' inside denotes a Hadamard check node (H-CN), and each filled circle denotes a degree-1 Hadamard variable node (D1H-VN). We assume that there are $n$ P-VNs and $m$ H-CNs.
The base matrix of the proposed PLDPC-Hadamard codes is then denoted by $\bm{B}_{m \times n} = \{b_{i,j}\}$ , where $b_{i,j}$ represents the number of edges connecting the $i$-th H-CN ($i = 0,1,\dots,m-1$) and the $j$-th P-VN ($j = 0,1,\dots,n - 1$).

Moreover, we denote the weight of the $i$-th row by ${d_{{c_i}}} = \sum\nolimits_{j = 0}^{n - 1} {{b_{i,j}}}$, which represents the total number of edges connecting the $i$-th H-CN to all P-VNs.
For example in Fig. \ref{fig:PLDPCH_protograph}, the number of edges connecting each of the three displayed H-CNs to all P-VNs is equal to $d_{c_i} = 6$.
These $d_{c_i}$ edges are considered as (input) information bits to the $i$-th Hadamard code while the connected D1H-VNs represent the corresponding (output) parity bits in the Hadamard code.
{\color{black}Recall that an order-$r$ Hadamard code contains $2^{r+1}$ codewords with 
each codeword containing  $r+1$ information bits.
Suppose a Hadamard code
of order-$(d_{c_i} - 1)$ is used to encode 
these $d_{c_i}$ inputs  
and generate $2^{(d_{c_i} - 1)}-d_{c_i}$ Hadamard parity-check bits.
As these $d_{c_i} = r + 1$ bits take part in the same parity-check equation of an LDPC code
and need to fulfill the SPC constraint, 
the number of possible combinations of these $d_{c_i}$ bits is only $2^{(d_{c_i} - 1)}$
and thus  $2^{(d_{c_i} - 1)}=2^r$ Hadamard codewords will be generated. 
In other words, only half the of $2^{r+1}$ available Hadamard codewords are used,
making the encoding process very inefficient.
Same as in LDPC-Hadamard codes \cite{Yue2007}, we utilize Hadamard codes with order $r = d_{c_i} - 2 \; (r > 2)$  in the proposed
PLDPC-Hadamard codes.
With such an arrangement, all possible Hadamard codewords, i.e., $2^{(d_{c_i} - 1)}=2^{r+1}$ can be utilized;
fewer Hadamard parity bits compared with the case of $r = d_{c_i} - 1$ need to be added
(only ($2^{(d_{c_i} - 2)}-d_{c_i}$) and ($2^{(d_{c_i} - 2)}-2$) Hadamard parity-check bits are generated for $r$ is even and odd, respectively); the encoding process becomes most efficient; 
 the overall code rate is increased; and the decoding performance is improved. 
 }
(Note that a Hadamard code with order $r = 2$ is equivalent to the $(4, 3)$ SPC code.
No extra parity-check bits (i.e., D1H-VNs) will be generated if such an Hadamard code is used
in the PLDPC-Hadamard code. Thus Hadamard codes with order $r = 2$ are not considered.)
{\color{black}
In the following, we consider the cases when $r$ is even and odd separately. It is because 
systematic Hadamard encoding is possible when $r$ is even and non-systematic Hadamard encoding 
needs to be used when $r$ is odd.
}

\subsubsection{$r = d_{c_i} - 2$ is an even number}
\begin{figure}[t]
\centerline{
\includegraphics[width=3.5in]{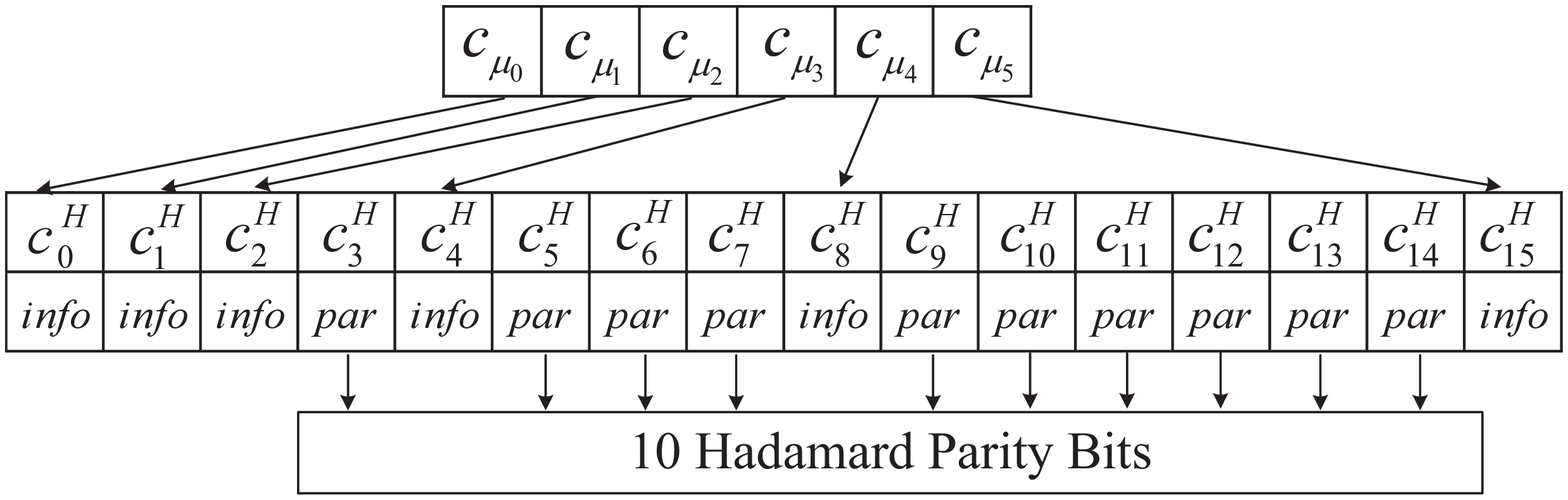}}
 \caption{Example of encoding a length-6 SPC codeword into a length-16 ($r=4$) Hadamard codeword.}
  \label{fig:r_even}
\end{figure}

We denote a Hadamard codeword by ${\bm{c}^H} = \left[ {c_0^H \ c_1^H \ \ldots c_{{2^r} - 1}^H} \right]$.
For $r$ being an even number, it has been shown that \cite{Yue2007}
\begin{equation}
\label{eq:SPC}
\left[ c_0^H  \oplus c_1^H  \oplus c_2^H \oplus \cdots\oplus  {\ c_{{2^{k-1}}}^H} \oplus \cdots  \oplus {\ c_{{2^{r-1}}}^H} \right] \oplus c_{{2^r} - 1}^H   = 0.
\end{equation}
Viewing from another perspective, if there is a length-$(r+2)$ SPC codeword denoted by ${\bm{c}_\mu } = [ c_{\mu _0} \ c_{\mu _1} \  \ldots \  c_{\mu_r} \  c_{\mu_{r + 1}} ]$, these bits can be used as inputs to a systematic Hadamard encoder and form a Hadamard codeword ${\bm{c}^H} = \left[ {c_0^H \ c_1^H \  \ldots \  c_{{2^r} - 1}^H} \right]$
where
\begin{eqnarray}
c_0^H&=&c_{\mu _0}  \nonumber\\
c_1^H &=& c_{\mu _1} \cr
&\vdots& \cr
c_{{2^{k-1}}}^H&=&c_{\mu_k} \label{eq:cH}\\
&\vdots& \cr
c_{{2^{r-1}}}^H&=&c_{\mu_r} \nonumber\\
c_{{2^{r}-1}}^H&=&c_{\mu_{r+1}} \nonumber
\end{eqnarray}
correspond to $r+2$ P-VNs and the remaining Hadamard parity bits in ${\bm{c}^H}$ correspond to $2^r-(r+2)$ D1H-VNs.
Fig.~\ref{fig:r_even} shows an example in which a $(6, 5)$ SPC codeword is encoded into a length-16 ($r=4$) Hadamard codeword.
In this case,
$c_0^H = {c_{{\mu _0}}},c_{\rm{1}}^H = {c_{{\mu _1}}},c_{\rm{2}}^H = {c_{{\mu _{\rm{2}}}}},c_{\rm{4}}^H = {c_{{\mu _{\rm{3}}}}},c_{\rm{8}}^H = {c_{{\mu _{\rm{4}}}}},c_{{\rm{15}}}^H = {c_{{\mu _{\rm{5}}}}}$ and the other $10$ ($=2^r - (r+2)$) code bits are
Hadamard parity bits.
We call this systematic Hadamard encoding when the original information bits can be exactly located in the codeword.

Referring to Fig.~\ref{fig:PLDPCH_protograph}, the links connecting the P-VNs to the $i$-th H-CN always form a SPC.
These links can make use of the above mechanism to derive the parity bits of the Hadamard code (denoted as D1H-VNs of the Hadamard check node in Fig.~\ref{fig:PLDPCH_protograph}) if $d_{c_i}$ is even.
In this case, the Hadamard code length equals $2^{d_{c_i}-2}$, and the number of D1H-VNs equals $2^{d_{c_i}-2} - d_{c_i}$.
Assuming $d_{c_i}$ is even for all $i=0,1,\ldots,m-1$, the total number of D1H-VNs is given by \begin{equation} \sum\limits_{i = 0}^{m-1} {\left( {{2^{{d_{{c_i}}} - 2}} - {d_{{c_i}}}} \right)}.
\end{equation} 
When all VNs are sent to the channel, the code rate of the
protograph given in Fig.~\ref{fig:PLDPCH_protograph} equals
\begin{equation}
R^{\rm even} = \frac{{n - m}}{{\sum\limits_{i = 0}^{m-1} {\left( {{2^{{d_{{c_i}}} - 2}} - {d_{{c_i}}}} \right)}  + n}}. 
\end{equation}
If we further assume that all rows in $\bm{B}_{m \times n}$ have the same weight  which is equal to $d$, i.e., $d_{c_i}=d$ for all $i$, the code rate is simplified to
\begin{equation}
R_{d_{c_i}=d}^{\rm even} = \frac{{n - m}}{{m \left( {{2^{{d} - 2}} - {d}} \right)}  + n}. \label{eq:even_R} 
\end{equation}
When $n_p (<n)$ P-VNs are punctured, the code rate further becomes
\begin{equation}
\label{eq:even_R_punc}
R_{\rm punctured}^{\rm even} = \frac{{n - m}}{{m\left( {{2^{d - 2}} - d} \right) + n - n_p }}.
\end{equation}

\subsubsection{$r = d_{c_i} - 2$ is an odd number}\label{sect:r=odd}
\begin{figure}[t]
\centerline{
\includegraphics[width=2.2in]{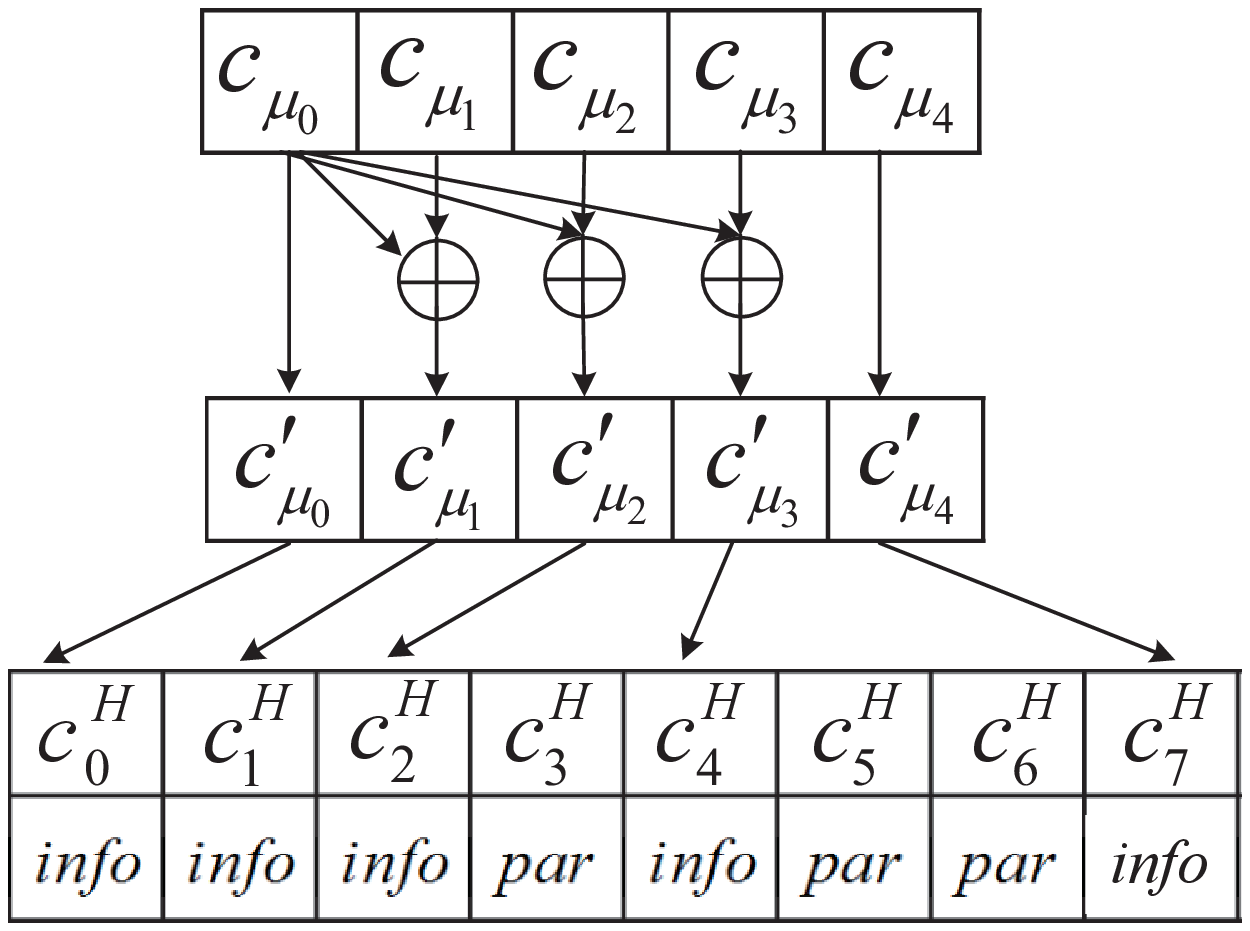}}
\caption{ Example of encoding a length-5 SPC codeword into a length-8 ($r=3$) Hadamard codeword.}
\label{fig:r_odd}
\end{figure}

For $r$ being an odd number, the $2^r$ Hadamard codewords in
$+\bm{H}_q$ can satisfy \eqref{eq:SPC} but all the $2^r$ Hadamard codewords in
$-\bm{H}_q$ do not satisfy \eqref{eq:SPC}.
We apply the same non-systematic encoding method in \cite{Yue2007} to encode the SPC codeword.
Supposing ${\bm{c}_\mu }$ is a SPC codeword, we preprocess
${\bm{c}_\mu } = \left[ {c_{{\mu _0}} \ {c_{{\mu _1}}}  \ldots \ {c_{{\mu _{r + 1}}}}} \right]$ to obtain
${\bm{c}'_\mu } = \left[ {{{c'}_{{\mu _0}}}\ {{c'}_{{\mu _1}}}\ldots \ {{c'}_{{\mu _{r + 1}}}}} \right],$
and then we perform Hadamard encoding for ${\bm{c}'_\mu }$ to obtain $\bm{c}^H = \left[ {c_0^H \ c_1^H \ \ldots c_{{2^r} - 1}^H} \right]$, where
\begin{eqnarray}
c_0^H\ =& c'_{\mu_0}&=\ c_{\mu_0}  \nonumber\\
c_1^H\ =& c'_{\mu_1} &=\ c_{\mu_1} \oplus c_{\mu_0} \cr
&\vdots& \cr
c_{{2^{k-1}}}^H\ =&c'_{\mu_k} &=\ c_{\mu_k} \oplus c_{\mu_0} \label{eq:c'}\\
&\vdots& \cr
c_{{2^{r-1}}}^H\ =&{c'_{\mu_r}} &=\ c_{\mu_r} \oplus c_{\mu_0} \nonumber\\
c_{{2^r-1}}^H\ =&c'_{\mu_{r+1}}&=\ c_{\mu_{r+1}}. \nonumber
\end{eqnarray}
%
\footnote{Note that there are other non-systematic encoding methods, e.g.,
preprocess ${\bm{c}_\mu } = \left[ {c_{{\mu _0}} \ {c_{{\mu _1}}}  \ldots \ {c_{{\mu _{r + 1}}}}} \right]$ to obtain
${\bm{c}'_\mu } = \left[ {{c'_{\mu_0}}\ {c'_{{\mu _1}}}\ldots \ {c'_{{\mu_{r + 1}}}}} \right]$,
where
$c'_{\mu_i}=c_{\mu _i} \; \text{for} \; i=0,1,2,\ldots,r$; and
${c'_{\mu_{r+1}}}={c_{\mu_{r+1}}} \oplus c_{\mu_0}$.}
Fig.~\ref{fig:r_odd} shows an example in which a (5,4) SPC codeword  is encoded into a length-8 ($r = 3$) Hadamard codeword.

{\color{black}
We transmit all the Hadamard code bits except the original information bits.
It can be seen from \eqref{eq:c'} that after the non-systematic encoding,
only the first and last code bits are the same as the original information bits, i.e., $c_0^H = c'_{{\mu _0}} = c_{{\mu _0}}$ and ${\ c_{{2^{r}-1}}^H} = c'_{{\mu _{r+1}}} = c_{{\mu _{r+1}}}$.
Thus we send the remaining code bits, i.e., $c_1^H$ to ${\ c_{{2^{r}-2}}^H}$,
%
to provide more channel observations for the decoder and the number of D1H-VNs equals $2^{{d_{{c_i}}} - 2} -2$.
For example, the code bits $\left[ {c_1^H \ c_2^H \ c_3^H \ c_4^H \ c_5^H \ c_6^H} \right]$  shown in Fig.~\ref{fig:r_odd} will be sent. 


Assuming  all the rows in $\bm{B}_{m \times n}$ have the same weight $d$, the code rate is given by
\begin{equation}
\label{eq:odd2_R}
R_{d_{c_i}=d}^{\rm odd} = \frac{{n - m}}{{m \left( {{2^{{d} - 2}} - 2} \right)}  + n}. 
\end{equation}
If $n_p \; (<n)$ P-VNs are punctured,
the code rate becomes
\begin{equation}
\label{eq:odd2_R_punc}
R_{\rm punctured}^{\rm odd} = \frac{{n - m}}{{m\left( {{2^{d - 2}} - 2} \right) + n - n_p }}.
\end{equation}
Note that for $k=1,2,\dots,r$,
\begin{itemize}
\item $c_{2^{k-1}}^H = {c'_{{\mu _k}}} = c_{\mu_k} \oplus c_0$ and hence $ c_{\mu_k}  = c_{2^{k-1}}^H \oplus c_0$;
\item $c_{\mu_k}$ is transmitted as P-VN; and
\item $c_{2^{k-1}}^H$ is transmitted as D1H-VN.
\end{itemize}
Thus the $r$ information bits $c_{\mu_k}$ can have both the \textit{a priori} information provided by the extrinsic information from P-VNs and the channel information of $c_{2^{k-1}}^H = {c'_{{\mu _k}}} = c_{\mu_k} \oplus c_0$ from D1H-VNs.
However, the two information bits $c_{\mu_0}$ and $c_{\mu_{r+1}}$ only have the \textit{a priori} information from P-VNs and the $2^r-(r+2)$ Hadamard parity bits only have the channel information from D1H-VNs.
Supposing for every H-CN,
$n_h \; (\le r)$ D1H-VNs corresponding to code bits $c_{2^{k-1}}^H$ ($k=1,2,\dots,r$) are also punctured.
The code rate further becomes
\begin{equation}
\label{eq:odd3_R_punc}
R_{\rm punctured \; D1H-VN}^{\rm odd} = \frac{{n - m}}{{m\left( {{2^{d - 2}} - 2 - n_h} \right) + n - n_p }}.
\end{equation}
}

\subsection{Decoder of the PLDPC-Hadamard Codes}\label{PLDPCH-dec}
\begin{figure}[t]
\centerline{
\includegraphics[width=0.5\columnwidth]{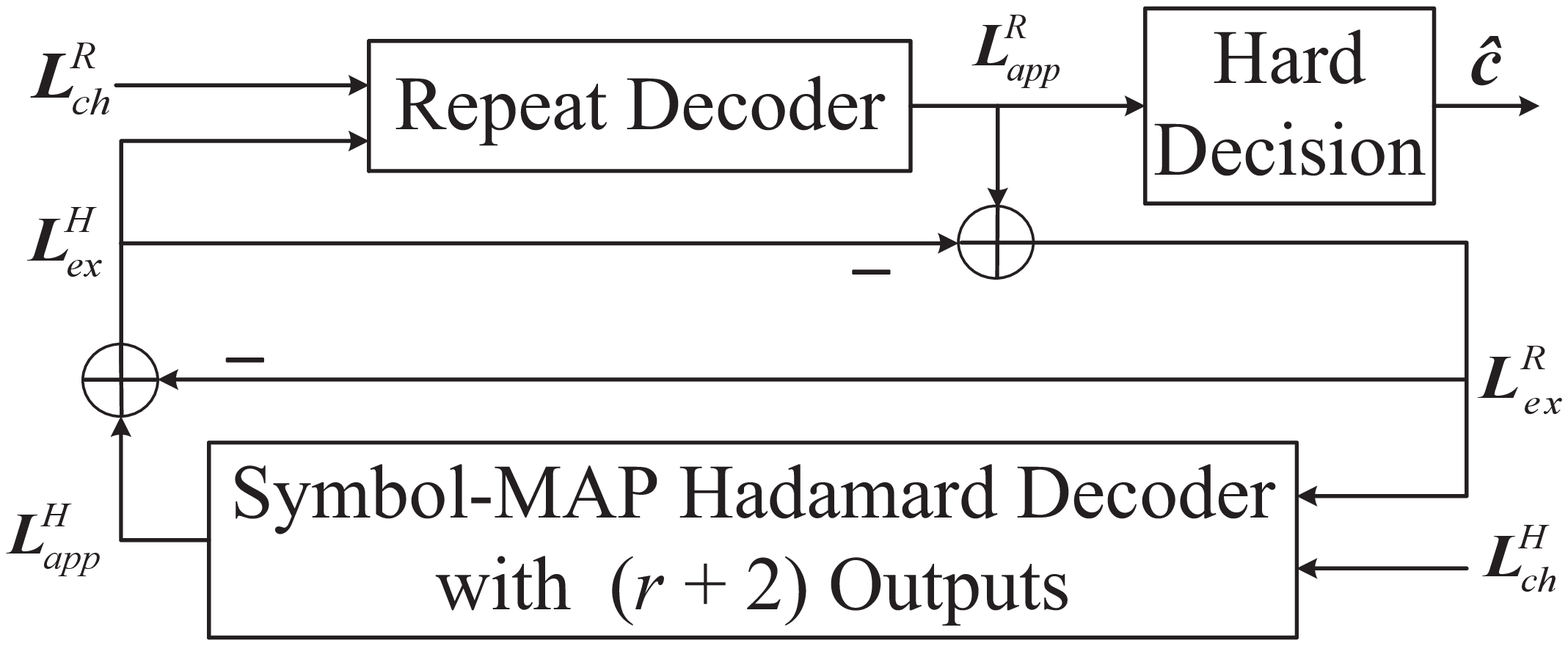}}
\caption{Block diagram of a PLDPC-Hadamard decoder. The repeat decoder is the same as the variable-node processor used in LDPC decoder.
{\color{black}
For the symbol-MAP Hadamard decoder, the number of outputs is always $r+2$; the number of inputs is $2^r$ when $r$ is even;
 the number of inputs is $2^r+r$ when $r$ is odd.}}
\label{fig:dec_PLDPCH}
\end{figure}

To evaluate the performance of PLDPC-Hadamard codes,
the iterative decoder shown in Fig.~\ref{fig:dec_PLDPCH} is used.
It consists of a repeat decoder and a symbol-MAP Hadamard decoder.
The repeat decoder is the same as the variable-node processor used in an LDPC decoder and is therefore not described here.

As described in the previous section, each H-CN with an  order-$r$ Hadamard constraint is  connected to $r+2$ P-VNs in the protograph of a PLDPC-Hadamard code.
The symbol-MAP Hadamard decoder of order-$r$ has a total of $2^r$ or $2^r+r$ inputs, among which $r+2$ come from the repeat decoder and are updated in each iteration;
and the remaining inputs come from the channel LLR information which do not change during the iterative process.
Moreover, the symbol-MAP Hadamard decoder will produce $r+2$ extrinsic LLR outputs which are fed back to the repeat decoder.
The iterative process between the repeat decoder and symbol-MAP Hadamard decoder continues until the information bits corresponding to all Hadamard codes (after hard decision) become valid SPCs or the maximum number of iterations has been reached.
In the following, we show the details of the operations of the symbol-MAP Hadamard decoder.

\subsubsection{$r$ is an even number}
\begin{figure}[t]
\centerline{
\includegraphics[width=0.6\columnwidth]{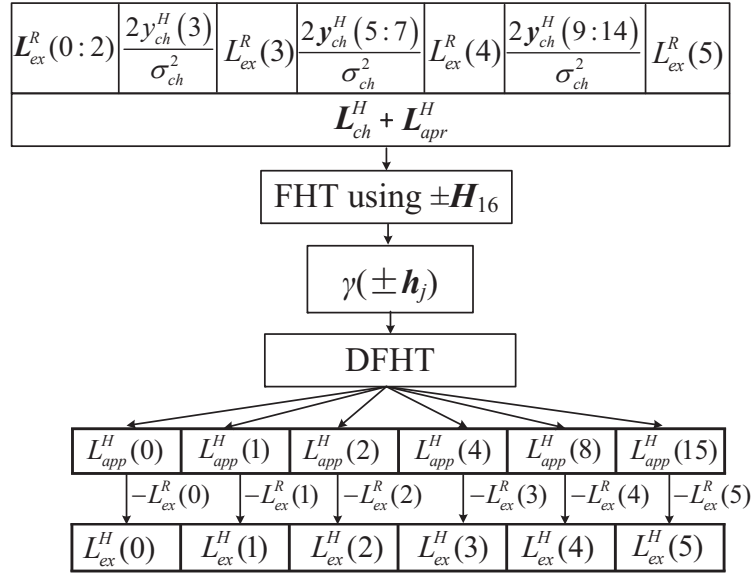}}
 \caption{Operations in the symbol-MAP Hadamard decoder for $r=4$, i.e., $16$ LLR inputs and $6$ output LLR values for the information bits.}
  \label{fig:dec_even}
\end{figure}

A H-CN has $r + 2$ links to P-VNs and is connected to $2^r - (r + 2)$ D1H-VNs.
Specifically, we denote
\begin{itemize}
\item $\bm{L}_{ex}^R = [ L_{ex}^R(0) \ L_{ex}^R(1) \ \cdots \ L_{ex}^R(r+1) ]^T$ as the $r+2$ extrinsic LLR information values coming from the repeat decoder (P-VNs),
\item $\bm{L}_{apr}^H = [L_{apr}^H(0) \ L_{apr}^H(1) \ \cdots \ L_{apr}^H(2^r - 1) ]^T$ as the $2^r$ \textit{a priori} LLR values of $\bm{c}^H$,
\item $\bm{y}_{ch}^H = [ y_{ch}^H(0) \ y_{ch}^H(1) \ \cdots \ y_{ch}^H(2^r  - 1) ]^T$ as the length-$2^r$ channel observation vector corresponding to $\bm{c}^H$ and is derived from the D1H-VNs
(note that $r+2$ channel observations are zero),
\item $\bm{L}_{ch}^H = [ L_{ch}^H(0) \ L_{ch}^H(1) \ \cdots \ L_{ch}^H(2^r  - 1) ]^T$ as the  length-$2^r$  channel LLR observations corresponding to $\bm{c}^H$.
\end{itemize}
Based on \eqref{eq:cH} and the transmission mechanism,  \textit{a priori} LLR values exist
only for the $r+2$ information bits in $\bm{c}^H$ and they are equal to
the extrinsic LLR values $\bm{L}_{ex}^R$.
Correspondingly,  channel LLR values only exist for the $2^r-r-2$ Hadamard parity bits in $\bm{c}^H$ and they are obtained from the received channel observations $\bm{y}_{ch}^H$.
{\color{black}In other words, only  $2^r-r-2$ entries in $\bm{y}_{ch}^H$ and also $\bm{L}_{ch}^H$ are non-zero.} 
Thus the entries of
$\bm{L}_{ch}^H$ and $\bm{L}_{apr}^H$ are assigned as
\begin{eqnarray}
&&\!\!\!\!\!\!\!\!\!\!\!\!\!
\left\{
\begin{array}{ll}
L_{apr}^H( k ) = L_{ex}^R(0) \nonumber \\[3pt]
L_{ch}^H( k ) = \frac{ 2y^H_{ch}(0) }{ \sigma_{ch}^2 }  = 0
\end{array}
{\rm for} \; k = 0\right.\\
&&\!\!\!\!\!\!\!\!\!\!\!\!\!
\left\{
\begin{array}{ll}
L_{apr}^H( k ) = L_{ex}^R(i) \nonumber \\[3pt]
L_{ch}^H( k ) = \frac{ 2y^H_{ch}( k ) }{ \sigma_{ch}^2 }  = 0
\end{array}
{\rm for} \; k = 1,2,\cdots,2^{i-1},\cdots, 2^{r-1}\right.\\
&&\!\!\!\!\!\!\!\!\!\!\!\!\!
\left\{
\begin{array}{ll}
L_{apr}^H( k ) = L_{ex}^R(r+1) \nonumber \\[3pt]
L_{ch}^H( k ) = \frac{ 2y^H_{ch}( k ) }{ \sigma_{ch}^2 }  = 0
\end{array}
{\rm for} \; k = 2^r-1\right.\\
&&\!\!\!\!\!\!\!\!\!\!\!\!\!
\left\{
\begin{array}{ll}
L_{apr}^H( k ) = 0  \nonumber \\[3pt]
L_{ch}^H( k ) = \frac{ 2y^H_{ch}( k ) }{ \sigma_{ch}^2 }
\end{array}
{\ \ \ \ \ \ \rm for\ the\ } 2^r-r-2 {\ \rm remaining } \; k. \right.\\
\end{eqnarray}
The symbol-MAP Hadamard decoder then computes the \textit{a posteriori} LLR ($\bm{L}_{app}^H$) of the code bits using \eqref{eq:gamma} and \eqref{eq:cpt_app}.
By subtracting the \textit{a priori} LLR values from the \textit{a posteriori} LLR values, the extrinsic LLR values ($\bm{L}_{ex}^H$) can be obtained.
Fig.~\ref{fig:dec_even} illustrates the flow of the computation of $\bm{L}_{app}^H$ and hence $\bm{L}_{ex}^H$
for $r=4$, which corresponds to $r+2=6$ information bits (and $2^r-(r+4)=10$ Hadamard parity bits).

{\color{black}
\subsubsection{$r$ is an odd number}
A H-CN is connected to $r+2$ P-VNs and $2^r - 2$ D1H-VNs, and
the bits corresponding to the $r+2$ P-VNs form a SPC codeword $\bm{c}_{\mu}$.
Similar to the ``$r$ is an even number" case, we denote
\begin{itemize}
\item $\bm{L}_{ex}^R = [ L_{ex}^R(0) \ L_{ex}^R(1) \ \cdots \ L_{ex}^R(r+1) ]^T$ as the $r+2$ extrinsic LLR information values coming from the repeat decoder (P-VNs),
\item $\bm{L}_{apr}^H = [L_{apr}^H(0) \ L_{apr}^H(1) \ \cdots \ L_{apr}^H(2^r - 1) ]^T$ as the $2^r$ \textit{a priori} LLR values of $\bm{c}^H$,
\item $\bm{y}_{ch}^H = [ y_{ch}^H(0) \ y_{ch}^H(1) \ \cdots \ y_{ch}^H(2^r  - 1) ]^T$ as the length-$2^r$ channel observation vector corresponding to $\bm{c}^H$ and is derived from the D1H-VNs
(note that the first and the last channel observations are zero),
\item $\bm{L}_{ch}^H = [ L_{ch}^H(0) \ L_{ch}^H(1) \ \cdots \ L_{ch}^H(2^r  - 1) ]^T$ as the  length-$2^r$  channel LLR observations corresponding to $\bm{c}^H$.
\end{itemize}
Since non-systematic Hadamard code is used,
$\bm{c}_{\mu}$ does not represent the information bits in $\bm{c}^H$ for {\color{black}$c_{\mu_0}=``1"$}.
Thus, we cannot directly apply \eqref{eq:cpt_app} to obtain the \textit{a posteriori} LLR of $\bm{c}_{\mu}$.
Here, we present the decoding steps when $r$ is odd.
Detailed derivations are shown in the Appendix \ref{app:a}.

Referring to \eqref{eq:c'}, the assignment of $\bm{L}_{apr}^H$ depends on $c_{\mu_0}$.
For convenience of explanation, we denote $\bm{L}_{apr}^{+H} / \bm{L}_{apr}^{-H}$  as the assignment of $\bm{L}_{apr}^H$ for $c_{\mu_0} = ``0" / ``1"$, respectively. We use \eqref{assign_apr} to assign $\bm{L}_{apr}^{\pm H}$ and \eqref{assign_ch} to assign $\bm{L}_{ch}^{H}$.
Since the first bit in all $+\bm{h}_j / -\bm{h}_j$ is $``0" / ``1"$ ($+1$ mapped to $``0"$ and $-1$ to $``1"$), we apply $\bm{L}_{apr}^{\pm H}$ and $\bm{L}_{ch}^H$ to compute $\gamma(\pm \bm{h}_j)$, i.e.,
\begin{equation}
\gamma \left( { \pm {\bm{h}_j}} \right) = \exp \left( {\frac{1}{2}\left\langle { \pm {\bm{h}_j},\bm{L}_{ch}^H+\bm{L}_{apr}^{\pm H}} \right\rangle } \right).
\label{eq:gamma_p}
\end{equation}
We define the $r+2$ \textit{a posteriori} LLR values ($\bm{L}_{app}^H$) of the original bits $\bm{c}_\mu$ by
\begin{eqnarray}\label{eq:L_app^H}
&&\!\!\!\!\!\!\!\!\!\!\!\!\!\!\!\!\!\!\!\!\!
\bm{L}_{app}^H = [ L_{app}^H(0) \ L_{app}^H(1) \ \cdots \ L_{app}^H(2^{i-1}) \ \cdots \nonumber \\
&&\qquad\qquad\qquad\qquad
L_{app}^H(2^{r-1})  \ L_{app}^H(2^r - 1) ]^T.
\end{eqnarray}
We use \eqref{eq:gamma} and \eqref{eq:cpt_app} to compute $L_{app}^H(0)$ and $L_{app}^H(2^r - 1)$;
\eqref{eq:rela_h2} to obtain $\gamma'(-\bm{h}_j)$ and then DFHT to compute \eqref{eq:cpt_app_k_2} to obtain $L_{app}^H(2^{k-1})$ $k = 1, 2, \cdots,r$.
Fig.~\ref{fig:gamma} illustrates  for the case $r=3$, the transformation from $\gamma(-\bm{h}_j)$ to $\gamma'(-\bm{h}_j)$, i.e., $\gamma'(-\bm{h}_j)=\gamma(-\bm{h}_{2^r-1-j})$.
\begin{figure}[t]
\centerline{
\includegraphics[width=0.6\columnwidth]{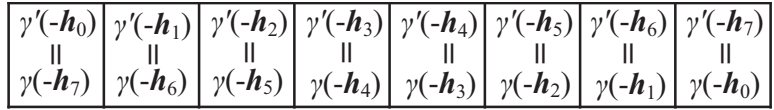}}
 \caption{Illustration of $\gamma'(-\bm{h}_j)=\gamma(-\bm{h}_{2^r-1-j})$ for $r=3$.}
  \label{fig:gamma}
\end{figure}
Then $\bm{L}_{ex}^H = \bm{L}_{app}^H -  \bm{L}_{ex}^R$ of length $r+2$ is computed and fed back to the repeat decoder.
The steps to compute $\bm{L}_{ex}^H $ for the case $r=3$ is shown in Fig.~\ref{fig:dec_odd_1}.
\begin{figure*}[t]
\centerline{
\includegraphics[width=0.9\columnwidth]{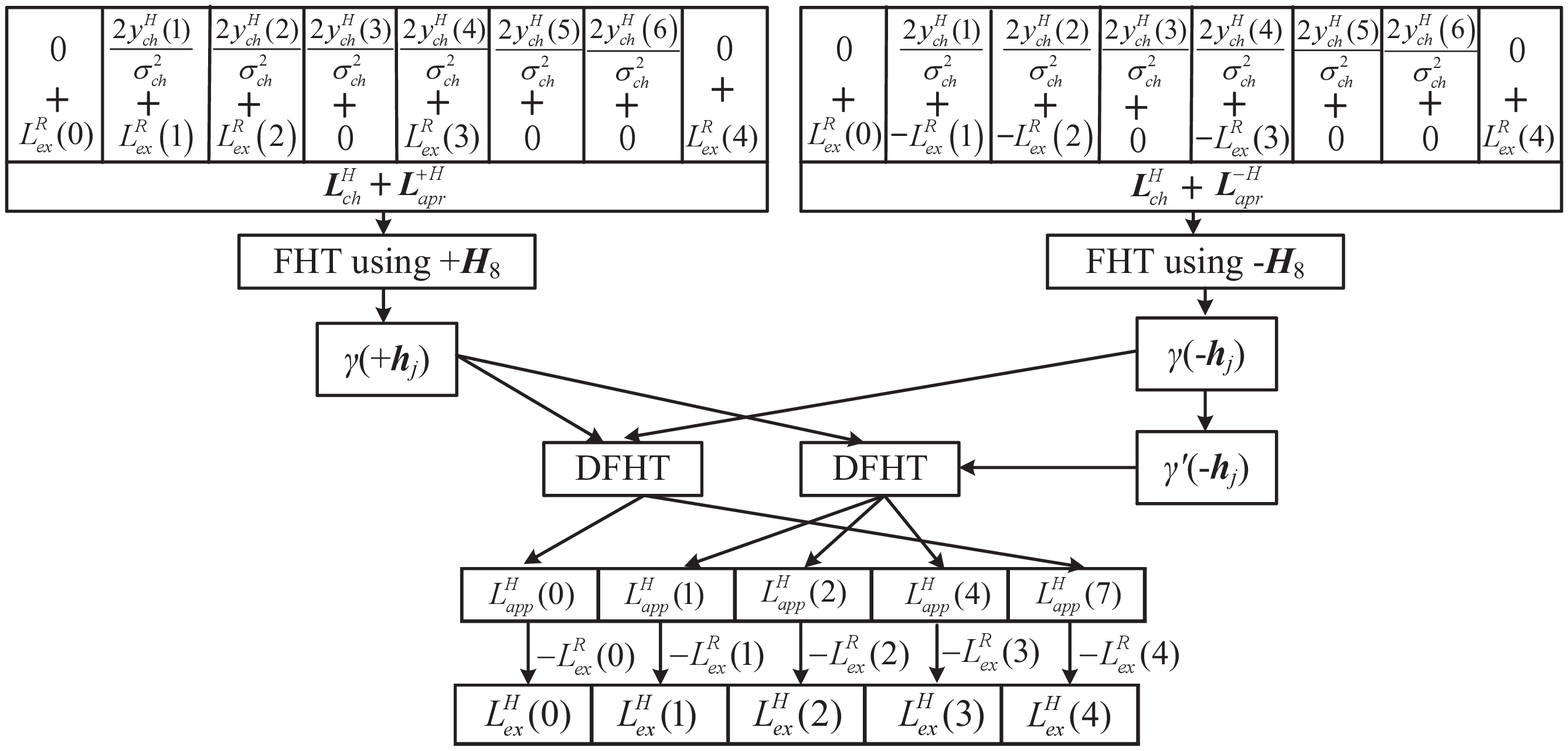}}
 \caption{Operations in the symbol-MAP Hadamard decoder for $r=3$, i.e., $11$ LLR inputs and $5$ output LLR values for the information bits.}
  \label{fig:dec_odd_1}
\end{figure*}


\noindent
\textit{Remark}: If some more bits in $c_{2^{k-1}}^H$ for $k = 1, 2, \ldots,r$ are punctured, the corresponding channel observation $y_{ch}^H(2^{k-1})$ and
LLR values of $L_{ch}^H(2^{k-1})$ are set to $0$ and the overall code rate will slightly increase.



\subsection{Code Design Optimization}

We propose a low-complexity PEXIT algorithm for analyzing PLDPC-Hadamard codes.
This technique can analyze protomatrices containing degree-$1$ and/or punctured VNs in the protographs.
It will also produce different analytical results for protomatrices with the same weight distribution but different structures. 
{\color{black}
Our low-complexity PEXIT algorithm uses the same MI updating method as the original PEXIT algorithm \cite{Liva2007} for the PVNs. However, our algorithm computes extrinsic MI for the symbol-MAP Hadamard decoder whereas the original PEXIT algorithm computes  extrinsic MI for the SPC decoder. 
We use Monte Carlo method in obtaining the extrinsic MI values of the symbol-MAP Hadamard decoder. The algorithm not only has a low complexity, but also is generic and applicable to analyzing both
systematic and non-systematic Hadamard codes.
}

We define the following symbols.
\begin{itemize}
\item $I_{av}(i,j)$: the \textit{a priori} mutual information (MI)  from the $i$-th H-CN to the  $j$-th P-VN
\item $I_{ev}(i,j)$: extrinsic MI from the $j$-th P-VN to the $i$-th H-CN
\item $I_{ah}(i,k)$: the \textit{a priori} MI of the $k$-th information bit in the $i$-th H-CN
\item $I_{eh}(i,k)$: extrinsic MI of the $k$-th information bit in the $i$-th H-CN
\item $I_{app}(j)$ the \textit{a posteriori} MI of the $j$-th P-VN
\end{itemize}

Referring to Fig.~\ref{fig:PLDPCH_protograph}, the channel LLR value $L_{ch}$ follows a normal distribution  $\mathcal{N}(\sigma_{L_{ch}}^{2}/2,\sigma_{L_{ch}}^{2})$ where $\sigma_{L_{ch}}^2 = 8R \cdot E_b/N_0$, $R$ is the code rate, and $E_b/N_0$ denotes the bit-energy-to-noise-power-spectral-density ratio.

When the output MI of a decoder is $I$, the corresponding LLR values of the extrinsic information obeys a Gaussian distribution of $(\pm \sigma^2/2, \sigma^2)$.
The relationship between $I$ and $\sigma$ can be approximately computed by functions $I = J(\sigma)$ and $\sigma = J^{-1}(I)$ \cite{Fang2015,Brink2001}, i.e.,
\begin{equation}\label{eq:J}
J(\sigma)=\left\{ \begin{array}{ll}
a_{1}\sigma^{3}+b_{1}\sigma^{2}+c_{1}\sigma, \!\!& \textrm{$0\leq \sigma\leq 1.6363$}\\
1 - e^{(a_{2}\sigma^{3}+b_{2}\sigma^{2}+c_{2}\sigma + d_{2})},   \!\!\!\!& \textrm{$1.6363<\sigma<10$}\\
1,&\textrm{$10  \le \sigma$}
\end{array} \right.
\end{equation}
and
\begin{equation}\label{eq:J1}
J^{-1}(I)=\left\{ \begin{array}{ll}
a'_{1}I^{2}+b'_{1}I+c'_{1}\sqrt{I}, & \textrm{$0\leq I\leq 0.3646$}\\
-a'_{2}\ln[b'_{2}(1-I)]-c'_{2}I,   & \textrm{$0.3646<I<1$}
\end{array} \right.
\end{equation}
where
\begin{itemize}
\item $a_{1} = -0.0421061$, $a_{2} = 0.00181491$,
$b_{1} = 0.209252$, $b_{2} = -0.142675$,
$c_{1} = -0.00640081$, $c_{2} = -0.0822054$, $d_{2} = 0.0549608$; and
\item  $a'_{1} = 1.09542$, $a'_{2} = 0.706692$,
$b'_{1} = 0.214217$, $b'_{2} = 0.386013$,
$c'_{1} = 2.33727$ and $c'_{2} = -1.75017$.
\end{itemize}



 \begin{figure*}[t]
 \begin{eqnarray}
\bm{B}_{3 \times 4} &=&  \left[ {\begin{array}{*{20}{c}}
{\rm{2}}&{\rm{0}}&{\rm{2}}&{\rm{2}}\\
{\rm{0}}&{\rm{2}}&{\rm{2}}&{\rm{2}}\\
{\rm{3}}&{\rm{2}}&{\rm{0}}&{\rm{1}}
\end{array}} \right] \label{eq:example} \\
I_{ev}(i,j) &=& J\left( {\sqrt {\sum\limits_{s \ne i} {{b_{s,j}}} {{\left( {{J^{ - 1}}\left( {{I_{av}}\left( {s,j} \right)} \right)} \right)}^2} + \left( {{b_{i,j}} - 1} \right) \cdot {{{\left( {{J^{ - 1}}\left( {{I_{av}}\left( {i,j} \right)} \right)} \right)}^2} + \sigma_{L_{ch}}^{2} } } }
 \right) \label{eq:I_ev} \\
I_{ev}&=&\left[ {\begin{array}{*{20}{c}}
{{I_{ev}}\left( {0,0} \right)}&{\rm{0}}&{{I_{ev}}\left( {0,2} \right)}&{{I_{ev}}\left( {0,3} \right)}\\
{\rm{0}}&{{I_{ev}}\left( {1,1} \right)}&{{I_{ev}}\left( {1,2} \right)}&{{I_{ev}}\left( {1,3} \right)}\\
{{I_{ev}}\left( {2,0} \right)}&{{I_{ev}}\left( {2,{\rm{1}}} \right)}&{\rm{0}}&{{I_{ev}}\left( {2,3} \right)}
\end{array}} \right]   \label{eq:I_ev_ex} \\
I_{ah} &=& \left[ {\begin{array}{*{20}{c}}
{{I_{ah}}\left( {0,0} \right)}&{{I_{ah}}\left( {0,1} \right)}&{{I_{ah}}\left( {0,2} \right)}&{{I_{ah}}\left( {0,3} \right)}&{{I_{ah}}\left( {0,4} \right)}&{{I_{ah}}\left( {0,5} \right)}\\
{{I_{ah}}\left( {1,0} \right)}&{{I_{ah}}\left( {1,1} \right)}&{{I_{ah}}\left( {1,2} \right)}&{{I_{ah}}\left( {1,3} \right)}&{{I_{ah}}\left( {1,4} \right)}&{{I_{ah}}\left( {1,5} \right)}\\
{{I_{ah}}\left( {2,0} \right)}&{{I_{ah}}\left( {2,1} \right)}&{{I_{ah}}\left( {2,2} \right)}&{{I_{ah}}\left( {2,3} \right)}&{{I_{ah}}\left( {2,4} \right)}&{{I_{ah}}\left( {2,5} \right)}
\end{array}} \right] \cr
&=&\left[ {\begin{array}{*{20}{c}}
{{I_{ev}}\left( {0,0} \right)}&{{I_{ev}}\left( {0,0} \right)}&{{I_{ev}}\left( {0,2} \right)}&{{I_{ev}}\left( {0,2} \right)}&{{I_{ev}}\left( {0,3} \right)}&{{I_{ev}}\left( {0,3} \right)}\\
{{I_{ev}}\left( {1,1} \right)}&{{I_{ev}}\left( {1,1} \right)}&{{I_{ev}}\left( {1,2} \right)}&{{I_{ev}}\left( {1,2} \right)}&{{I_{ev}}\left( {1,3} \right)}&{{I_{ev}}\left( {1,3} \right)}\\
{{I_{ev}}\left( {2,0} \right)}&{{I_{ev}}\left( {2,0} \right)}&{{I_{ev}}\left( {2,0} \right)}&{{I_{ev}}\left( {2,1} \right)}&{{I_{ev}}\left( {2,1} \right)}&{{I_{ev}}\left( {2,3} \right)}
\end{array}} \right]  \label{eq:I_ah} \\
{I_E} &=& \frac{1}{2}\sum\limits_{x \in \{ 0, 1\} } {\int_{ - \infty }^\infty  {{p_e}\left( {\xi |X = x} \right)  } }
{{{\log }_2}\frac{{2 \cdot {p_e}\left( {\xi |X = x} \right)}}{{{p_e}\left( {\xi |X =  ``0"} \right) + {p_e}\left( {\xi |X = ``1"} \right)}}}d\xi  \label{eq:I_E} \\
 I_{eh}
&=& \left[ {\begin{array}{*{20}{c}}
{{I_{eh}}\left( {0,0} \right)}&{{I_{eh}}\left( {0,1} \right)}&{{I_{eh}}\left( {0,2} \right)}&{{I_{eh}}\left( {0,3} \right)}&{{I_{eh}}\left( {0,4} \right)} &{{I_{eh}}\left( {0,5} \right)}\\
{{I_{eh}}\left( {1,0} \right)}&{{I_{eh}}\left( {1,1} \right)}&{{I_{eh}}\left( {1,2} \right)}&{{I_{eh}}\left( {1,3} \right)}&{{I_{eh}}\left( {1,4} \right)} &{{I_{eh}}\left( {1,5} \right)}\\
{{I_{eh}}\left( {2,0} \right)}&{{I_{eh}}\left( {2,1} \right)}&{{I_{eh}}\left( {2,2} \right)}&{{I_{eh}}\left( {2,3} \right)}&{{I_{eh}}\left( {2,4} \right)} &{{I_{eh}}\left( {2,5} \right)}
\end{array}} \right] \label{I_eh}  \\
I_{av}&=&\left[ {\begin{array}{*{20}{c}}
{{I_{av}}\left( {0,0} \right)}&{\rm{0}}&{{I_{av}}\left( {0,2} \right)}&{{I_{av}}\left( {0,3} \right)}\\
{\rm{0}}&{{I_{av}}\left( {1,1} \right)}&{{I_{av}}\left( {1,2} \right)}&{{I_{av}}\left( {1,3} \right)}\\
{{I_{av}}\left( {2,0} \right)}&{{I_{av}}\left( {2,{\rm{1}}} \right)}&{\rm{0}}&{{I_{av}}\left( {2,3} \right)}
\end{array}} \right]  \cr
&=&
 \left[ {\begin{array}{*{20}{c}}
\frac{1}{2}\sum\limits_{k = 0}^1 I_{eh}(0,k)&0&\frac{1}{2}\sum\limits_{k = 2}^3 I_{eh}(0,k)&\frac{1}{2}\sum\limits_{k = 4}^5 I_{eh}(0,k)\\
0&\frac{1}{2}\sum\limits_{k = 0}^1 I_{eh}(1,k)&\frac{1}{2}\sum\limits_{k = 2}^3 I_{eh}(1,k)&\frac{1}{2}\sum\limits_{k = 4}^5 I_{eh}(1,k)\\
\frac{1}{3}\sum\limits_{k = 0}^2 I_{eh}(2,k)&\frac{1}{2}\sum\limits_{k = 3}^4 I_{eh}(2,k) &0&I_{eh}(2,5)
\end{array}} \right]
\label{eq:I_av}
  \end{eqnarray}
 \hrulefill
 \end{figure*}

\subsubsection{PEXIT Algorithm}
To generate the PEXIT curves for the repeat decoder and symbol-MAP Hadamard decoder,
we apply the following steps for a given set of protomatrix $\bm{B}_{m \times n}$, code rate $R$ and
$E_b/N_0$ in dB (denoted as $E_b/N_0({\rm dB})$).
\begin{enumerate}
\renewcommand{\labelenumi}{\roman{enumi})}
\item Compute $\sigma_{L_{ch}} = {(8\cdot R \cdot 10^{(E_b/N_0({\rm dB})) / 10})}^{1/2}$ {\color{black} for ${L_{ch}}$}.
\item For $ i=0,1,\ldots,m-1 $ and $ j=0,1,\ldots,n-1$, set 
${I_{av}}\left( {i,j} \right)=0$. 
%
\item \label{step:start} For $ i=0,1,\ldots,m-1 $ and $ j=0,1,\ldots,n-1$, compute \eqref{eq:I_ev}
 if  $b_{i,j} > 0$;
 else set  $I_{ev}({i,j}) = 0$.
 Taking the $3 \times 4$ protomatrix in \eqref{eq:example} as an example,
the weight of each row  is $d=6$ and hence $r+2=6 \Rightarrow r=4$.
After analyzing the MI of the P-VNs,
 the corresponding $3 \times 4$ $\{I_{ev}(i, j)\}$ MI matrix  can be represented by
\eqref{eq:I_ev_ex}.
\item Convert the $m \times n$ $\{I_{ev}(i, j)\}$ MI matrix into an $m \times d$
$\{I_{ah}(i, k)\}$ MI matrix by eliminating the $0$ entries and  repeating
$\{I_{ev}(i, j)\}$ $b_{i,j} (\ge 1)$ times in the same row. Using the previous example,
the $3 \times 4$ $\{I_{ev}(i, j)\}$ MI matrix is converted into the $3 \times 6$
$\{I_{ah}(i, k)\}$ MI matrix shown in \eqref{eq:I_ah}.

\item For $i=0,1,\ldots,m-1$, using the $d$ entries in the $i$-th row of $I_{ah}$ and $\sigma_{L_{ch}}^{2}$, generate a large number of sets of LLR values as inputs to the symbol-MAP Hadamard decoder and record the output extrinsic LLR values of the $k$-th information bit ($k=0,1,\ldots,d-1)$.
    Compute the extrinsic MI of the information bit using \eqref{eq:I_E},
where $p_e( {\xi |X = x} )$ denotes the PDF of the LLR values given
the bit $x$ being ``$0$'' or ``$1$''.
Form the extrinsic MI matrix $\{I_{eh}(i,k)\}$ of size $m \times d$.
(Details of the method is shown in Appendix \ref{app:b}.)
Using the previous example, the matrix is represented by \eqref{I_eh}. \\
%
\textit{Remark:} Our technique makes use of multiple  \textit{a priori} MI values ($\{I_{ah}(i, k)\}$) and produces multiple extrinsic MI values  ($\{I_{eh}(i, k)\}$).
In \cite{Lik2008}, an EXIT function of symbol-MAP Hadamard decoder under the AWGN channel is obtained. However, the function involves very high computational complexity, which increases rapidly with an increase of the Hadamard order $r$. The function also cannot be used for analyzing non-systematic Hadamard codes.
In \cite{Yue2007}, simulation is used to characterize the symbol-MAP Hadamard decoder but the method is based on a single \textit{a priori} MI value and produces only one output extrinsic MI.
\item \label{step:end}  Convert the $m \times d$ $\{I_{eh}(i, k)\}$ MI matrix into an $m \times n$ $\{I_{av}(i, j)\}$ MI matrix.
For $i=0,1,\ldots,m-1$ and $ j=0,1,\ldots,n-1$;
if $b_{i,j} > 0$, set the value of $I_{av}(i, j)$ as the average of the corresponding $b_{i,j}$ MI values in the $i$-th row of $\{I_{eh}(i, k)\}$;
else set  $I_{av}(i, j)=0$.
In the above example, $\{I_{av}(i, j)\}$ becomes \eqref{eq:I_av}.
\item Repeat Steps iii) to vi) 
until the maximum number of iterations is reached; or
when $I_{app}(j)=1$ for all $ j=0,1,\ldots,n-1$
where ${I_{app}}\left( j \right) =
J\left( \sqrt {\sum\limits_{i=0}^{m-1} {{b_{i,j}}} {{\left( {{J^{ - 1}}\left( {{I_{av}}\left( {i,j} \right)} \right)} \right)}^2} + \sigma_{L_{ch}}^2 }\right).$
\end{enumerate}
Note that our PEXIT algorithm can be used to analyze PLDPC-Hadamard designs with
degree-$1$ and/or punctured VNs. In case of puncturing,
the corresponding channel LLR values in the analysis will be set to zero.

\subsubsection{Optimization Criterion}
For a given code rate, our objective is to find a protograph of the PLDPC-Hadamard code such that it achieves $I_{app}(j)=1 \; \forall j$ within a fixed number of iterations and with the lowest threshold $E_b/N_0$.
To reduce the search space, we impose the following constraints.
First, the weights of all rows in the protomatrix are fixed at $d$.
Second, the maximum column weight, the minimum column weight,
and the maximum value of each entry in protomatrix are preset
 according to the code rate and  order of the Hadamard code.
Third, the maximum number of iterations used in the PEXIT algorithm is set to $300$.
Fourth, we set a target threshold below $-1.40$~dB. \footnote{We only consider low-rate codes with the corresponding Shannon limits below $-1.40$~dB.}
Thus, we start our analysis with an $E_b/N_0$ requirement of $-1.40$~dB.


Algorithm \ref{alg:PEXIT_PLDPCH} shows the steps to find a protomatrix with a low threshold.
A protomatrix is first randomly generated according to the constraints above.
Then it is iteratively analyzed by the PEXIT algorithm
 to see if the corresponding PEXIT curves converge under the current $E_b/N_0$~(dB).
If the protomatrix is found satisfying $I_{app}(j)=1$ for all $j$, $E_b/N_0$~(dB) is reduced by $0.01$~dB and the protomatrix is analyzed again.
If the number of iterations reaches $300$ and the condition $I_{app}(j)=1$ for all $j$ is not satisfied,
 the analysis is terminated and the $E_b/N_0$ threshold is determined. 
The process is repeated until a protomatrix with a satisfactory $E_b/N_0$ threshold is found.
{\color{black}(On average, the PEXIT algorithm takes $35$s (for $r=4$) to $120$s (for $r=10$) to determine the threshold of a protomatrix. Using annealing approaches or genetic algorithms to generate the protomatrices would speed up the search and is part of our on-going research effort.)   
}

\begin{algorithm}[t]
  \caption{Searching $\bm{B}_{m\times n}$ with a low threshold }
  \label{alg:PEXIT_PLDPCH}
  {
    Generate a {\color{black}random} base matrix $\bm{B}_{m \times n}$ according to the corresponding constraints\;
    ${E_b/N_0}\rm{(dB)} = -1.40\ dB$\;
    \While{${E_b/N_0}\rm{(dB)} > -1.59\ dB$}
    {
        $\sigma_{L_{ch}} = \sqrt{8\cdot R \cdot 10^{(E_b/N_0\rm{(dB)}) / 10}}$\;
        $I_{ch}(j) = J(\sigma_{L_{ch}})$ for $\forall j$\;
        $I_{av}(i, j) = 0$ for $\forall i, j$\;
        $It = 0$\;
        \While{$It < 300$}
        {
            Use the proposed PEXIT algorithm to analyze $\bm{B}_{m \times n}$ and obtain $I_{app}(j)$ for $ j=0,1,\ldots,n-1$\;
            \If{$I_{app}(j) = 1$ \rm{for} $\forall j$}
            {
                ${E_b/N_0}$\rm{(dB)} = $E_b/N_0$\rm{(dB)} $- \; 0.01$~dB;
                Goto line 3\;
            }
            $It = It + 1$\;
        }
        \textbf{Break};
    }
  }
  Threshold equals $E_b/N_0\rm{(dB)} + 0.01\ dB$.
\end{algorithm}

%


\section{Simulation Results}\label{sect:sim_re}

In this section, we report our simulation results. Once a protomatrix with low threshold is found,
we use a two-step lifting mechanism together with the PEG  method \cite{Hu2005}
to construct an LDPC code. 
{\color{black}(See Appendix \ref{app:c} for details of the lifting process.)} Subsequently, each CN will be replaced by a Hadamard CN connected
to an appropriate number of D1H-VNs.
Without loss of generality, we transmit
all-zero codewords.
Moreover, the code bits are modulated using binary phase shift keying and sent through an AWGN channel.
The maximum number of iterations performed by the decoder is $300$.
At a particular $E_b/N_0$, we run the simulation until $100$ frame errors are collected.
Then we record the corresponding  bit error rate (BER), frame error rate (FER) and average number of iterations per decoded frame.


\subsection{PLDPC-Hadamard Codes}


\subsubsection{${r = 4}$ and $d=r+2=6$}
We attempt to find a PLDPC-Hadamard code
with a target code rate of approximately $0.05$.
We substitute  $R\approx 0.05$ and $d=6$ into \eqref{eq:even_R}, and obtain $\frac{m}{n}\approx 0.63$.
We therefore select a protomatrix $\bm{B}_{7 \times 11}$ of size $7 \times 11$, i.e., $m = 7$ and $n = 11$, and hence the code rate equals $R = 0.0494$. Moreover, we set the
minimum column weight to $1$,
maximum column weight to $9$, and
maximum entry value to $3$.
The overall constraints of the protomatrix are listed as follows:
\begin{itemize}
\item size equals $7 \times 11$,
\item row weight equals $\sum\nolimits_{j = 0}^{10} {{b_{i,j}}} = d = 6$,
\item minimum column weight equals $\sum\nolimits_{i = 0}^{6} {{b_{i,j}}} = 1$,
\item maximum column weight equals $\sum\nolimits_{i = 0}^{6} {{b_{i,j}}} = 9$, and
\item maximum entry value in $\bm{B}_{7 \times 11}$ equals $3$.
\end{itemize}

%
%
%


Using the proposed analytical method under the constraints above, we find the following protomatrix which has
a theoretical threshold of $-1.42$ dB.
\begin{equation}\label{mat:r=4}
{\bm{B}_{7 \times 11}} = \left[ {\begin{array}{*{11}{c}}
1&0&0&0&0&0&1&0&3&0&1\\
0&1&2&0&0&0&0&0&0&2&1\\
2&1&0&0&1&1&0&0&0&0&1\\
0&1&0&3&0&0&0&0&0&2&0\\
2&0&0&0&0&0&0&1&0&3&0\\
3&0&0&2&0&0&1&0&0&0&0\\
1&0&0&1&1&0&0&0&1&2&0
\end{array}} \right]
\end{equation}
Fig.~\ref{fig:pexit_0_05} plots the PEXIT curves of the repeat decoder and the symbol-MAP Hadamard decoder under $E_b/N_0 = -1.42$ dB.
It can be observed that the two curves are matched.
By lifting the protomatrix with factors of $z_1 = 32$ and $z_2 = 512$,
we obtain a PLDPC-Hadamard code with information length $k =z_1 z_2(n-m)  = 65,536$ and  code length $N_{total} =z_1 z_2 [ m ( 2^{d - 2} - d )  + n] =1,327,104$.
{\color{black}(See Appendix \ref{app:c} for details of the code structure after the lifting process.)} 

\begin{figure}[htbp]
\centerline{
\includegraphics[width=0.6\columnwidth]{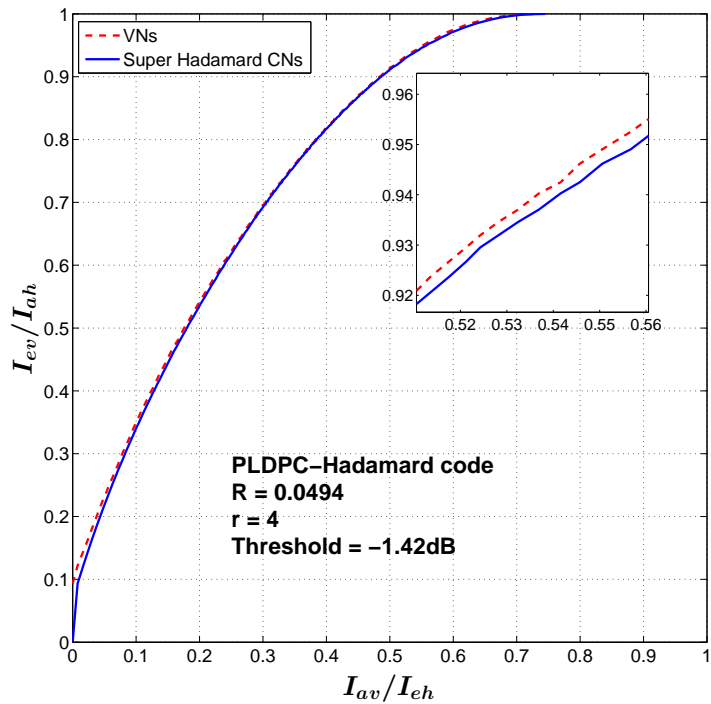}}
 \caption{The PEXIT chart of the PLDPC-Hadamard code given in \eqref{mat:r=4}
with $R = 0.0494$ and $r = 4$. }
  \label{fig:pexit_0_05}
\end{figure}

\begin{figure}[htbp]
\centerline{
\includegraphics[width=0.6\columnwidth]{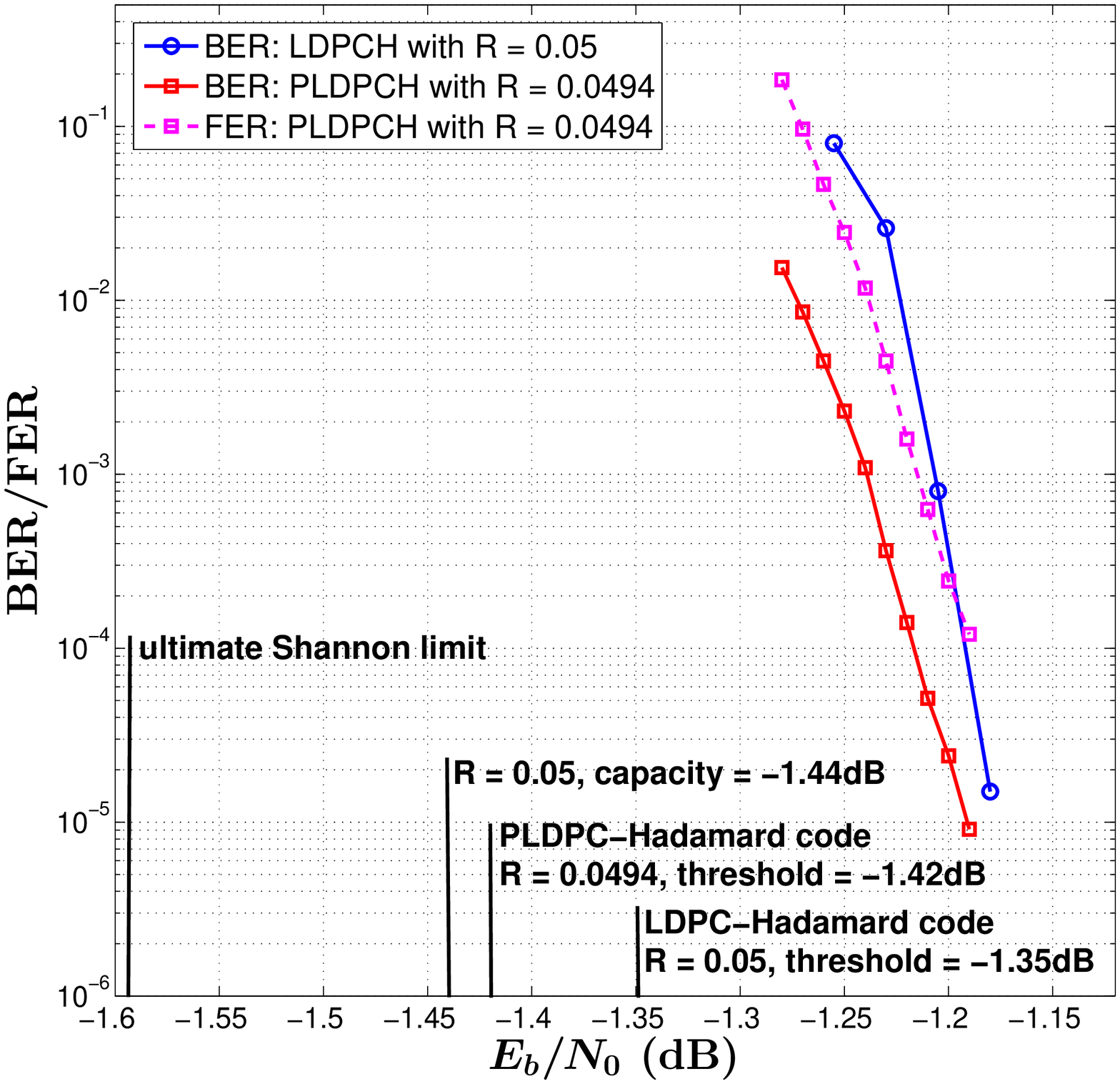}}
 \caption{BER (red curve) and FER (pink curve) performance of the proposed PLDPC-Hadamard code compared with the BER of the LDPC-Hadamard code (blue curve) in \cite{Yue2007}. $r = 4$ and $k=65,536$. }
  \label{fig:ber_0_05}
\end{figure}

\begin{table}[t]
\centering\caption{Detail results achieving a BER of $10^{-5}$ for $r = 4$ PLDPC-Hadamard code with rate-$0.494$ until 100 frame errors are reached.}\label{tb:r4} 
\begin{center} \small
 \begin{tabular}{|c|c|}
\hline
$E_b/N_0$ & $-1.19$ dB \\
\hline
 No. of  frame sent &	$832,056$ \\
\hline
FER	& $1.2 \times 10^{-4}$ \\
\hline
BER	& $9.1 \times 10^{-6}$ \\
\hline
Av. no. of iterations&	$127$\\
\hline
\end{tabular}
\end{center}
\end{table}

\begin{figure}[htbp]
\centerline{
\includegraphics[width=0.6\columnwidth]{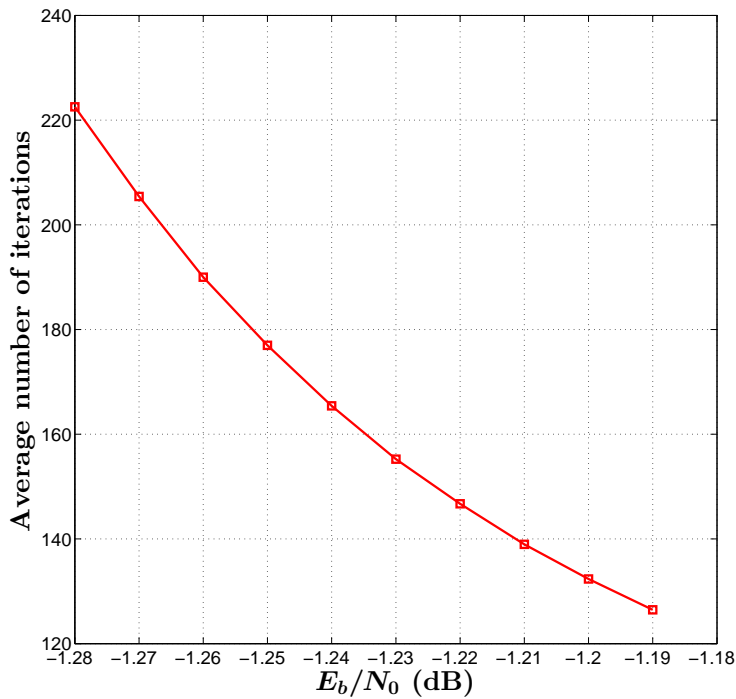}}
 \caption{ Average number of iterations required to decode the PLDPC-Hadamard code versus $E_b/N_0$ with  $r = 4$ and $k = 65,536$. }
  \label{fig:it_r4_original}
\end{figure}

The BER and FER results of the PLDPC-Hadamard code found are plotted in Fig.~\ref{fig:ber_0_05}.
Our code achieves a BER of $10^{-5}$ at $E_b/N_0 = -1.19$ dB, which is $0.23$~dB from the threshold.
Table \ref{tb:r4} lists the detailed results at $-1.19$ dB.
A total of $832,056$ frames need to be sent before $100$ frame errors are collected.
Hence a FER of $1.2 \times 10^{-4}$ is achieved.
Fig.~\ref{fig:it_r4_original} plots the average number of iterations versus $E_b/N_0$.
Our code requires an average of $127$ iterations for decoding at $E_b/N_0 = -1.19$ dB.
At a BER of $10^{-5}$, the gaps of our rate-0.0494 PLDPC-Hadamard code to the Shannon capacity for $R = 0.05$ and to the ultimate Shannon limit are $0.25$~dB and $0.40$~dB, respectively.
The comparison of gaps is also listed in Table \ref{tb:gap_r4}.

\begin{figure}[t]
\centerline{
\includegraphics[width=0.6\columnwidth]{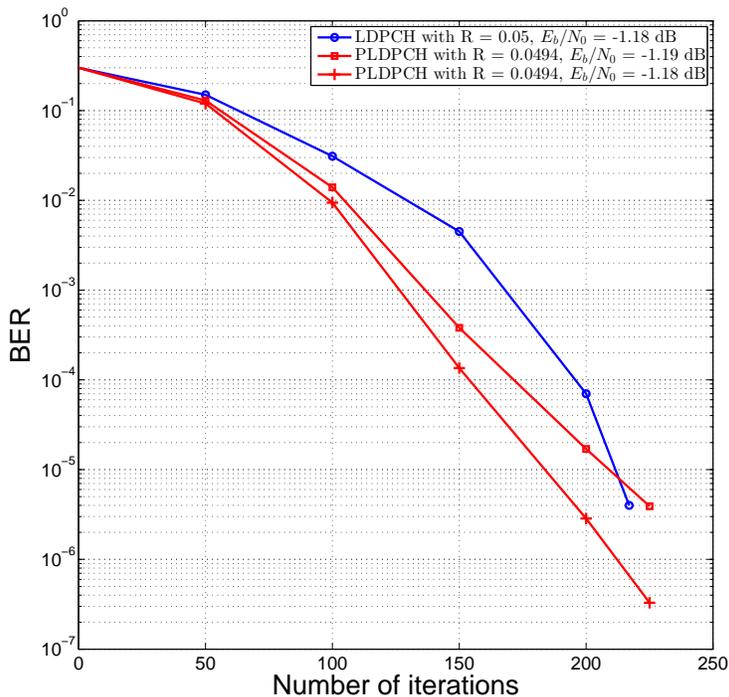}}
 \caption{ {\color{black}BER performance versus number of iterations for the LDPC-Hadamard code in  \cite{Yue2007} ($E_b/N_0=-1.18$ dB) and PLDPC-Hadamard code ($E_b/N_0=-1.18$ and $-1.19$ dB). $r = 4$ and $k = 65,536$.} }
  \label{fig:ber_num_it}
\end{figure}

{\color{black}In Fig.~\ref{fig:ber_num_it}, we further compare the BER results of the rate-$0.05$ LDPC-Hadamard code  in  \cite{Yue2007} at $E_b/N_0 = -1.18$~dB and our rate-$0.0494$ PLDPC-Hadamard code at $E_b/N_0 = -1.18$~dB and $-1.19$~dB under different number of iterations. Note that the result of the LDPC-Hadamard code is the average from 20 simulations \cite{Yue2007}, whereas our result is the average from $10,000$ simulations. In other words, our simulation results are statistically very accurate due to the large number of simulations involved. For the same number of iterations and at $E_b/N_0 = -1.18$~dB, our PLDPC-Hadamard code produces a lower BER compared with the LDPC-Hadamard code in  \cite{Yue2007}. When our proposed PLDPC-Hadamard code operates at a slightly lower $E_b/N_0$, i.e.,  $-1.19$~dB, the BER of the proposed code still outperforms the conventional code except for iteration numbers beyond $200$. Thus, we conclude that the proposed code achieves a faster convergence rate compared with the conventional code. In particular, our results are more precise because $10,000$ simulations are used for our code compared with only 20 simulations used for the conventional code in  \cite{Yue2007}. }

Compared with the LDPC-Hadamard code in \cite{Yue2007} which uses $R = 0.05$ and $r = 4$, our proposed PLDPC-Hadamard code has a slight performance improvement.
The relatively advantage of our proposed PLDPC-Hadamard code over the LDPC-Hadamard code is probably due to degree-$1$ VNs in the protograph.
Such degree-$1$ VNs are regarded as a kind of precoding structure which
 can increase the linear minimum distance 
 \cite{Fang2015,Abbasfar2007}.

\begin{table}[t]
\newcommand{\tabincell}[2]{\begin{tabular}{@{}#1@{}}#2\end{tabular}}
\caption{Gaps to theoretical threshold, rate-$0.05$ Shannon limit and ultimate Shannon limit for $r = 4$ LDPC-Hadamard and PLDPC-Hadamard codes at a BER of $10^{-5}$.}\label{tb:gap_r4} \normalsize \small
\begin{center}
\begin{tabular}{|c|c|c|}
 \hline
\multirow{2}{*}{\tabincell{c}}{Type of Code} & {\tabincell{c}{ \cite{Yue2007} Rate-$0.05$ \\ LDPCH code   } } & {\tabincell{c}{ Rate-$0.0494$ \\ PLDPCH code  } } \\
\hline
{\tabincell{c}{ Theoretical \\ threshold }} &  {\tabincell{c}{ $-1.35$ dB by \\ EXIT chart  } }  & {\tabincell{c}{ $-1.42$ dB by \\ PEXIT chart  } } \\
\hline
{\tabincell{c}{ $E_b/N_0$ at a\\ BER of $10^{-5}$ }} &  $-1.18$ dB  & $-1.19$ dB \\
\hline
{\tabincell{c}{ Gap to theoretical \\ threshold  } } &  $0.17$ dB  & $0.23$ dB \\
\hline
{\tabincell{c}{ Gap to rate-$0.05$ \\ Shannon limit \\ ($-1.44$ dB) } } &  $0.26$ dB  & $0.25$ dB \\
\hline
{\tabincell{c}{ Gap to ultimate \\ Shannon limit \\ ($-1.59$ dB) } } &  $0.41$ dB  & $0.40$ dB \\
\hline
\end{tabular}
\end{center}
\end{table}

\subsubsection{${r = 5}$ and $d=7$}
We attempt to search a PLDPC-Hadamard code with a target code rate of approximately $R\approx0.02$. Using \eqref{eq:odd2_R}, we obtain $m = 6$ and $n = 10$ and $\frac{m}{n}\approx 0.61$.
Hence the actual code rate is $R = 0.021$. The constraints of the protomatrix are as follows:

%
%

\begin{itemize}
\item size equals $6 \times 10$,
\item row weight equals $\sum\nolimits_{j = 0}^{9} {{b_{i,j}}} = d = 7$,
\item minimum column weight equals $\sum\nolimits_{i = 0}^{5} {{b_{i,j}}} = 1$,
\item maximum column weight equals $\sum\nolimits_{i = 0}^{5} {{b_{i,j}}} = 9$,
\item maximum entry value in $\bm{B}_{6 \times 10}$ equals $3$.
\end{itemize}
The following protomatrix with a threshold of $-1.51$ dB is found.
\begin{equation}\label{mat:r=5}
{\bm{B}_{6 \times 10}}{\rm{ = }}\left[ {\begin{array}{*{10}{c}}
{\rm{3}}&{\rm{2}}&{\rm{0}}&{\rm{0}}&{\rm{1}}&{\rm{0}}&{\rm{0}}&{\rm{0}}&{\rm{1}}&{\rm{0}}\\
{\rm{0}}&{\rm{0}}&{\rm{2}}&{\rm{0}}&{\rm{0}}&{\rm{2}}&{\rm{1}}&{\rm{2}}&{\rm{0}}&{\rm{0}}\\
{\rm{0}}&{\rm{0}}&{\rm{0}}&{\rm{3}}&{\rm{1}}&{\rm{0}}&{\rm{0}}&{\rm{1}}&{\rm{0}}&{\rm{2}}\\
{\rm{0}}&{\rm{1}}&{\rm{0}}&{\rm{1}}&{\rm{0}}&{\rm{0}}&{\rm{0}}&{\rm{2}}&{\rm{0}}&{\rm{3}}\\
{\rm{0}}&{\rm{0}}&{\rm{0}}&{\rm{2}}&{\rm{0}}&{\rm{0}}&{\rm{1}}&{\rm{2}}&{\rm{0}}&{\rm{2}}\\
{\rm{2}}&{\rm{0}}&{\rm{1}}&{\rm{1}}&{\rm{0}}&{\rm{0}}&{\rm{0}}&{\rm{2}}&{\rm{0}}&{\rm{1}}
\end{array}} \right].
\end{equation}
The same lifting factors  $z_1 = 32$ and $z_2 = 512$ are used to expand $\bm{B}_{6 \times 10}$.
The rate-$0.021$ PLDPC-Hadamard code has an information length of $k =z_1 z_2(n-m)  = 65,536$ and a code length of $N_{total} =z_1 z_2 [ m ( 2^{d - 2} - 2 )  + n] = 3,112,960$.
Fig. \ref{fig:pexit_0_02} shows the PEXIT chart of the code at $E_b/N_0 = -1.51$ dB.
We can observe that the two curves do not crossed and are matched.

\begin{figure}[t]
\centerline{
\includegraphics[width=0.6\columnwidth]{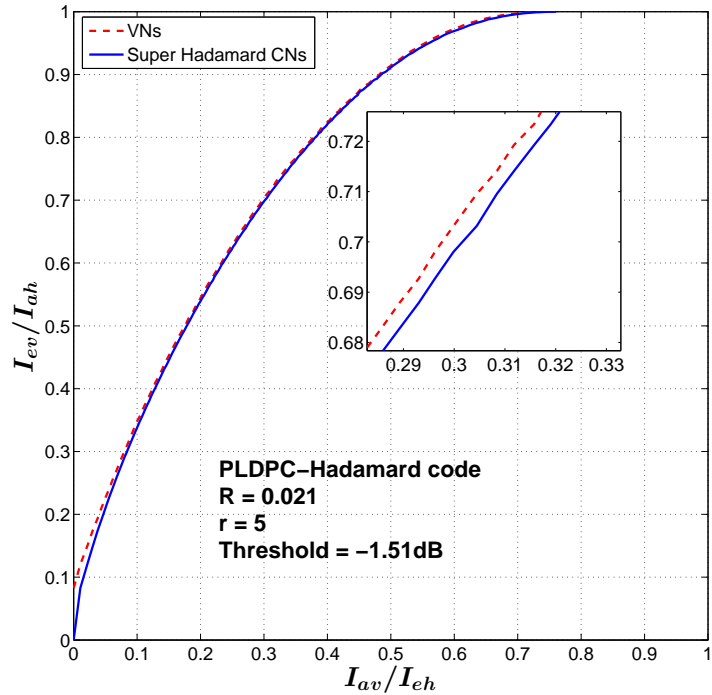}}
 \caption{
 The PEXIT chart of the PLDPC-Hadamard code given in \eqref{mat:r=5}
 with $R = 0.021$ and $r = 5$. }
  \label{fig:pexit_0_02}
\end{figure}

Fig. \ref{fig:ber_0_02} plots the BER and FER performance of the PLDPC-Hadamard code.
The code achieves a BER of $ 1.4 \times 10^{-5}$ and a FER of $1.3 \times 10^{-4}$ at $E_b/N_0 = -1.24$~dB (red curve), which is $0.27$~dB away from the designed threshold.
Compared with the BER curve (blue curve) of the rate-$0.022$ LDPC-Hadamard code in \cite{Yue2007}, our PLDPC-Hadamard code can achieve comparable results. 
Table \ref{tb:r5} shows that at $E_b/N_0 = -1.24$ dB, 
$780,660$ frames have been decoded with an average of
 $119$ decoding iterations per frame.
Fig. \ref{fig:it_r5_original} can obtain other average number of iterations for the code.
At a BER of $10^{-5}$, the gaps to the Shannon capacity of $R = 0.020$ and to the ultimate Shannon limit are $0.29$ dB and $0.35$ dB, respectively, which are listed in Table \ref{tb:gap_r5}.

\begin{figure}[htbp]
\centerline{
\includegraphics[width=0.6\columnwidth]{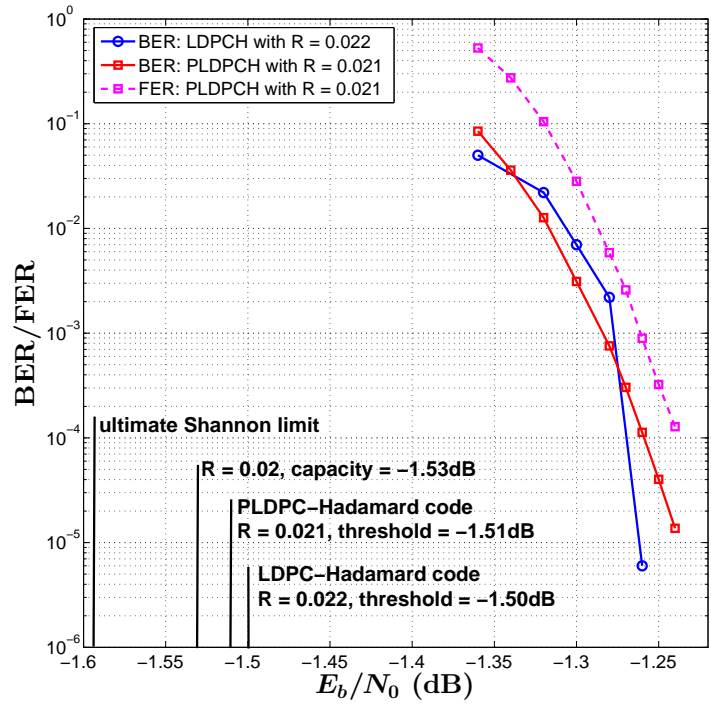}}
 \caption{BER (red curve) and FER (pink curve) performance of the proposed PLDPC-Hadamard code compared with the BER of the LDPC-Hadamard code (blue curve) in \cite{Yue2007}.
 $r = 5$ and $k = 65,536$. }
  \label{fig:ber_0_02}
\end{figure}

\begin{table}[htbp]
\centering\caption{Detail results achieving a BER of $10^{-5}$ for $r = 5$ PLDPC-Hadamard code until 100 frame errors are reached.}\label{tb:r5}
\begin{center} \small
 \begin{tabular}{|c|c|}
\hline
$E_b/N_0$ &	$-1.24$ dB  \\
\hline
 No. of  frame sent &	$780,660$ \\
\hline
FER	& $1.3 \times 10^{-4}$\\
\hline
BER	& $1.4 \times 10^{-5}$\\
\hline
Av. no. of iterations&	$119$ \\
\hline
\end{tabular}
\end{center}
\end{table}

 \begin{figure}[htbp]
\centerline{
\includegraphics[width=0.61\columnwidth]{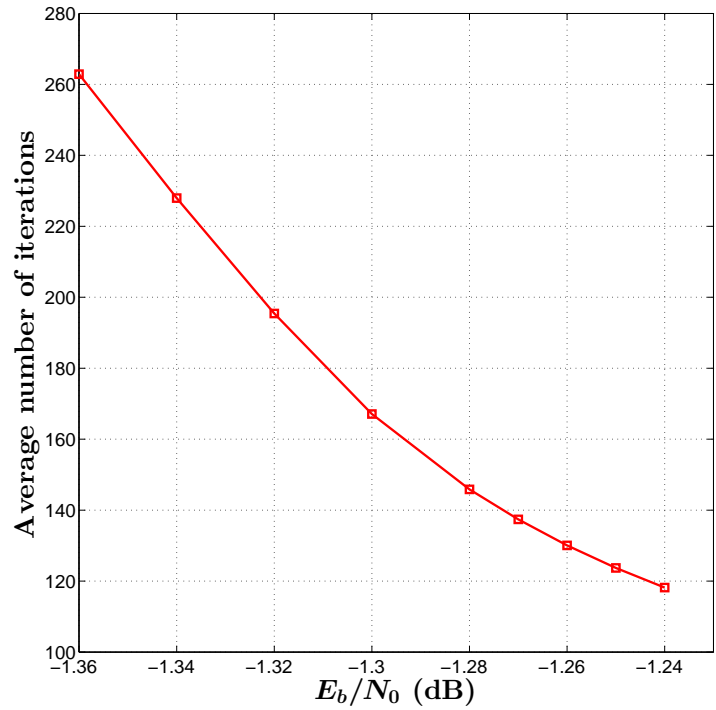}}
 \caption{Average number of iterations required to decode the PLDPC-Hadamard code with $r = 5$ and $k = 65,536$. }
  \label{fig:it_r5_original}
\end{figure}

\begin{table}[htbp]
\newcommand{\tabincell}[2]{\begin{tabular}{@{}#1@{}}#2\end{tabular}}
\caption{Gaps to theoretical threshold, rate-$0.02$ Shannon limit and ultimate Shannon limit for $r = 5$ LDPC-Hadamard and PLDPC-Hadamard codes at a BER of $10^{-5}$.}\label{tb:gap_r5} \normalsize \small
\begin{center}
\begin{tabular}{|c|c|c|}
 \hline
\multirow{2}{*}{\tabincell{c}}{Type of Code} & {\tabincell{c}{ \cite{Yue2007} Rate-$0.022$\\ LDPCH code } } & {\tabincell{c}{ Rate-$0.021$\\ PLDPCH code } }  \\
\hline
{\tabincell{c}{ Theoretical \\ Threshold }} &  {\tabincell{c}{ $-1.50$ dB by \\ EXIT chart  } }  & {\tabincell{c}{ $-1.51$ dB by \\ PEXIT chart  } } \\
\hline
{\tabincell{c}{ $E_b/N_0$ at a\\ BER of $10^{-5}$ }} &  $-1.26$ dB  & $-1.24$ dB \\
\hline
{\tabincell{c}{ Gap to theoretical \\ Threshold  } } &  $0.24$ dB  & $0.27$ dB \\
\hline
{\tabincell{c}{ Gap to rate-$0.020$ \\ Shannon limit \\ ($-1.53$ dB) } } &  $0.27$ dB  & $0.29$ dB \\
\hline
{\tabincell{c}{ Gap to ultimate \\ Shannon limit \\ ($-1.59$ dB) } } &  $0.33$ dB  & $0.35$ dB \\
\hline
\end{tabular}
\end{center}
\end{table}




\subsubsection{${r = 8}$ and $d=10$}
A rate-$0.008$ PLDPC-Hadamard code is constructed using $m = 5$ and $n = 15$.
%
%
The constraints of the protomatrix are as follows:
\begin{itemize}
\item size equals $5 \times 15$,
\item row weight equals $\sum\nolimits_{j = 0}^{14} {{b_{i,j}}} = d = 10$,
\item minimum column weight equals $\sum\nolimits_{i = 0}^{4} {{b_{i,j}}} = 1$,
\item maximum column weight equals $\sum\nolimits_{i = 0}^{4} {{b_{i,j}}} = 11$,
\item maximum entry value in $\bm{B}_{5 \times 15}$ equals $3$.
\end{itemize}
Compared with the constraints for low-order ($r = 4$ or $5$) PLDPC-Hadamard protomatrices, the maximum column weight is increased to $11$.
Based on these constraints and using our proposed analytical method, the following protomatrix is found with a threshold of $-1.53$ dB. 
\begin{equation}\label{mat:r=8}
\begin{split}
&{\bm{B}_{5 \times 15}} = \\
&\left[ {\begin{array}{*{15}{c}}
2&0&1&0&0&0&0&3&2&0&0&1&0&0&1\\
0&2&0&1&1&0&0&0&0&0&0&3&0&3&0\\
0&0&1&0&0&2&2&0&0&1&1&2&1&0&0\\
0&0&0&2&2&0&0&0&0&1&0&3&0&0&2\\
0&0&0&0&0&1&1&0&1&1&1&2&3&0&0
\end{array}} \right]
\end{split}
\end{equation}

\begin{figure}[t]
\centerline{
\includegraphics[width=0.6\columnwidth]{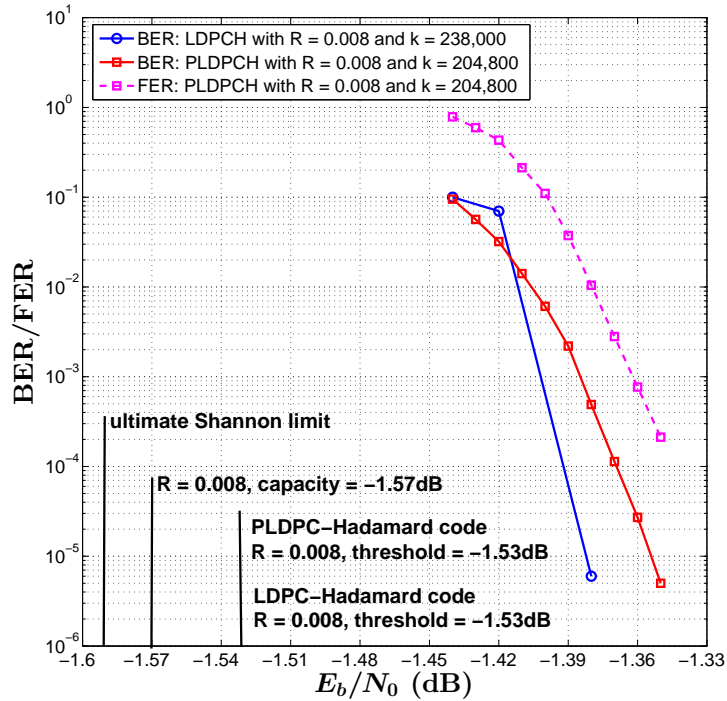}}
 \caption{BER (red curve) and FER (pink curve) performance of the proposed PLDPC-Hadamard code compared with the BER of the LDPC-Hadamard code (blue curve) in \cite{Yue2007}.
 $r = 8$. }
  \label{fig:ber_0_008}
\end{figure}

\begin{figure}[t]
\centerline{
\includegraphics[width=0.6\columnwidth]{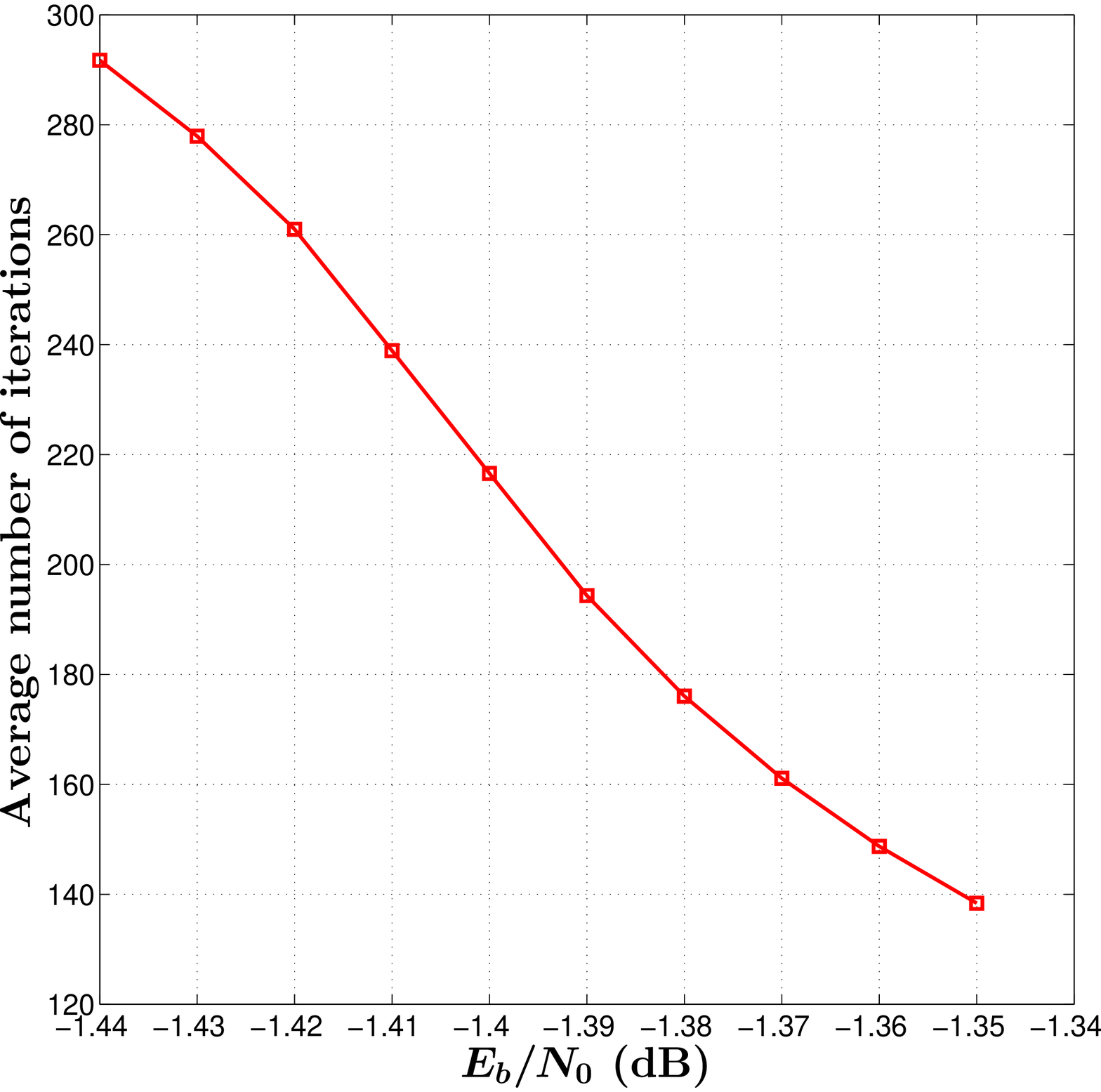}}
 \caption{Average number of iterations required to decode the PLDPC-Hadamard code with $r = 8$ and $k = 204,800$. }
  \label{fig:it_r8_original}
\end{figure}


We use lifting factors of $z_1 = 16$ and $z_2 = 1280$.
The rate-$0.008$ PLDPC-Hadamard code thus has an information length of $k = 204,800$ and a code length of $N_{total} = 25,497,600$.
This code has the same theoretical threshold as the rate-$0.008$ LDPC-Hadamard code in \cite{Yue2007}.
The FER and BER curves of our code and the BER of the code in \cite{Yue2007} are plotted in Fig. \ref{fig:ber_0_008}.
At $E_b/N_0 = -1.35$ dB, the PLDPC-Hadamard code achieves a FER of $2.1 \times 10^{-4}$, and a BER of $3.8 \times 10^{-6}$ which is $0.18$ dB away from the designed threshold.
Compared with the BER curve of \cite{Yue2007}, there is a performance gap of about $0.03$ dB at a BER of $10^{-5}$. 
At the same BER, the gaps of our code to the rate-0.008 Shannon limit and to the ultimate Shannon limit are $0.22$ dB and $0.24$ dB, respectively, as shown in Table \ref{tb:gap_r8}.
The convergence rate of the code can be found in Fig. \ref{fig:it_r8_original} and the average number of iterations is $139$ at $E_b/N_0 = -1.35$ dB.

\begin{table}[t]
\newcommand{\tabincell}[2]{\begin{tabular}{@{}#1@{}}#2\end{tabular}}
\caption{Gaps to theoretical threshold, rate-$0.008$ Shannon limit and ultimate Shannon limit for $r = 8$ LDPC-Hadamard and PLDPC-Hadamard codes at a BER of $10^{-5}$.}\label{tb:gap_r8} \normalsize \small
\begin{center}
\begin{tabular}{|c|c|c|}
 \hline
\multirow{2}{*}{\tabincell{c}}{Type of Code} & {\tabincell{c}{ \cite{Yue2007} Rate-$0.008$ \\ LDPCH code  } } & {\tabincell{c}{ Rate-$0.008$ \\ PLDPCH code  } } \\
\hline
{\tabincell{c}{ Theoretical \\ Threshold }} &  {\tabincell{c}{ $-1.53$ dB by \\ EXIT chart  } }  & {\tabincell{c}{ $-1.53$ dB by \\ PEXIT chart  } } \\
\hline
{\tabincell{c}{ $E_b/N_0$ at a\\ BER of $10^{-5}$ }} &  $-1.38$ dB  & $-1.35$ dB \\
\hline
{\tabincell{c}{ Gap to theoretical \\ Threshold  } } &  $0.15$ dB  & $0.18$ dB \\
\hline
{\tabincell{c}{ Gap to rate-$0.008$ \\ Shannon limit \\ ($-1.57$ dB) } } &  $0.19$ dB  & $0.22$ dB \\
\hline
{\tabincell{c}{ Gap to ultimate \\ Shannon limit \\ ($-1.59$ dB) } } &  $0.21$ dB  & $0.24$ dB \\
\hline
\end{tabular}
\end{center}
\end{table}

\subsubsection{${r = 10}$ and $d=12$}
A rate-$0.00295$ PLDPC-Hadamard code is constructed using $m = 6$ and $n = 24$.
(The target rate is approximately $0.003$.)
The constraints of the protomatrix are as follows:
%
%
\begin{itemize}
\item size equals $6 \times 24$,
\item row weight equals $\sum\nolimits_{j = 0}^{23} {{b_{i,j}}} = d = 12$,
\item minimum column weight equals $\sum\nolimits_{i = 0}^{5} {{b_{i,j}}} = 1$,
\item maximum column weight equals $\sum\nolimits_{i = 0}^{5} {{b_{i,j}}} = 11$,
\item maximum entry value in $\bm{B}_{6 \times 24}$ equals $4$.
\end{itemize}
\begin{figure*}[t]
\begin{equation}\label{mat:r=10}
{\bm{B}_{{\rm{6}} \times {\rm{24}}}} = \left[ {\begin{array}{*{24}{c}}
{\rm{1}}&{\rm{0}}&{\rm{0}}&{\rm{0}}&{\rm{0}}&{\rm{0}}&{\rm{3}}&{\rm{0}}&{\rm{0}}&{\rm{0}}&{\rm{0}}&{\rm{0}}&{\rm{0}}&{\rm{0}}&{\rm{2}}&{\rm{0}}&{\rm{0}}&{\rm{0}}&{\rm{0}}&{\rm{1}}&{\rm{4}}&{\rm{0}}&{\rm{1}}&{\rm{0}}\\
{\rm{0}}&{\rm{0}}&{\rm{0}}&{\rm{3}}&{\rm{2}}&{\rm{0}}&{\rm{0}}&{\rm{0}}&{\rm{1}}&{\rm{1}}&{\rm{0}}&{\rm{0}}&{\rm{1}}&{\rm{0}}&{\rm{0}}&{\rm{0}}&{\rm{3}}&{\rm{1}}&{\rm{0}}&{\rm{0}}&{\rm{0}}&{\rm{0}}&{\rm{0}}&{\rm{0}}\\
{\rm{0}}&{\rm{1}}&{\rm{2}}&{\rm{0}}&{\rm{0}}&{\rm{1}}&{\rm{0}}&{\rm{0}}&{\rm{0}}&{\rm{0}}&{\rm{0}}&{\rm{0}}&{\rm{0}}&{\rm{0}}&{\rm{0}}&{\rm{1}}&{\rm{0}}&{\rm{0}}&{\rm{3}}&{\rm{0}}&{\rm{4}}&{\rm{0}}&{\rm{0}}&{\rm{0}}\\
{\rm{0}}&{\rm{0}}&{\rm{0}}&{\rm{0}}&{\rm{0}}&{\rm{0}}&{\rm{0}}&{\rm{1}}&{\rm{0}}&{\rm{0}}&{\rm{3}}&{\rm{3}}&{\rm{0}}&{\rm{0}}&{\rm{0}}&{\rm{0}}&{\rm{0}}&{\rm{0}}&{\rm{0}}&{\rm{0}}&{\rm{2}}&{\rm{2}}&{\rm{0}}&{\rm{1}}\\
{\rm{2}}&{\rm{2}}&{\rm{0}}&{\rm{0}}&{\rm{0}}&{\rm{0}}&{\rm{0}}&{\rm{0}}&{\rm{0}}&{\rm{0}}&{\rm{0}}&{\rm{0}}&{\rm{0}}&{\rm{1}}&{\rm{0}}&{\rm{1}}&{\rm{0}}&{\rm{1}}&{\rm{0}}&{\rm{3}}&{\rm{0}}&{\rm{0}}&{\rm{1}}&{\rm{1}}\\
{\rm{0}}&{\rm{0}}&{\rm{0}}&{\rm{0}}&{\rm{0}}&{\rm{1}}&{\rm{0}}&{\rm{3}}&{\rm{3}}&{\rm{2}}&{\rm{0}}&{\rm{0}}&{\rm{1}}&{\rm{1}}&{\rm{0}}&{\rm{0}}&{\rm{0}}&{\rm{0}}&{\rm{0}}&{\rm{0}}&{\rm{1}}&{\rm{0}}&{\rm{0}}&{\rm{0}}
\end{array}} \right]
\end{equation}
\end{figure*}
Note that the maximum value of the entries in $\bm{B}_{6 \times 24}$ is increased to $4$.
For the rate-$0.00295$ PLDPC-Hadamard code, the protomatrix in (\ref{mat:r=10}) with a theoretical threshold of $-1.53$ dB is found by the proposed analytical method.
The threshold is slightly higher ($0.02$ dB) than that of the LDPC-Hadamard code in\cite{Yue2007}.
We use lifting factors of $z_1 = 20$ and $z_2 = 1280$. The information length equals $k = 460,800$ and the code length equals $N_{total} = 156,057,600$.

\begin{figure}[htbp]
\centerline{
\includegraphics[width=0.6\columnwidth]{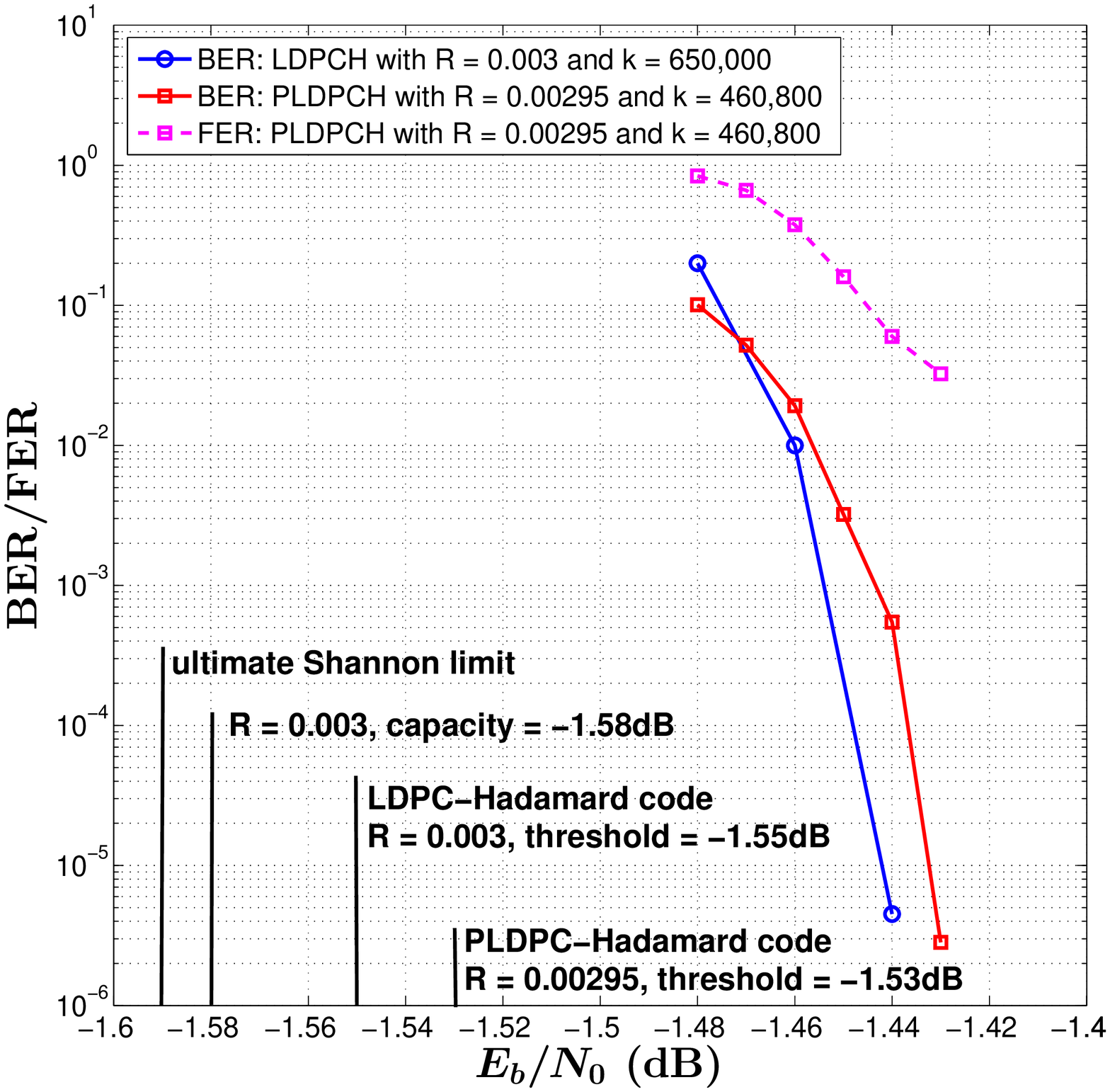}}
 \caption{BER (red curve) and FER (pink curve) performance of the proposed PLDPC-Hadamard code compared with the BER of the LDPC-Hadamard code (blue curve) in \cite{Yue2007}.
 $r = 10$. }
  \label{fig:ber_0_003}
\end{figure}

\begin{figure}[htbp]
\centerline{
\includegraphics[width=0.6\columnwidth]{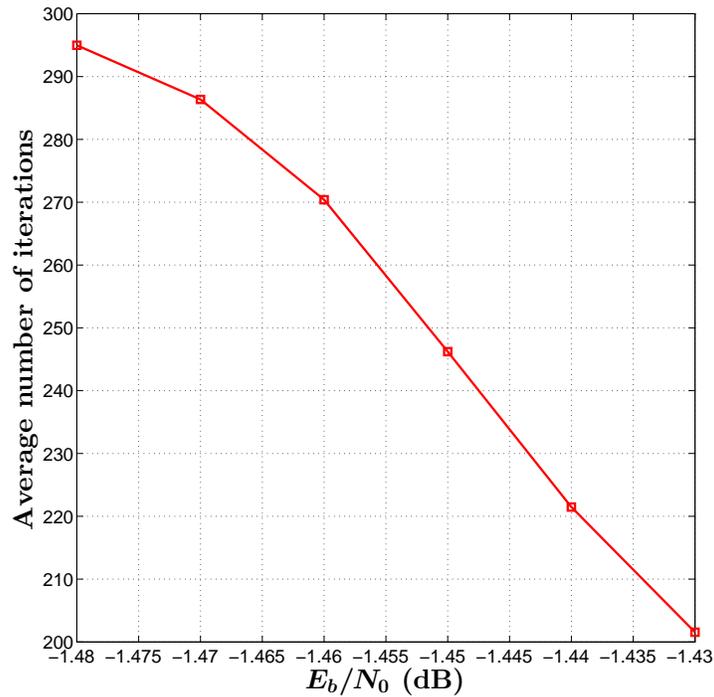}}
 \caption{Average number of iterations required to decode the PLDPC-Hadamard code with $r = 10$ and $k = 460,800$. }
  \label{fig:it_r10_original}
\end{figure}



Fig. \ref{fig:ber_0_003} plots the error performance of our constructed code and that in \cite{Yue2007}.
Our PLDPC-Hadamard code achieves a BER of $2.8 \times 10^{-6}$ at $E_b/N_0 = -1.43$ dB, which is $0.01$ dB higher compared with the LDPC-Hadamard code in \cite{Yue2007}.
However, our code is $29.11\%$ shorter compared with the code in \cite{Yue2007}.
This performance has $0.10$ dB gap from the designed threshold.
The gaps of our code to the rate-$0.003$ Shannon limit and to the ultimate Shannon limit are $0.15$ dB and $0.16$ dB, respectively.
Table \ref{tb:gap_r10} lists the detailed comparison.
The convergence rate of the code can be found in Fig. \ref{fig:it_r10_original} and the average number of iterations is $202$ at $E_b/N_0 = -1.43$ dB.

{\color{black} \textit{Remark}: For cases with $r=5, 8$ and $10$ (Figs. \ref{fig:ber_0_02}, \ref{fig:ber_0_008} and \ref{fig:ber_0_003}), the BER results may appear that our proposed PLDPC-Hadamard codes are slightly outperformed by the LDPC-Hadamard codes in \cite{Yue2007} at the high $E_b/N_0$ region. For our codes, we keep running the simulations until 100 block errors are recorded. Thus our reported results have a high degree of accuracy. However, the stopping criterion of the LDPC-Hadamard code simulation in \cite{Yue2007}  is not known. If an inadequate number of simulations are performed, there could be some statistical difference between the actual error performance and the reported results. } 

\begin{table}[t]
\newcommand{\tabincell}[2]{\begin{tabular}{@{}#1@{}}#2\end{tabular}}
\caption{Gaps to theoretical threshold, rate-0.003 Shannon limit and ultimate Shannon limit for $r = 10$ LDPC-Hadamard and PLDPC-Hadamard codes at a BER of $10^{-5}$.}\label{tb:gap_r10} \normalsize \small
\begin{center}
\begin{tabular}{|c|c|c|}
 \hline
\multirow{2}{*}{\tabincell{c}}{Type of Code} & {\tabincell{c}{ \cite{Yue2007} Rate-$0.003$ \\ LDPCH code  } } & {\tabincell{c}{ Rate-$0.00295$ \\ PLDPCH code  } } \\
\hline
{\tabincell{c}{ Theoretical \\ Threshold }} &  {\tabincell{c}{ $-1.55$ dB by \\ EXIT chart } }  & {\tabincell{c}{ $-1.53$ dB by \\ PEXIT chart  } } \\
\hline
{\tabincell{c}{ $E_b/N_0$ at a\\ BER of $10^{-5}$ }} &  $-1.44$ dB  & $-1.43$ dB \\
\hline
{\tabincell{c}{ Gap to theoretical \\ Threshold  } } &  $0.11$ dB  & $0.10$ dB \\
\hline
{\tabincell{c}{ Gap to rate-0.003 \\ Shannon limit \\ ($-1.58$ dB) } } &  $0.14$ dB  & $0.15$ dB \\
\hline
{\tabincell{c}{ Gap to ultimate \\ Shannon limit \\ ($-1.59$ dB) } } &  $0.15$ dB  & $0.16$ dB \\
\hline
\end{tabular}
\end{center}
\end{table}
\subsection{Punctured PLDPC-Hadamard Codes}
When a code is punctured, the code rate increases.
The signals corresponding to the punctured variable nodes are
not sent to the receiver and hence their channel LLR values 
are initialized to zero.
In this section, we evaluate the performance of
the PLDPC-Hadamard codes designed in the previous section when the codes
are punctured.
(Note that our proposed PEXIT chart method can be used to design
good punctured PLDPC-Hadamard codes.)

We use $\alpha$ to denote a column number in a protomatrix and $\beta$ to denote the weight of a column.
For example in the protomatrix shown in \eqref{mat:r=4}, $[4, 6]$ refers to the $4$-th column $[0\ 0\ 0\ 3\ 0\ 2\ 1]^T$ which has a column weight of $6$. Thus we use ``punctured $[\alpha, \beta]$''  to denote a PLDPC-Hadamard code in which the
P-VN corresponding to the $\alpha$-th column in the protomatrix is punctured. Moreover,
the punctured P-VN has a degree of $\beta$.

\subsubsection{${r = 4}$}
We first consider the rate-$0.0494$ PLDPC-Hadamard code shown in  \eqref{mat:r=4}
and puncture one P-VN with the largest degree (i.e., $9$) or lowest degree (i.e., $1$).
Four cases are therefore considered, i.e., $[1,9]$, $[10, 9]$, $[6,1]$ and $[8,1]$.
After puncturing, all codes have a rate of $0.0500$ (by applying \eqref{eq:even_R_punc}).
Fig.~\ref{fig:ber_r4_puncture} shows that at a BER of $10^{-4}$, punctured
$[10, 9]$, $[1,9]$, $[6,1]$ and $[8,1]$ have performance losses of about $0.075$ dB, $0.065$ dB,
$0.012$ dB and $0.004$ dB, respectively, compared with the unpunctured code.
Fig.~\ref{fig:fer_r4_puncture} plots the FER of the unpunctured/punctured codes
and it shows a similar relative error performance.
We also simulate the code when both $[6,1]$ and $[8,1]$ P-VNs are punctured.
The code rate is further increased to $0.0506$.
The error performance of the code, as shown in Figs.~\ref{fig:ber_r4_puncture}  and
Fig.~\ref{fig:fer_r4_puncture}, is found to be between punctured $[6,1]$ and $[8,1]$.

Fig.~\ref{fig:it_r4_puncture} plots the average number of iterations required to decode a codeword at different $E_b/N_0$. Punctured $[8,1]$
has the fastest convergence speed compared with other punctured codes and has almost the convergence speed as the unpunctured code.
Table \ref{tb:punc_r4} lists (i) the number of frame sent, (ii) BER, (iii) FER and (iv) average number of iterations until $100$ frame errors are reached,  at $E_b/N_0 = -1.19$ dB for the unpunctured/punctured PLDPC-Hadamard codes with $r = 4$.
The aforementioned results conclude that punctured $[8,1]$ outperforms
other punctured codes being considered and has a very similar performance as the unpunctured code.



\begin{table*}[t]
\newcommand{\tabincell}[2]{\begin{tabular}{@{}#1@{}}#2\end{tabular}}
\centering\caption{Performance of unpunctured/punctured PLDPC-Hadamard codes with $r = 4$ at $E_b / N_0 = -1.19$ dB.}\label{tb:punc_r4} 
\begin{center} \small
 \begin{tabular}{|c|c|c|c|c|c|c|c|}
\hline
Punctured P-VN(s) &Unpunctured  & $[1,9]$  & $[10,9]$  & $[6,1]$  & $[8,1]$  & $[6,1] \& [8,1]$\\
\hline
Code rate &	0.0494 &	0.0500 &	0.0500 &	0.0500 &	0.0500 &	0.0506\\
\hline
FER	& $1.2 \times 10^{-4}$ & $1.1 \times 10^{-2}$ & $1.4 \times 10^{-2}$& $1.7 \times 10^{-3}$ & $1.4 \times 10^{-4}$ & $8.0 \times 10^{-4}$\\
\hline
BER	& $9.1\times 10^{-6}$ & $2.8 \times 10^{-3}$ & $3.9 \times 10^{-3}$& $4.1 \times 10^{-5}$ & $1.2 \times 10^{-5}$ & $1.7 \times 10^{-5}$\\
\hline
{\tabincell{c}{ No. of frames sent}} &832,056 &9,314	 &6,962	&59,564	 &727,997	 &124,672	\\
\hline
{\tabincell{c}{Av. no. of iterations}}&127	&135	&137	&138	&127	&135	\\
\hline
\end{tabular}
\end{center}
\end{table*}

\begin{figure}[htbp]
\centerline{
\includegraphics[width=0.6\columnwidth]{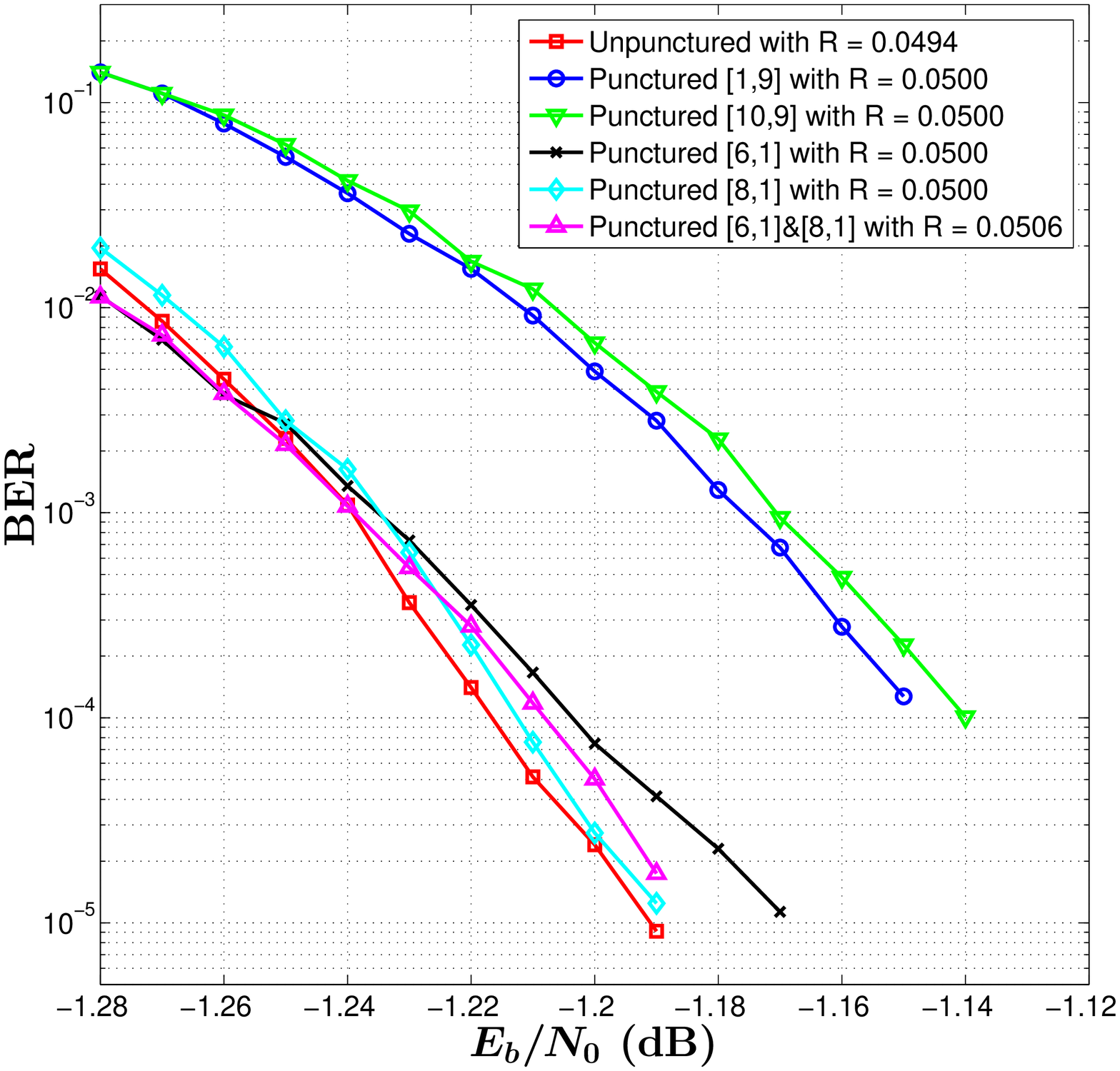}}
 \caption{BER performance of unpunctured/punctured PLDPC-Hadamard codes. One or two P-VNs is/are punctured. $r = 4$ and $k = 65,536$. }
  \label{fig:ber_r4_puncture}
\end{figure}

\begin{figure}[htbp]
\centerline{
\includegraphics[width=0.6\columnwidth]{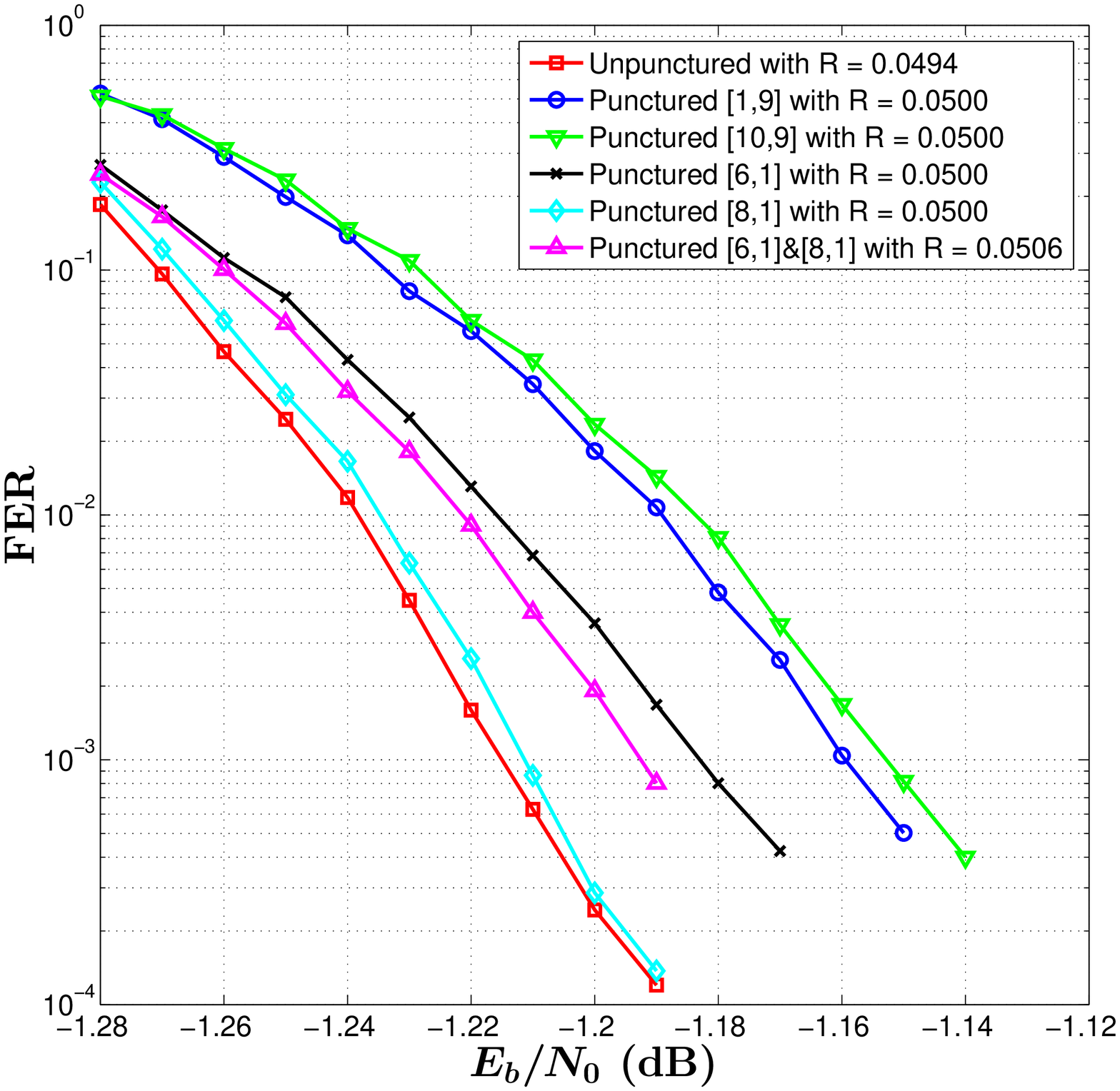}}
 \caption{FER performance of unpunctured/punctured PLDPC-Hadamard codes. One or two P-VNs is/are punctured. $r = 4$ and $k = 65,536$. }
  \label{fig:fer_r4_puncture}
\end{figure}

\begin{figure}[htbp]
\centerline{
\includegraphics[width=5in]{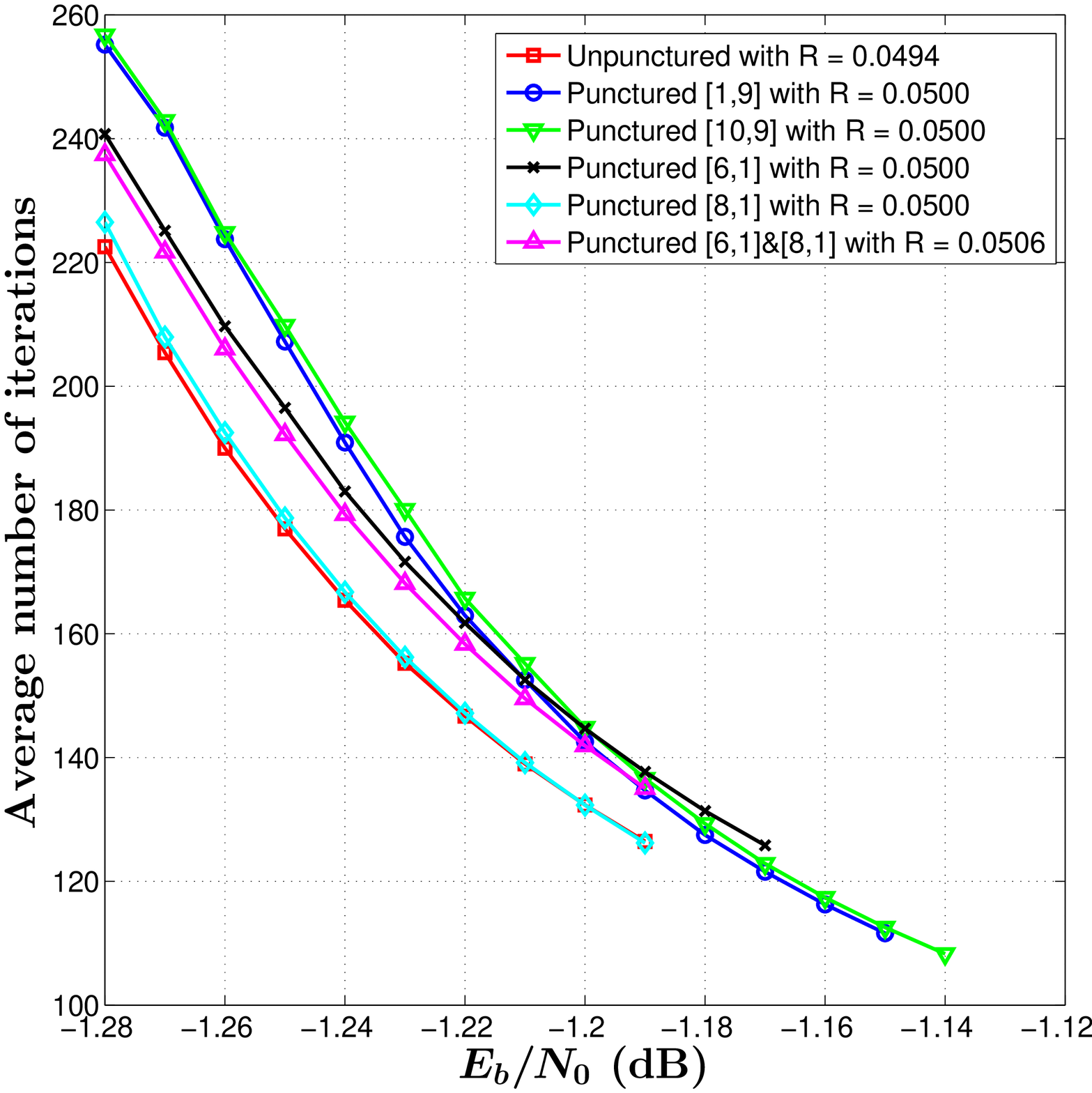}}
 \caption{Average number of iterations required to decode unpunctured/punctured PLDPC-Hadamard codes. One or two P-VNs is/are punctured. $r = 4$ and $k = 65,536$. }
  \label{fig:it_r4_puncture}
\end{figure}


\subsubsection{${r = 5}$}



We consider the rate-$0.021$ PLDPC-Hadamard code shown in  \eqref{mat:r=5};
and puncture $[8, 9]$ (largest degree) and $[9, 1]$ (lowest degree), respectively.
%
After puncturing, both codes have a rate of $0.02116$ (by applying \eqref{eq:odd2_R_punc}).
Fig. \ref{fig:fer_ber_r5_puncture} shows that punctured $[9, 1]$ achieves the lowest BER
while punctured $[8, 9]$ achieves the lowest FER.
Fig.~\ref{fig:it_r5_puncture} plots the average number of decoding iterations.
The results indicate that punctured $[8, 9]$ converges faster than the unpunctured code, which in turn converges faster than punctured $[9, 1]$.

%
\begin{figure}[t]
\centerline{
\includegraphics[width=0.6\columnwidth]{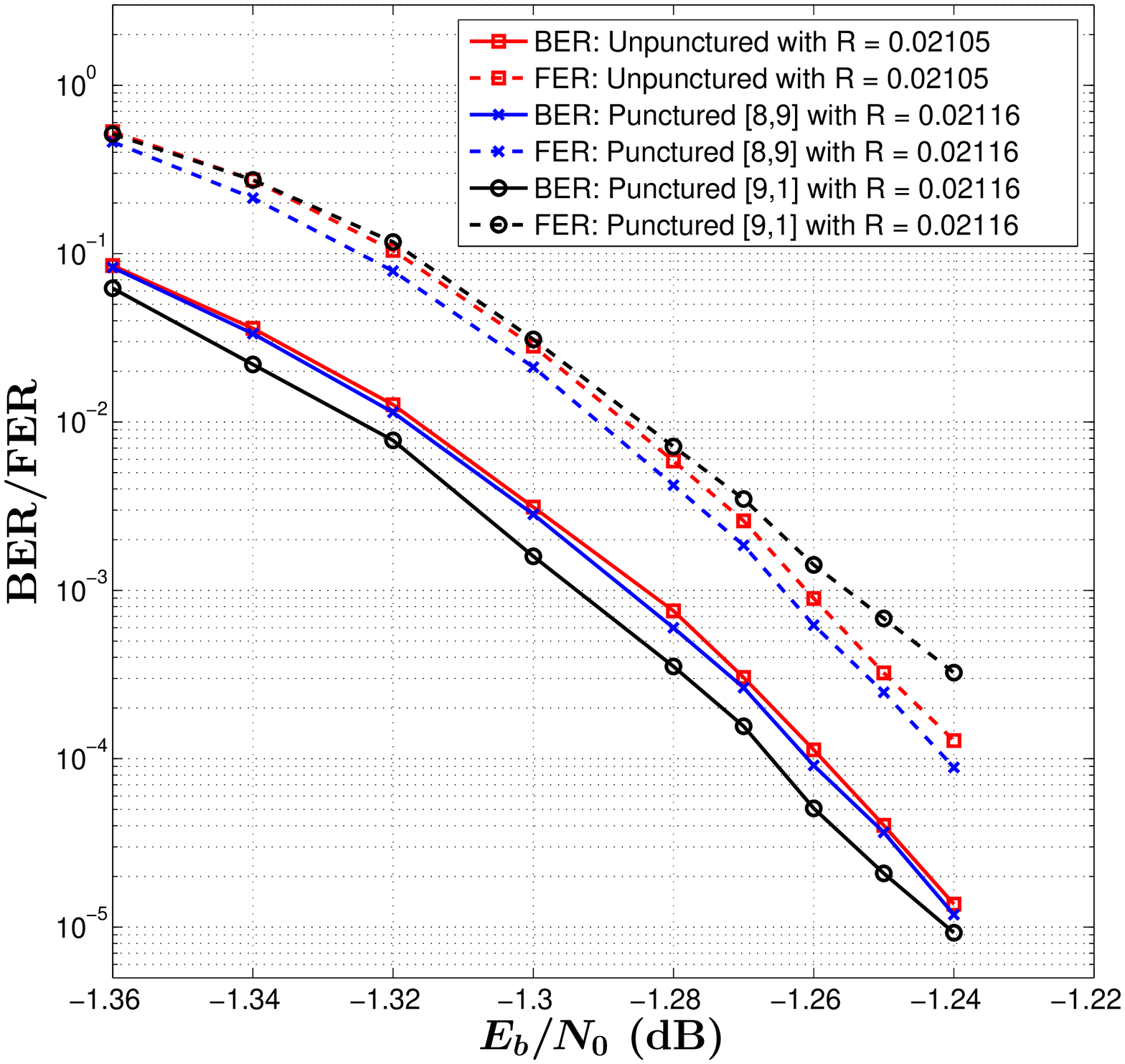}}
 \caption{BER/FER performance of unpunctured/punctured PLDPC-Hadamard codes. One P-VN is punctured.  $r = 5$ and $k = 65,536$. }
  \label{fig:fer_ber_r5_puncture}
\end{figure}

\begin{figure}[htbp]
\centerline{
\includegraphics[width=0.6\columnwidth]{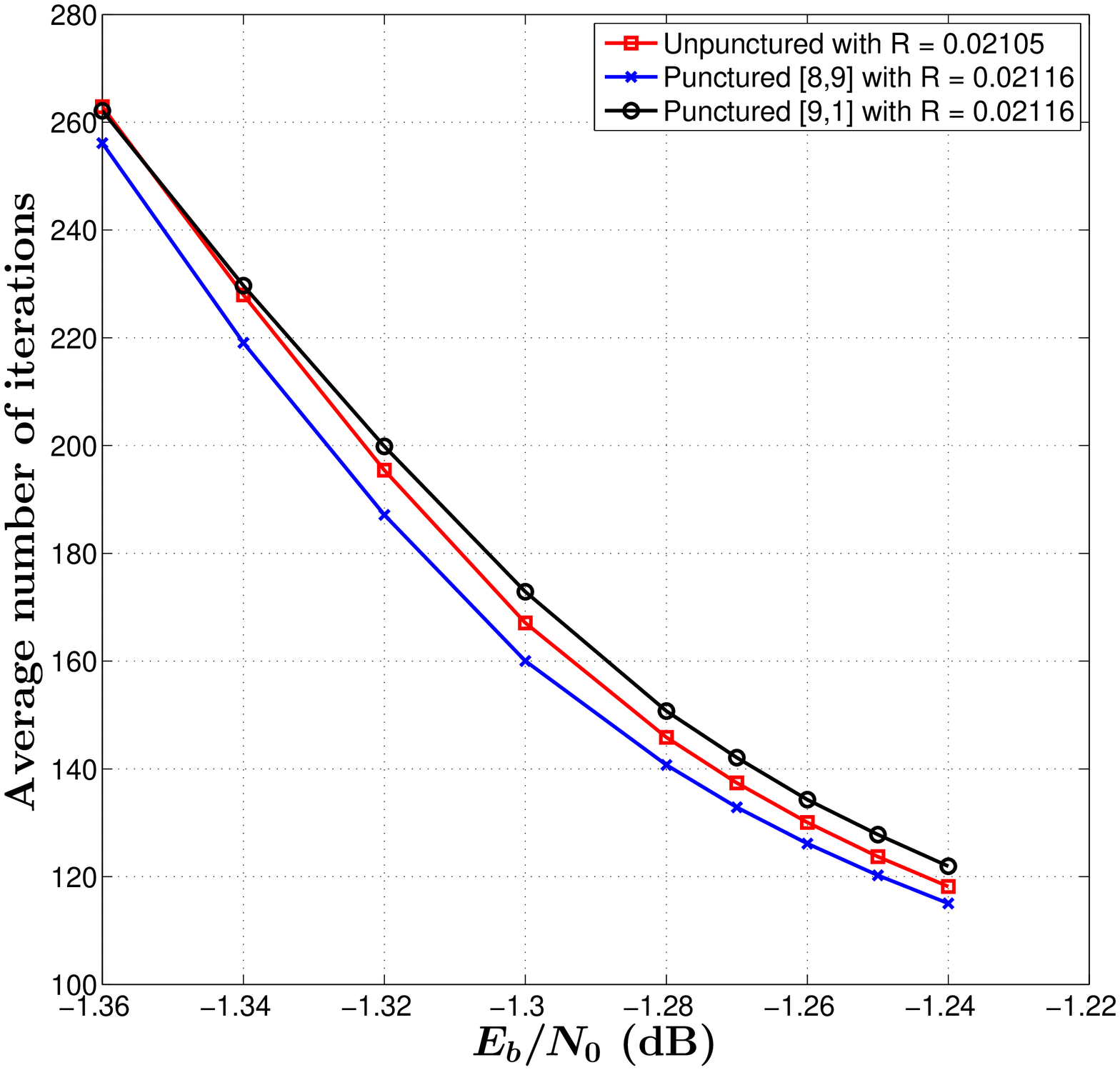}}
 \caption{Average number of iterations required to decode unpunctured/punctured PLDPC-Hadamard codes. One P-VN is punctured. $r = 5$ and $k = 65,536$. }
  \label{fig:it_r5_puncture}
\end{figure}

We further consider puncturing D1H-VNs
corresponding to code bits $c_{2^{k-1}}^H$ ($k=1,2,\dots,r$) for every H-CN.
The rate of such punctured codes is computed using \eqref{eq:odd3_R_punc}.
%
We use $[c_{1}^H\ c_{2}^H\ \cdots c_{2^{k-1}}^H  ]$ ($1 \le k \le r$) to denote the set of bits being punctured.
Three sets of punctured bits are being considered. They are
$[c_{8}^H\ c_{16}^H]$, $[c_{2}^H\ c_{4}^H\ c_{8}^H\ c_{16}^H]$ and $[c_{1}^H\ c_{2}^H\ c_{4}^H\ c_{8}^H\ c_{16}^H]$; and their corresponding rates are $0.022$, $0.024$ and $0.025$, respectively.
Fig. \ref{fig:fer_ber_r5_puncture_2} shows that  in terms of BER and FER, all the punctured codes are degraded compared with the unpunctured rate-$0.022$ PLDPC Hadamard code.
Particularly compared with the unpunctured  code,
punctured $[c_{8}^H\ c_{16}^H]$ has a $0.02$ dB performance loss at a BER of $3.6\times 10^{-5}$;
punctured $[c_{2}^H\ c_{4}^H\ c_{8}^H\ c_{16}^H]$ has a $0.03$ dB performance loss
at a BER of $4.7\times 10^{-5}$;
and punctured $[c_{1}^H\ c_{2}^H\ c_{4}^H\ c_{8}^H\ c_{16}^H]$ has a $0.04$ dB performance loss  at a BER of $1.4\times 10^{-5}$.
The BER/FER results indicate that the channel observations corresponding to these D1H-VNs
 provide very useful information for the decoder to decode successfully.
Fig. \ref{fig:it_r5_puncture_2} plots the average number of decoding iterations.
It shows that the unpunctured code requires the lowest number of decoding iterations.


\begin{figure}[t]
\centerline{
\includegraphics[width=0.6\columnwidth]{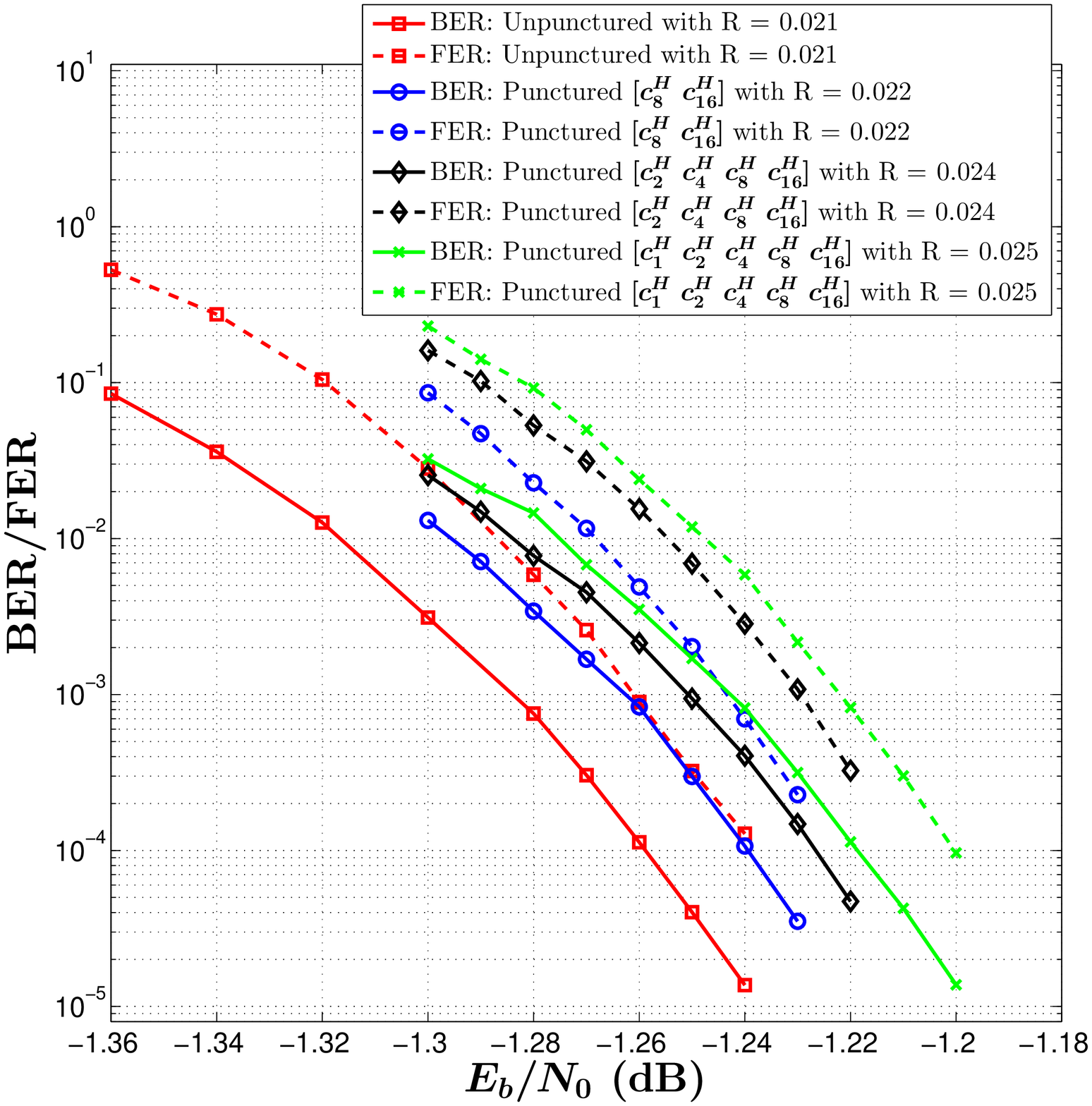}}
 \caption{BER/FER performance of unpunctured/punctured PLDPC-Hadamard codes. Two, four and five D1H-VNs are punctured.  $r = 5$ and $k = 65,536$.
}
  \label{fig:fer_ber_r5_puncture_2}
\end{figure}

\begin{figure}[htbp]
\centerline{
\includegraphics[width=0.6\columnwidth]{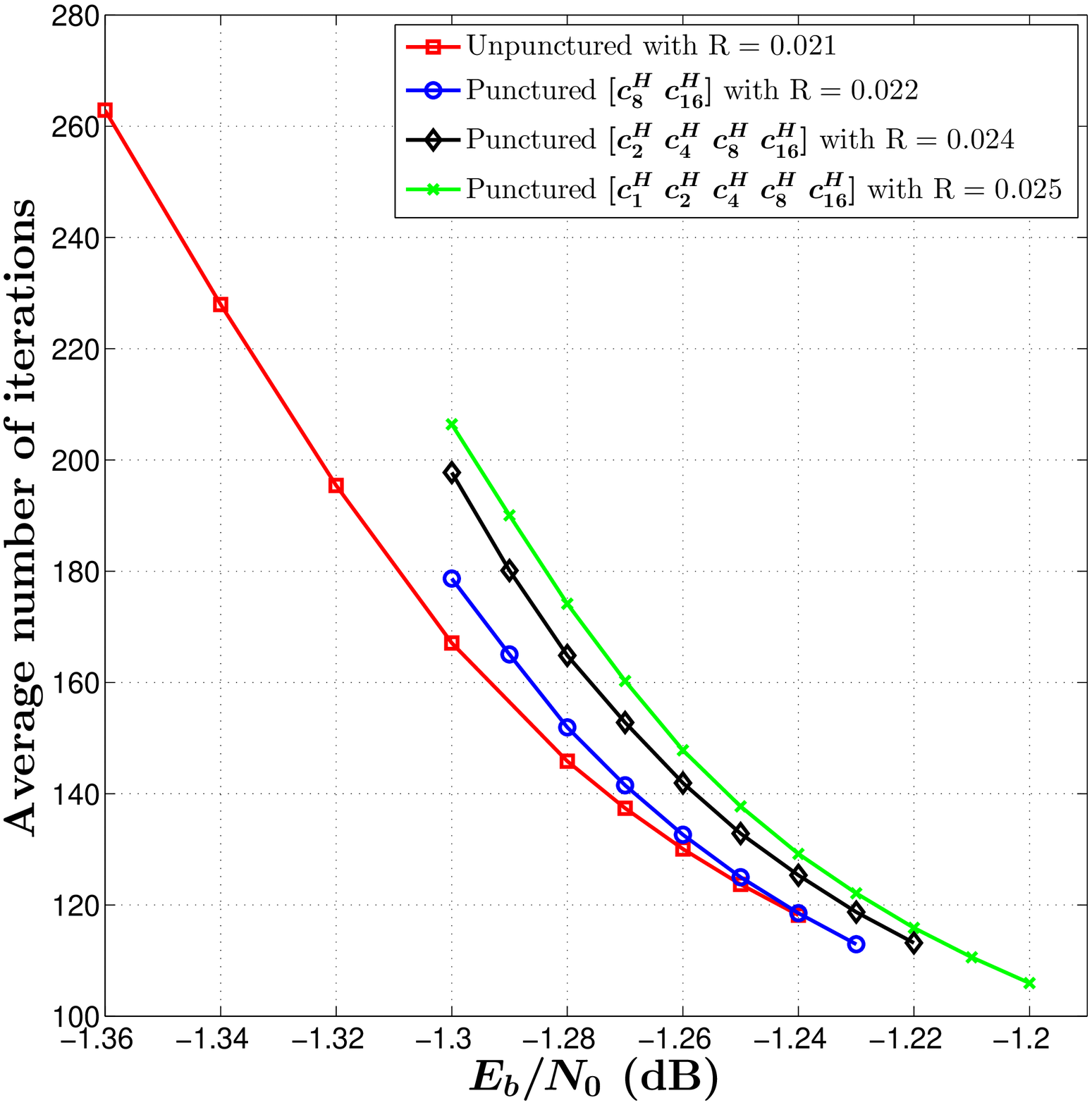}}
 \caption{ Average number of iterations required to decode unpunctured/punctured PLDPC-Hadamard codes.
 Two, four and five D1H-VNs are punctured.
 $r = 5$ and $k = 65,536$. }
  \label{fig:it_r5_puncture_2}
\end{figure}

\subsubsection{${r = 8}$}
We consider the rate-$0.008$ PLDPC-Hadamard code shown in  \eqref{mat:r=8};
and puncture $[12, 11]$ (largest degree) and $[2, 2]$ (lowest degree), respectively. 
The code rate is increased slightly from $0.008032$ to $0.008038$. 
Figs. \ref{fig:fer_ber_r8_puncture} and \ref{fig:it_r8_puncture} show that
compared with the unpunctured code,
the punctured ones are degraded only very slightly  in terms of BER/FER and
have almost the same convergence rates.

\begin{figure}[t]
\centerline{
\includegraphics[width=0.6\columnwidth]{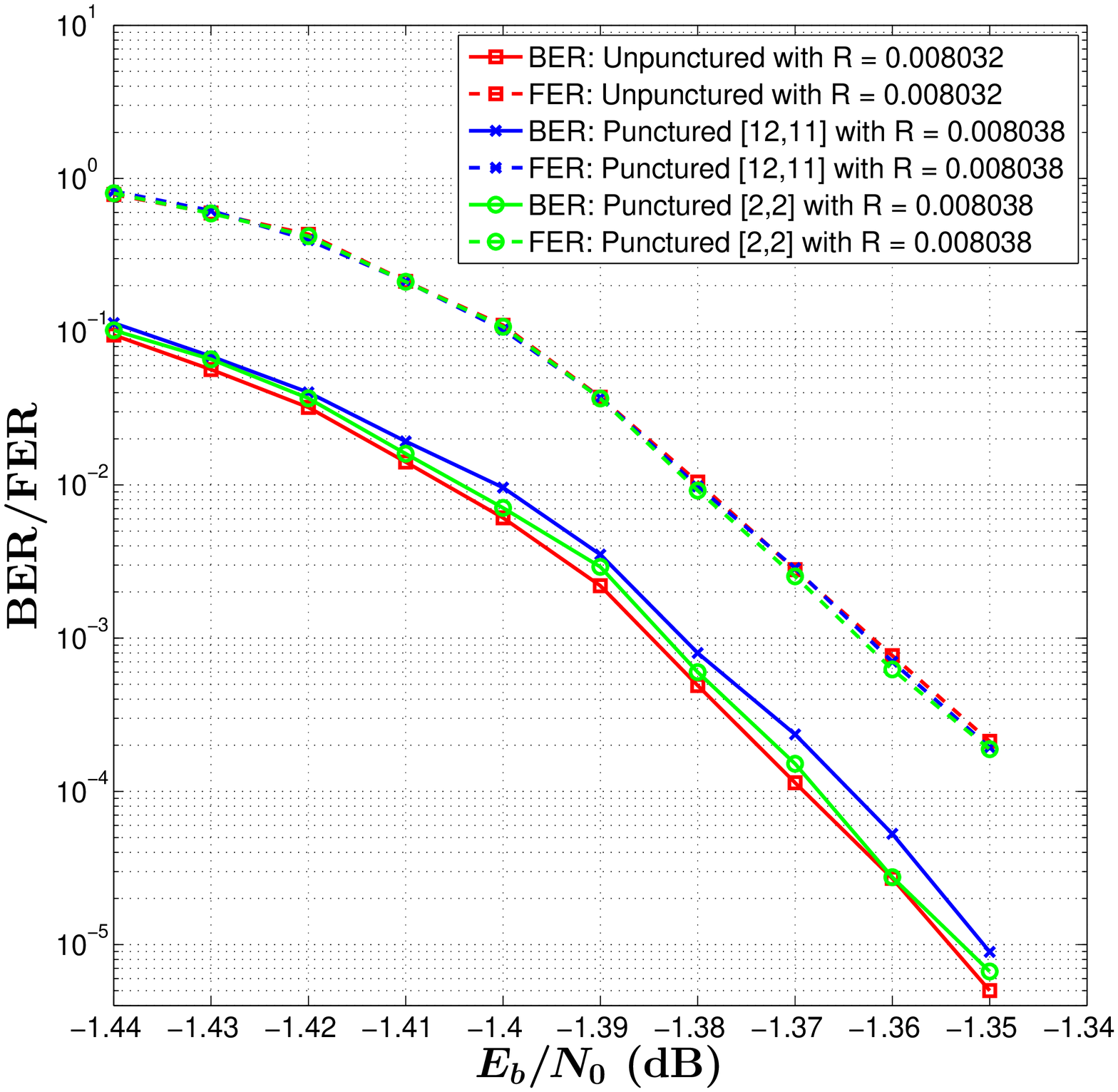}}
 \caption{BER/FER performance of unpunctured/punctured PLDPC-Hadamard codes. One P-VN is punctured. 
 $r = 8$ and $k = 204,800$. }
  \label{fig:fer_ber_r8_puncture}
\end{figure}

\begin{figure}[t]
\centerline{
\includegraphics[width=0.6\columnwidth]{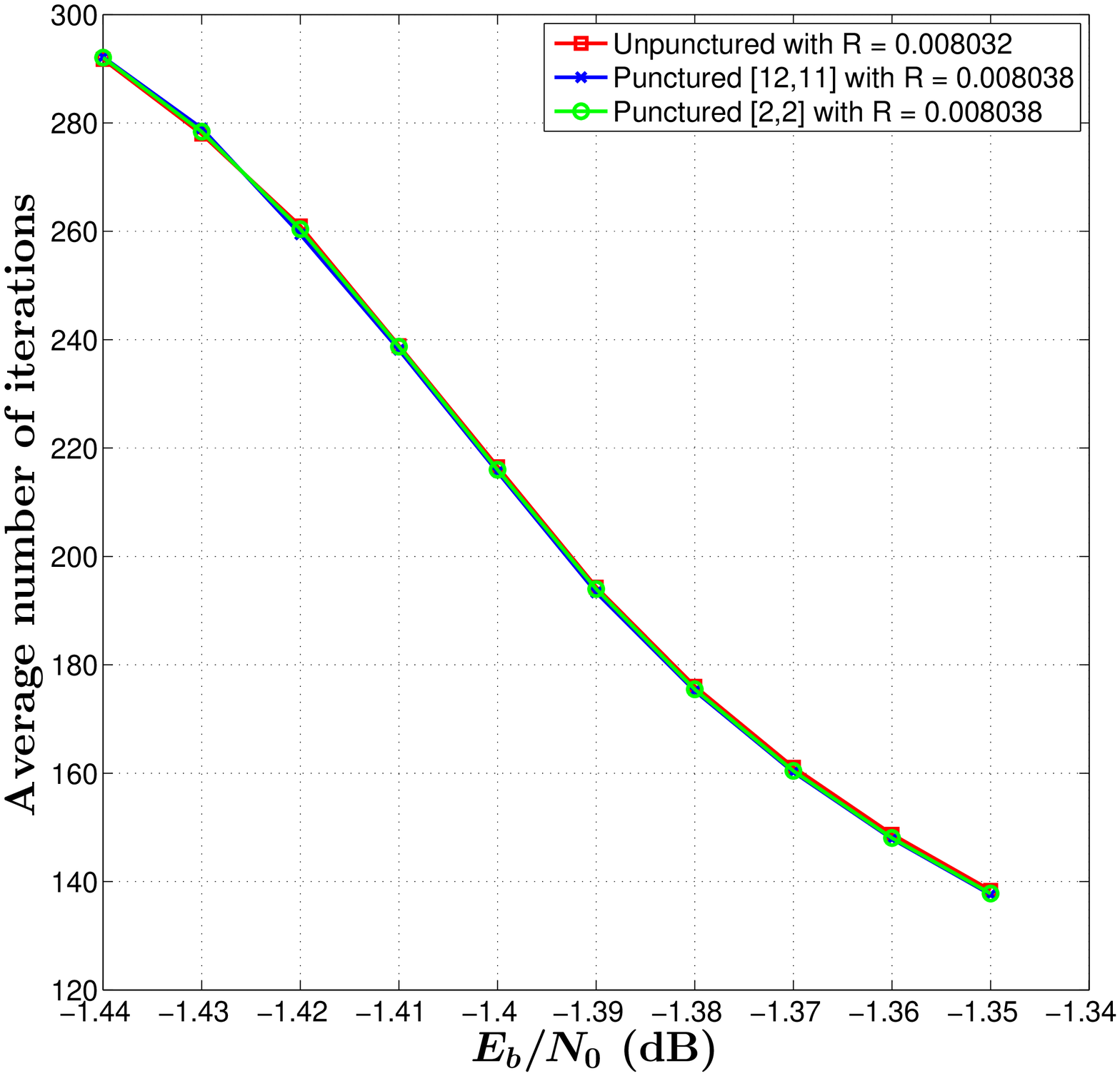}}
 \caption{ Average number of iterations required to decode unpunctured/punctured PLDPC-Hadamard codes. One P-VN is punctured.
 $r = 8$ and $k = 204,800$. }
  \label{fig:it_r8_puncture}
\end{figure}

\subsubsection{${r = 10}$}
We consider the rate-$0.002950$ PLDPC-Hadamard code shown in  \eqref{mat:r=10};
and puncture $[21, 11]$ (largest degree) and $[3, 2]$ (lowest degree), respectively.
The code rate is increased slightly from $0.002950$ to $0.002953$.
Figs. \ref{fig:fer_ber_r10_puncture} and \ref{fig:it_r10_puncture} show that
compared with the unpunctured code,
the punctured ones have almost the same performance in terms of BER/FER and
 convergence rate.


\begin{figure}[t]
\centerline{
\includegraphics[width=0.6\columnwidth]{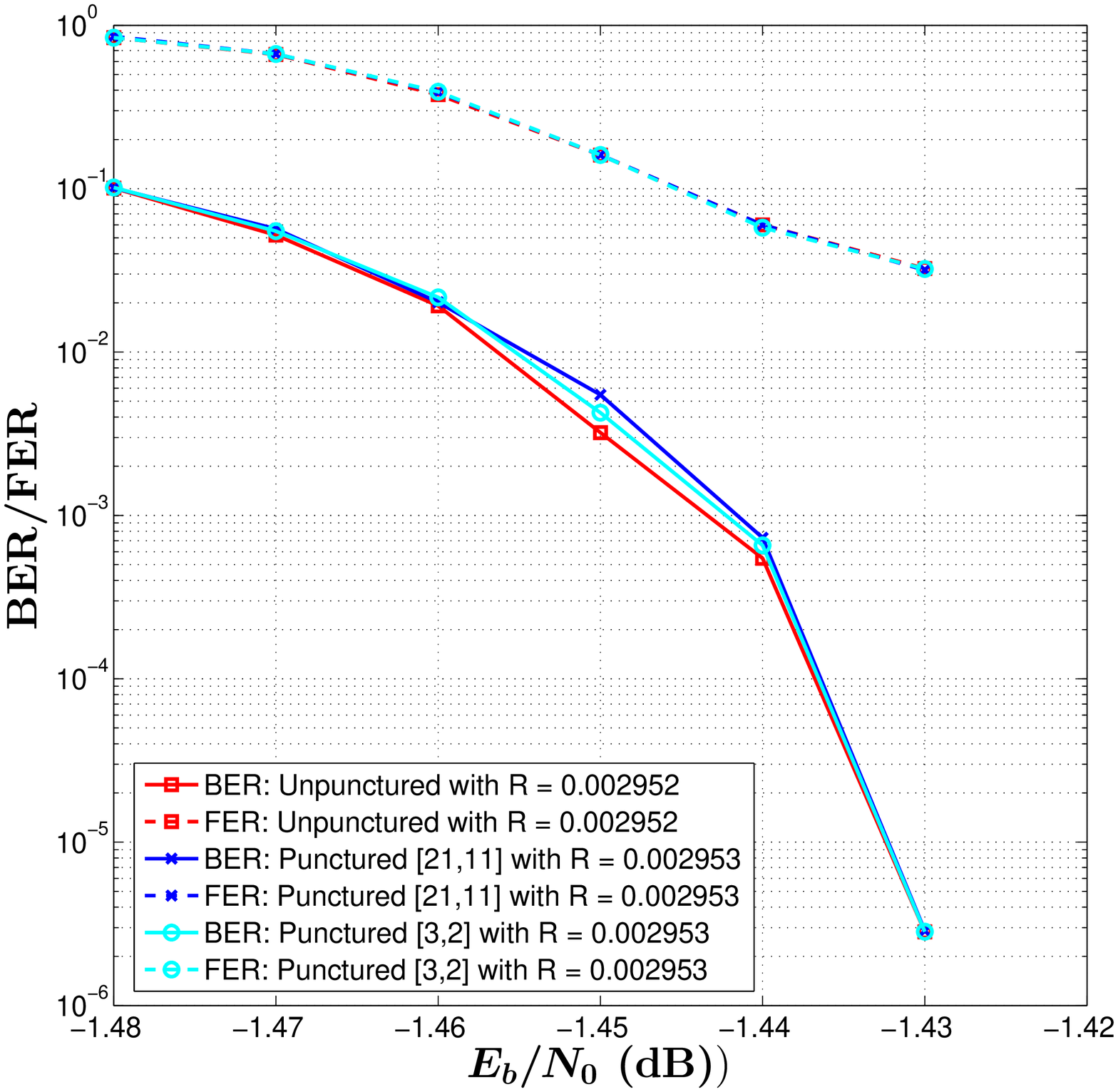}}
 \caption{BER/FER performance of unpunctured/punctured PLDPC-Hadamard codes. One P-VN is punctured. 
 $r = 10$ and $k = 460,800$. }
  \label{fig:fer_ber_r10_puncture}
\end{figure}

\begin{figure}[t]
\centerline{
\includegraphics[width=0.6\columnwidth]{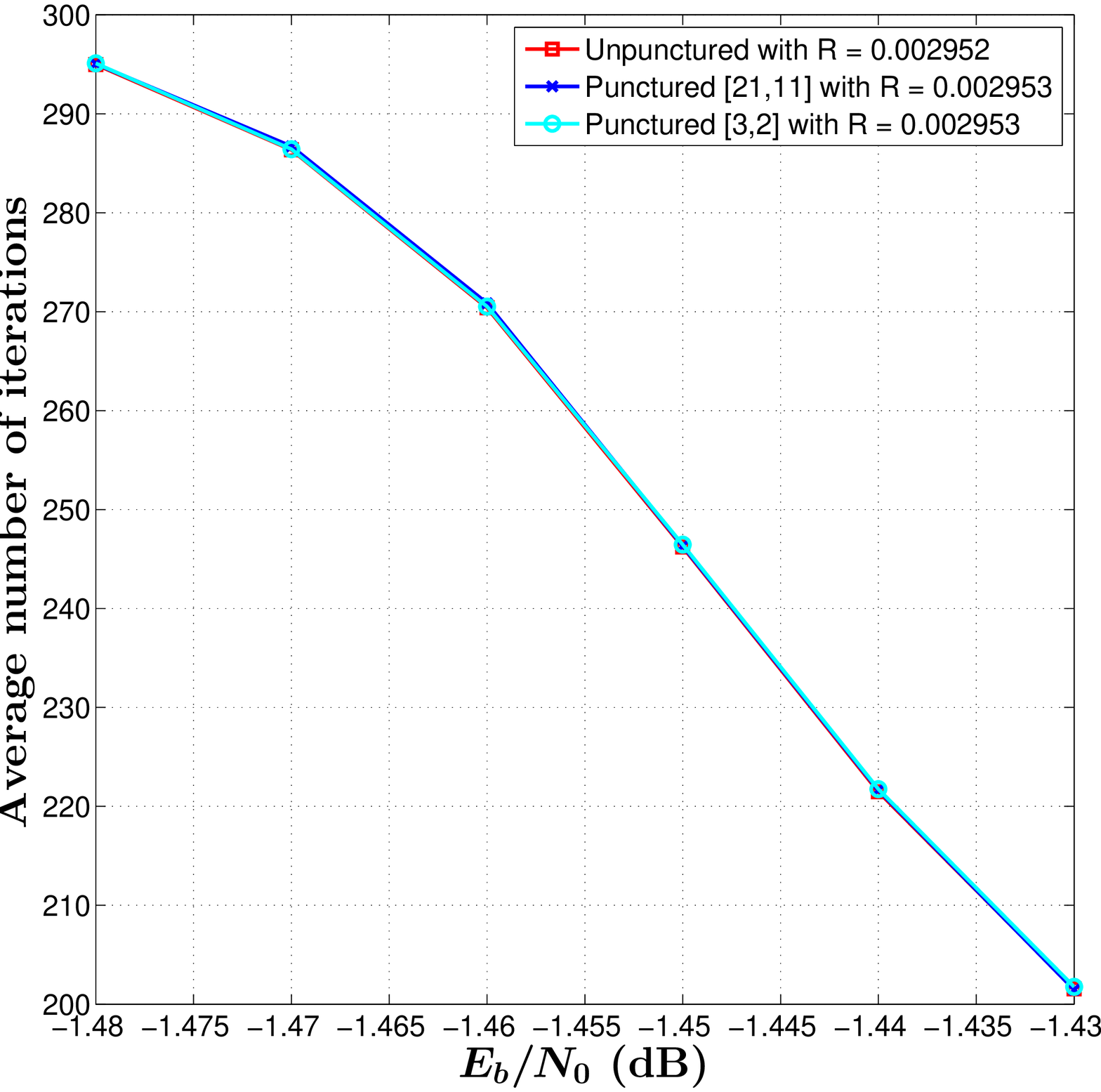}}
 \caption{ Average number of iterations required to decode unpunctured/punctured PLDPC-Hadamard codes. One P-VN is punctured.
 $r = 10$ and $k = 460,800$. }
  \label{fig:it_r10_puncture}
\end{figure}

\section{Conclusion}\label{sect:conclusion}
In this paper, we have proposed an alternate method of designing
ultimate-Shannon-limit-approaching LDPC-Hadamard codes ---- protograph-based LDPC-Hadamard (PLDPC-Hadamard) codes.
By appending degree-$1$ Hadamard
variable nodes (D1H-VNs) to the protograph of LDPC codes,
a generalized protograph can be formed to characterize the structure of PLDPC-Hadamard codes.
We have also proposed a low-complexity PEXIT algorithm to analyze the threshold of the codes, which is valid for PLDPC-Hadamard protographs with degree-$1$ variable nodes and/or punctured variable nodes/D1H-VNs.
Based on the proposed analysis method, we have found good PLDPC-Hadamard codes with different code rates and
have provided the corresponding protomatrices with very low thresholds ($<-1.40$ dB).

Reliable BER, FER and average number of decoding iterations are derived by running simulations until $100$ frame errors are obtained.
At a BER of $10^{-5}$, the gaps of our codes to the ultimate-Shannon-limit
range from $0.40$~dB (for rate = $0.0494$) to $0.16$~dB (for rate = $0.003$).
Moreover, the error performance of our codes is comparable to that of the traditional
 LDPC-Hadamard codes.
We have also investigated punctured PLDPC-Hadamard codes.
When the order of the Hadamard code $r=4$, puncturing different variable nodes in the protograph produce quite different BER/FER performance degradation compared with the unpunctured code.
When  $r=5$, puncturing one VN can actually improve the BER/FER performance slightly.
When  $r=8$ or $10$, puncturing one VN does not seem to have any effect.
Moreover, we conclude that when $r=5$,
puncturing the extra D1H-VNs provided by the non-systematic Hadamard code
degrades the error performance quite significantly.

Finally, we have made use of our proposed analytical technique to find optimal PLDPC-Hadamard code designs. By puncturing these PLDPC-Hadamard codes, punctured codes are obtained and simulated. However, these punctured codes, strictly speaking, are not optimized. In the future, we plan to apply
the proposed analytical technique to find optimal PLDPC-Hadamard codes with punctured VNs and compare their results with those presented in this work.
We will also investigate annealing approaches or genetic algorithms to 
speed up the search for optimal protomatrices under some given constraints.
We will consider spatially-coupled PLDPC-Hadamard codes, and compare their theoretical 
thresholds and error performance with the optimal PLDPC-Hadamard (block) codes. 

\clearpage
\newpage
\appendices
\section{Computing the APP LLRs of the Information Bits in a Non-Systemic Hadamard Code} \label{app:a}
\setcounter{equation}{0}
\renewcommand{\theequation}{\thesection\arabic{equation}}
We convert the extrinsic LLR values $\bm{L}^H_{ex}$ of the SPC code bits $\bm{c}_{\mu}$ to the \textit{a priori} LLR values of the information bits $\bm{c}'_{\mu}$ in $\bm{c}^H$.
In \eqref{eq:c'}, $c_0^H = c'_{{\mu _0}} = c_{{\mu _0}}$ and $c_{{2^{r}-1}}^H = c'_{{\mu _{r+1}}} = c_{{\mu _{r+1}}}$.
Hence, the \textit{a priori} information for $c_0^H = c'_{{\mu _0}}$ and $c_{2^{r}-1}^H = c'_{\mu _{r+1}}$ equals the extrinsic information for $c_{\mu_0}$ and $c_{\mu_{r+1}}$, that is,
\begin{eqnarray}
L_{apr}^H( 0 ) &=& L_{ex}^R(0); \nonumber \\
L_{apr}^H( 2^r-1 ) &=& L_{ex}^R(r+1).
\end{eqnarray}
For $k=1,2,\ldots,r$, \eqref{eq:c'} shows that $c^H_{2^{k-1}} = c'_{\mu_k} = c_{\mu_k} \oplus c_{\mu_0}$.
$L_{ex}^R(k)$ is the \textit{a priori} information for $c_{\mu_k}$, i.e.,
\begin{equation}
L_{ex}^R(k) = \ln \frac{\Pr( c_{\mu_k}= ``0" )}{\Pr( c_{\mu_k} = ``1" )}
= \ln \frac{\Pr(c'_{\mu_k}\oplus c_{\mu_0}= ``0" )}{\Pr( c'_{\mu_k}\oplus c_{\mu_0} = ``1" )}.
\end{equation}
Alternatively,
\begin{eqnarray}
&&L_{apr}^H( 2^{k-1} ) = \ln \frac{\Pr( c^H_{ 2^{k-1} }= ``0" )}{\Pr( c^H_{ 2^{k-1} } = ``1" )} \nonumber\\ 
&&= \ln \frac{\Pr( c'_{\mu_k}= ``0" )}{\Pr( c'_{\mu_k} = ``1" )}= \ln \frac{\Pr(c_{\mu_k}\oplus c_{\mu_0}= ``0" )}{\Pr( c_{\mu_k}\oplus c_{\mu_0} = ``1" )}\nonumber\\ 
&&= \left\{
\begin{array}{ll}
L_{ex}^R(k)  & {\rm if} \; c_{\mu_0}= ``0" \\[3pt]
-L_{ex}^R(k)  &  {\rm if} \; c_{\mu_0}= ``1"
\end{array}.
\right.
\end{eqnarray}
The $2^r-r-2$ remaining ${L}_{apr}^H$ values should be $0$. 
Thus, the assignment of $\bm{L}_{apr}^{+H}$ (if $c_{\mu_0}=``0"$) and $\bm{L}_{apr}^{-H}$ (if $c_{\mu_0}=``1"$) is as follows:
\begin{eqnarray}\label{assign_apr}
&&\!\!\!\!\!\!\!\!\!\!\!\!
L_{apr}^{\pm H}( k ) = L_{ex}^R(0) \ \ \ \ \ \ \ {\rm for} \; k = 0;\nonumber \\
&&\!\!\!\!\!\!\!\!\!\!\!\!\!\!\!\!\!\!\!\!
\left\{
\begin{array}{ll}
L_{apr}^{+ H}( k ) = L_{ex}^R(i) \nonumber\\[3pt]
L_{apr}^{- H}( k ) = -L_{ex}^R(i)  \nonumber
\end{array}
{ \ \ \ \; \rm for} \; k = 1,2,\cdots,2^{i-1},\cdots, 2^{r-1}; \right.\\
&&\!\!\!\!\!\!\!\!\!\!\!\!
L_{apr}^{\pm H}( k ) = L_{ex}^R(r+1) \ \; {\rm for} \; k = 2^r-1;  \nonumber\\
&&\!\!\!\!\!\!\!\!\!\!\!\!
L_{apr}^{\pm H}( k ) = 0\  {\ \ \ \ \ \ \ \ \ \ \ \ \rm for\ the\ } 2^r-r-2 {\ \rm remaining } \; k.\nonumber \\
\end{eqnarray}
The $2^r-2$ channel observations corresponding to the code bits $c_1^H$ to ${\ c_{{2^{r}-2}}^H}$ are received and the assignment of $\bm{L}_{ch}^H$ is as follows:
\begin{equation}\label{assign_ch}
\left\{
\begin{array}{ll}
L_{ch}^H( k ) = \frac{ 2y^H_{ch}(k) }{ \sigma_{ch}^2 }  &{\rm for} \; k = 1,2,\cdots,2^r-2; \\[3pt]
L_{ch}^H( k ) = \frac{ 2y^H_{ch}(k) }{ \sigma_{ch}^2 } = 0 &{\rm for} \;k = 0,2^r-1.
\end{array}
\right.
\end{equation}
\noindent \underline{$c_{\mu _0}$ and $c_{\mu _{r+1}}$}:
Since $c_{{\mu _0}}= c'_{{\mu _0}}= c^H_0$ and $ c_{{\mu _{r+1}}} =c'_{{\mu _{r+1}}}= c^H_{2^r-1}$ in \eqref{eq:c'},
we can apply DFHT directly to \eqref{eq:cpt_app} to obtain the \textit{a posteriori} LLR values $L_{app}^H(0)$ for $c_{\mu _0}$ and $L_{app}^H(2^r-1)$
for $c_{\mu _{r+1}}$.

\noindent \underline{$c_{\mu _k}$ for $k=1, 2, \cdots, r$}:
Based on the relationship between  $c_{\mu _k}$ and $c'_{\mu _k}$,
we derive \eqref{eq:cpt_app_k} for computing ${L_{app}^H}(2^{k-1})$.
Note that the first element in $+ \bm{h}_j$ is always $+1$  (corresponds to bit ``$c^H_0 = ``0"$').
Thus, the term $\sum_{+ H\left[ {2^{k-1},j} \right] =  + 1}\Pr( \bm{c}^H = +\bm{h}_j \mid \bm{y}_{ch}^H )$ can be used to compute $\Pr(c^H_{2^{k-1}} = ``0",c^H_0 = ``0" \mid \bm{y}_{ch}^H )$ in \eqref{eq:cpt_app_k}. Using a similar argument,
we arrive at the other three summation terms.
\begin{figure*}[htbp]
\begin{eqnarray}\label{eq:cpt_app_k}
&&
{L_{app}^H}(2^{k-1}) = \ln \frac{\Pr(c_{\mu_k} = ``0"\mid \bm{y}_{ch}^H)}{\Pr(c_{\mu_k} = ``1"\mid \bm{y}_{ch}^H)}
= \ln \frac{\Pr(c'_{\mu_k}\oplus c_{\mu_0} = ``0"\mid  \bm{y}_{ch}^H)}{\Pr(c'_{\mu_k}\oplus c_{\mu_0}  = ``1"\mid \bm{y}_{ch}^H)} \nonumber\\
&&
=\ln \frac{\Pr(c'_{\mu_k} = ``0",c_{\mu_0} = ``0" \mid \bm{y}_{ch}^H )+\Pr(c'_{\mu_k} = ``1", c_{\mu_0} = ``1"\mid \bm{y}_{ch}^H )}
{\Pr(c'_{\mu_k} = ``1", c_{\mu_0} = ``0"\mid \bm{y}_{ch}^H )+\Pr(c'_{\mu_k} = ``0", c_{\mu_0} = ``1"\mid \bm{y}_{ch}^H )}\nonumber\\
&&
=\ln \frac{\Pr(c^H_{2^{k-1}} = ``0",c^H_0 = ``0" \mid \bm{y}_{ch}^H )+\Pr(c^H_{2^{k-1}} = ``1", c^H_0 = ``1"\mid \bm{y}_{ch}^H )}
{\Pr(c^H_{2^{k-1}} = ``1", c^H_0 = ``0"\mid \bm{y}_{ch}^H )+\Pr(c^H_{2^{k-1}} = ``0", c^H_0 = ``1"\mid \bm{y}_{ch}^H )}\nonumber\\
&&
= \ln \frac{{\sum\limits_{+ H\left[ {2^{k-1},j} \right] =  + 1}\Pr( \bm{c}^H = +\bm{h}_j \mid \bm{y}_{ch}^H ) }+{\sum\limits_{- H\left[ {2^{k-1},j} \right] =  - 1}\Pr( \bm{c}^H = -\bm{h}_j \mid \bm{y}_{ch}^H ) }}
{{\sum\limits_{+ H\left[ {2^{k-1},j} \right] =  - 1}\Pr( \bm{c}^H = +\bm{h}_j \mid \bm{y}_{ch}^H ) }+{\sum\limits_{- H\left[ {2^{k-1},j} \right] =  + 1}\Pr( \bm{c}^H = -\bm{h}_j \mid \bm{y}_{ch}^H ) }}\nonumber\\
&&
= \ln \frac{{\sum\limits_{+ H\left[ {2^{k-1},j} \right] =  + 1} {\gamma \left( { + {\bm{h}_j}} \right)} }+{\sum\limits_{- H\left[ {2^{k-1},j} \right] =  - 1} {\gamma \left( { - {\bm{h}_j}} \right)} }}
{{\sum\limits_{+ H\left[ {2^{k-1},j} \right] =  - 1} {\gamma \left( { + {\bm{h}_j}} \right)} }+{\sum\limits_{- H\left[ {2^{k-1},j} \right] =  + 1} {\gamma \left( { - {\bm{h}_j}} \right)} }}.
\end{eqnarray}
\end{figure*}

In \eqref{eq:cpt_app}, the numerator only needs to consider the case $\pm H\left[ {i,j} \right] =  + 1$ while  the denominator only needs to consider the case $\pm H\left[ {i,j} \right] =  -1$.
However, in  \eqref{eq:cpt_app_k}, both
the numerator and denominator need to consider both
$\pm H\left[ {i,j} \right] =  + 1$ and $\pm H\left[ {i,j} \right] =  -1$; and thus
DFHT cannot be used directly to compute ${L_{app}^H}(2^{k-1})$.
To apply DFHT, the following simple transformation is required.

{\color{black}Considering the $2^{k-1}$-th row ($k=1, 2, \cdots, r$) of an
order-$r$ Hadamard matrix, there are $2^{r-1}$ entries with $-H[2^{k-1}, j] = +1$ and $2^{r-1}$ entries with $-H[2^{k-1}, j] = -1$.}
(In other words, there are $2^{r-1}$ $-\bm{h}_j$'s in which the 
$2^{k-1}$-th entry ($k=1, 2, \cdots, r$) equals $+1$; and 
there are $2^{r-1}$ $-\bm{h}_j$'s in which the 
$2^{k-1}$-th entry ($k=1, 2, \cdots, r$) equals $-1$.)
We denote
\begin{itemize}
\item $\bm{J}^k_{+1}$ as the set of column indexes s.t. the element $-H[2^{k-1}, j]= +1$
\item $\bm{J}^k_{-1}$ as the set of column indexes s.t. the element $-H[2^{k-1}, j]= -1$
\end{itemize}
It can also be readily proven that if $-H[2^{k-1}, j] = \pm1$ in $-\bm{h}_j$ ($j=0,1,\ldots,2^r-1$), then
  $-H[2^{k-1}, 2^r-1-j] = \mp1$ in $-\bm{h}_{2^r-1-j}$.
Thus we have 
$$\bm{J}^k_{+1} = \{2^r-1-j\mid j\in \bm{J}^k_{-1}\}$$ and
$$\bm{J}^k_{-1} = \{2^r-1-j'\mid j'\in \bm{J}^k_{+1}\}.$$
It means that
\begin{equation}
\sum_{j\in\bm{J}^k_{+1}} {\gamma \left( { - {\bm{h}_{j}}} \right)} =
\sum_{j'\in\bm{J}^k_{-1}} {\gamma \left( { - {\bm{h}_{2^r-1-{j'}}}} \right)}
\end{equation}
and
\begin{equation}
\sum_{j\in\bm{J}^k_{-1}} {\gamma \left( { - {\bm{h}_{j}}} \right)} =
\sum_{j'\in\bm{J}^k_{+1}} {\gamma \left( { - {\bm{h}_{2^r-1-{j'}}}} \right)}.
\end{equation}
By using the simple transformation
\begin{equation}
\label{eq:rela_h2}
\gamma'( - {\bm{h}_j}) = \gamma ( - {\bm{h}_{{2^r} - 1 - j}}), \;\;
 j= 0,1, \ldots, {2^r} - 1;
\end{equation}
\eqref{eq:cpt_app_k} can be rewritten as
\begin{eqnarray}\label{eq:cpt_app_k_2}
&&\!\!\!\!\!\!\!\!\!\!\!
 \ln \frac{{\sum\limits_{+ H\left[ {2^{k-1},j} \right] =  + 1} {\gamma \left( { + {\bm{h}_j}} \right)} }+{\sum\limits_{- H\left[ {2^{k-1},j} \right] =  + 1} {\gamma \left( { - {\bm{h}_{2^r-1-j}}} \right)} }}
{{\sum\limits_{+ H\left[ {2^{k-1},j} \right] =  - 1} {\gamma \left( { + {\bm{h}_j}} \right)} }+{\sum\limits_{- H\left[ {2^{k-1},j} \right] =  - 1} {\gamma \left( { - {\bm{h}_{2^r-1-j}}} \right)} }}\nonumber\\
&&\!\!\!\!\!\!\!\!\!\!\!
= \ln \frac{{\sum\limits_{+ H\left[ {2^{k-1},j} \right] =  + 1} {\gamma \left( { + {\bm{h}_j}} \right)} }+{\sum\limits_{- H\left[ {2^{k-1},j} \right] =  + 1} {\gamma' \left( { - {\bm{h}_j}} \right)} }}
{{\sum\limits_{+ H\left[ {2^{k-1},j} \right] =  - 1} {\gamma \left( { + {\bm{h}_j}} \right)} }+{\sum\limits_{- H\left[ {2^{k-1},j} \right] =  - 1} {\gamma' \left( { - {\bm{h}_j}} \right)} }}.\nonumber\\
\end{eqnarray}
As the numerator only needs to consider the case $\pm H\left[ {i,j} \right] =  + 1$ while  the denominator only needs to consider the case $\pm H\left[ {i,j} \right] =  -1$, DFHT can be readily applied to compute \eqref{eq:cpt_app_k_2}.

\section{Monte Carlo Method for Forming the ${m\times d}$ MI matrix  $\{I_{eh}(i,k)\}$} \label{app:b}
\setcounter{equation}{0}
\renewcommand{\theequation}{\thesection\arabic{equation}}
We define the following symbols:
\begin{itemize}
\item $\bm{\sigma}_\mu= [ \sigma_{\mu _0} \ \sigma_{\mu _1} \  \ldots \  \sigma_{\mu_{d - 1}} ]$: $d$ $(=r+2)$ noise standard deviations;
\item ${\bm{c}_\mu } = [ c_{\mu _0} \ c_{\mu _1} \  \ldots \  c_{\mu_{d - 1}} ]$: a length-$d$ SPC codeword;
\item ${\bm{c}_p } = [ c_{p _0} \ c_{p_1} \  \ldots \  c_{p_{g- 1}} ]$: $g$ Hadamard parity bits generated based on the SPC ${\bm{c}_\mu }$; $g=2^r-d$ and $g = 2^r-2$, respectively, for systematic ($r$ is even) and non-systematic coding ($r$ is odd);
\item $\bm{n}_{\mu}= [ n_{\mu _0} \ n_{\mu _1} \  \ldots \  n_{\mu_{d - 1}} ]$: $d$ AWGN samples;
\item $\bm{n}_p= [ n_{p _0} \ n_{p _1} \  \ldots \  n_{p_{g - 1}} ]$: $g$ AWGN samples;
\item ${\bm{L}_\mu } = [ L_{\mu _0} \ L_{\mu _1} \  \ldots \  L_{\mu_{d - 1}} ]$: $d$ LLR values corresponding to the SPC codeword ${\bm{c}_\mu }$;
\item ${\bm{L}_p } = [ L_{p _0} \ L_{p _1} \  \ldots \  L_{p_{g - 1}} ]$: $g$ channel LLR values corresponding to the Hadamard parity bits ${\bm{c}_p }$;
\item ${\bm{L}_e } = [ L_{e _0} \ L_{e _1} \  \ldots \  L_{e_{d - 1}} ]$: $d$ extrinsic LLR values generated by the Hadamard decoder;
\item $\bm{U}$: a $w \times d$ matrix in which each row represents a length-$d$ SPC codeword; and the $k$-th column  ($k=0,1,\ldots,d-1$) corresponds to the $k$-th bit ($c_{\mu _k}$) of the SPC codeword;
\item $\bm{V}$: a $w \times d$ matrix in which each row represents a set of ($d$) extrinsic LLR values generated by  the Hadamard decoder;
and  the $k$-th column  ($k=0,1,\ldots,d-1$) corresponds to the extrinsic LLR value for
the $k$-th bit ($c_{\mu _k}$) of the SPC codeword;
\item $\bm{p}_{e_0}=[p_e( {\xi |c_{\mu _0} = ``0"} )\; p_e( {\xi |c_{\mu _1} =  ``0"} )\;  \cdots $  $p_e( {\xi |c_{\mu _{d-1}} =  ``0"} )]$: PDFs for $c_{\mu _k} = ``0"$  ($k=0,1,\ldots,d-1$);
\item $\bm{p}_{e_1}=[p_e( {\xi |c_{\mu _0} = ``1"} )\; p_e( {\xi |c_{\mu _1} =  ``1"} )\;  \cdots $  $p_e( {\xi |c_{\mu _{d-1}} =  ``1"} )]$: PDFs for $c_{\mu _k} = ``1"$  ($k=0,1,\ldots,d-1$).
\end{itemize}
%
%
%
%
The ${m\times d}$ MI matrix $\{I_{eh}(i,k)\}$ is then  updated with following steps.
\begin{enumerate}
\renewcommand{\labelenumi}{\roman{enumi})}
\item Fix the standard deviation of channel noise as $\sigma_{ch}$.
\item Set $i=0$.
\item For the $i$-th row in the MI matrix $\{I_{ah}(i, k)\}$, use \eqref{eq:J1} to compute
the standard deviation $\sigma_{\mu_k}=J^{-1}(I_{ah}(i, k))$ for $k=0,1,\ldots,d-1$.
\item Set $j=0$.
\item \label{s1} Randomly generate a length-$d$ SPC codeword ${\bm{c}_\mu }$;
further encode ${\bm{c}_\mu }$ into a Hadamard codeword using
systematic (when $r=d-2$ is even) or non-systematic (when $r$ is odd) coding and
generate the $g$ Hadamard parity bits ${\bm{c}_p }$.
\item Randomly generate a AWGN sample vector $\bm{n}_{\mu}$ where each $n_{\mu_k}$ ($k=0,1,\ldots,d-1$) follows a different  Gaussian distribution $\mathcal{N}(\sigma_{\mu_k}^{2}/2,\sigma_{\mu_k}^{2})$.
\item Randomly generate a AWGN sample vector $\bm{n}_p$ where all $n_{p_{k'}}$'s  ($k'=0,1,\ldots,g-1$)
follow the same Gaussian distribution $\mathcal{N}(\sigma_{ch}^{2}/2,\sigma_{ch}^{2})$.
\item For $k=0,1,\ldots,d-1$,
set ${L}_{\mu_k}=+n_{\mu_k}$ if $c_{\mu _k}=``0$'';
otherwise set ${L}_{\mu_k}=-n_{\mu_k}$  if $c_{\mu _k}=``1$''.
\item For $k'=0,1,\ldots,g-1$, set ${L}_{p_{k'}}=+n_{p_{k'}}$ if $c_{p _{k'}}=``0$'';
 otherwise set ${L}_{p_{k'}}=-n_{p_{k'}}$  if $c_{p _{k'}}=``1$''.
\item Input $\bm{L}_\mu$ and $\bm{L}_p$, respectively, as the \textit{a priori} and channel LLRs to the Hadamard decoder.
Use the decoding algorithm described in Sect.~\ref{PLDPCH-dec} to compute
the $d$ output extrinsic LLR values $\bm{L}_e$.
\item Assign ${\bm{c}_\mu }$ to the $j$-th row of ${\bm{U} }$
and assign $\bm{L}_e$ to the $j$-th row of $\bm{V}$.
\item Set $j = j + 1$. If $j < w$, go to Step  v). (We set $w = 10,000$.)
\item The $k$-th column ($k=0,1,\ldots,d-1$) of both
${\bm{U} }$ and $\bm{V}$ correspond to bit $c_{\mu _k}$.
Obtain the PDFs $p_e( {\xi |c_{\mu_k} = ``0"} )$
and $p_e( {\xi |c_{\mu_k} = ``1"} )$ ($k=0,1,\ldots,d-1$) based on ${\bm{U} }$ and $\bm{V}$.
\item Use $p_e( {\xi |c_{\mu_k} = ``0"} )$
and $p_e( {\xi |c_{\mu_k} = ``1"} )$ to
compute (\ref{eq:I_E}) and hence $I_{eh}(i, k)$ ($k=0,1,\ldots,d-1$).
\item Set $i = i + 1$. If $i < m$, go to step iii).
\end{enumerate}

{\color{black}
\section{Two-Step Lifting of a Base Matrix} \label{app:c}
In the first step, we ``lift'' a base matrix $\{b(i,j)\}$ by replacing each non-zero entry $b(i,j)$ with a summation of $b(i,j)$ different $z_1 \times z_1$ permutation matrices and replacing each zero entry with the  $z_1 \times z_1$  zero matrix. After the first lifting process, all entries in the lifted matrix are either ``0'' or ``1''. 
In the second step, We lift the resultant matrix again by replacing each entry ``1'' with a  $z_2 \times z_2$ circulant permutation matrix (CPM), and replacing each entry ``0'' with the $z_2 \times z_2$ zero matrix. 
As can be seen, the final connection matrix can be easily represented by a series of CPMs. 
Note that in each lifting step, the permutation matrices and CPMs are selected using the PEG algorithm \cite{Hu2005}  such that the 
girth (shortest cycle) in the resultant matrix can be maximized. 

Take the PLDPC-Hadamard code with code rate $R=0.0494$ and Hadamard code order $r = 4$ as an example, i.e.,
\begin{equation}\label{mat:r=4app}
{\bm{B}_{7 \times 11}} = \left[ {\begin{array}{*{11}{c}}
1&0&0&0&0&0&1&0&3&0&1\\
0&1&2&0&0&0&0&0&0&2&1\\
2&1&0&0&1&1&0&0&0&0&1\\
0&1&0&3&0&0&0&0&0&2&0\\
2&0&0&0&0&0&0&1&0&3&0\\
3&0&0&2&0&0&1&0&0&0&0\\
1&0&0&1&1&0&0&0&1&2&0
\end{array}} \right].
\end{equation}
The size of the optimized base matrix is $7\times11$. After lifting the base matrix twice with factors $z_1 = 32$ and $z_2 = 512$, respectively, we can obtain a $114,688(=7 \times 32 \times 512)$ by $180,224(=11\times 32\times 512)$ connection matrix between the variable nodes and the Hadamard check nodes. The $114,688$ by $180,224$ connection matrix can simply be represented by a $224(=7\times32)$ by $352(=11\times32)$ matrix whose entries are CPMs. 
Such a connection matrix is a structured quasi-cyclic (QC) matrix which greatly facilitates parallel encoding/decoding and enhances the throughput. 
In this example, there are only $6(=r+2)$ non-zero CPMs in each row. To simplify the representation, we only record the positions of these non-zero CPMs in each row and their ``cyclic-shift'' values.

 In Table \ref{tab:final_mat}, we show the details of the structured QC matrix.
Besides the header row, there is a total of 224 rows. In each row, there are 6 entries each represented as $(c,s)$. The symbol $c$ denotes the column index where the non-zero CPM locates and it ranges from $1$ to $352$; while the symbol $s$ denotes the ``cyclic-shift'' value of this non-zero CPM and ranges from $0$ to $511$. For example, the entry $(20, 379)$ in the first row shows that in the first row, (i) there is a non-zero CPM in the $20$-th column and (ii) this CPM is constructed by cyclically left-shifting the $512 \times 512$ identity matrix by $379$ columns.

\begin{center}\footnotesize
\setlength{\tabcolsep}{0.5mm}{
\begin{longtable}{|c|c|c|c|c|c|c|}
\caption{QC matrix for rate-$0.0494$ PLDPC-Hadamard code} \label{tab:final_mat}\\
\hline
ROW& ( COL, CPM ) & ( COL, CPM ) & ( COL, CPM ) & ( COL, CPM ) & ( COL, CPM ) & ( COL, CPM ) \\
\hline
$1$& $(  20, 379 )$ & $( 211, 194 )$ & $( 261, 380 )$ & $( 267, 266 )$ & $( 278, 320 )$ & $( 345, 449 )$ \\
\hline
$2$& $(   5,  85 )$ & $( 210, 114 )$ & $( 263,  47 )$ & $( 275, 313 )$ & $( 282,  78 )$ & $( 335, 369 )$ \\
\hline
$3$& $(  21, 220 )$ & $( 224, 422 )$ & $( 258, 237 )$ & $( 268, 304 )$ & $( 287, 344 )$ & $( 348, 120 )$ \\
\hline
$4$& $(   4, 212 )$ & $( 222, 342 )$ & $( 258, 435 )$ & $( 273, 374 )$ & $( 286, 164 )$ & $( 343, 269 )$ \\
\hline
$5$& $(  32,  40 )$ & $( 218, 107 )$ & $( 259,  41 )$ & $( 271, 216 )$ & $( 283, 337 )$ & $( 339, 226 )$ \\
\hline
$6$& $(  10, 205 )$ & $( 200, 292 )$ & $( 262, 156 )$ & $( 278,  94 )$ & $( 280, 501 )$ & $( 340,  57 )$ \\
\hline
$7$& $(  13, 390 )$ & $( 217, 294 )$ & $( 266, 214 )$ & $( 276,  26 )$ & $( 286,  46 )$ & $( 326, 176 )$ \\
\hline
$8$& $(  14, 102 )$ & $( 219, 122 )$ & $( 260, 356 )$ & $( 269, 363 )$ & $( 287, 230 )$ & $( 323, 464 )$ \\
\hline
$9$& $(   3,   7 )$ & $( 198,  71 )$ & $( 264, 323 )$ & $( 276, 218 )$ & $( 277, 307 )$ & $( 338, 206 )$ \\
\hline
$10$& $(  11, 364 )$ & $( 199, 230 )$ & $( 258, 475 )$ & $( 274, 101 )$ & $( 283,  82 )$ & $( 346, 238 )$ \\
\hline
$11$& $(  15, 388 )$ & $( 221, 109 )$ & $( 263, 284 )$ & $( 272, 475 )$ & $( 284, 459 )$ & $( 333,  75 )$ \\
\hline
$12$& $(   8, 436 )$ & $( 214, 193 )$ & $( 262, 385 )$ & $( 272, 509 )$ & $( 281, 412 )$ & $( 349, 491 )$ \\
\hline
$13$& $(  18, 236 )$ & $( 202, 472 )$ & $( 265, 372 )$ & $( 273,  62 )$ & $( 285, 438 )$ & $( 330, 324 )$ \\
\hline
$14$& $(   2, 353 )$ & $( 207, 431 )$ & $( 259, 132 )$ & $( 278, 446 )$ & $( 279, 272 )$ & $( 336, 328 )$ \\
\hline
$15$& $(  28, 173 )$ & $( 212, 297 )$ & $( 262, 349 )$ & $( 271, 250 )$ & $( 288,  44 )$ & $( 324, 475 )$ \\
\hline
$16$& $(  27,  30 )$ & $( 194, 170 )$ & $( 267, 448 )$ & $( 269, 405 )$ & $( 282, 453 )$ & $( 342, 493 )$ \\
\hline
$17$& $(  29,  33 )$ & $( 196, 319 )$ & $( 267,  46 )$ & $( 274, 453 )$ & $( 285, 113 )$ & $( 321, 102 )$ \\
\hline
$18$& $(  19,  84 )$ & $( 193, 318 )$ & $( 260,  84 )$ & $( 270, 127 )$ & $( 286,  63 )$ & $( 329, 388 )$ \\
\hline
$19$& $(  24, 332 )$ & $( 223,  28 )$ & $( 265,  18 )$ & $( 271, 320 )$ & $( 287, 149 )$ & $( 327, 174 )$ \\
\hline
$20$& $(   7,  99 )$ & $( 195, 444 )$ & $( 263,  15 )$ & $( 277, 297 )$ & $( 283, 183 )$ & $( 337, 148 )$ \\
\hline
$21$& $(  25, 387 )$ & $( 204, 465 )$ & $( 257, 276 )$ & $( 277,  43 )$ & $( 280, 474 )$ & $( 347, 490 )$ \\
\hline
$22$& $(  22, 338 )$ & $( 215, 362 )$ & $( 261,  48 )$ & $( 273, 253 )$ & $( 284, 195 )$ & $( 331, 413 )$ \\
\hline
$23$& $(  12,  50 )$ & $( 209, 231 )$ & $( 266, 227 )$ & $( 272, 159 )$ & $( 285, 384 )$ & $( 344, 279 )$ \\
\hline
$24$& $(  16, 383 )$ & $( 216,  91 )$ & $( 259,  69 )$ & $( 275, 115 )$ & $( 288, 265 )$ & $( 351, 180 )$ \\
\hline
$25$& $(  31,  37 )$ & $( 201, 143 )$ & $( 265, 223 )$ & $( 276,  81 )$ & $( 282, 450 )$ & $( 325, 442 )$ \\
\hline
$26$& $(  17, 212 )$ & $( 205,  37 )$ & $( 257, 453 )$ & $( 269, 231 )$ & $( 281, 376 )$ & $( 350, 507 )$ \\
\hline
$27$& $(   6, 436 )$ & $( 203, 110 )$ & $( 261, 272 )$ & $( 270, 275 )$ & $( 288, 197 )$ & $( 332, 280 )$ \\
\hline
$28$& $(  30, 252 )$ & $( 206,   5 )$ & $( 266,  85 )$ & $( 268, 347 )$ & $( 279, 379 )$ & $( 341, 432 )$ \\
\hline
$29$& $(   1, 247 )$ & $( 213, 143 )$ & $( 264, 125 )$ & $( 270, 325 )$ & $( 279, 465 )$ & $( 322,  16 )$ \\
\hline
$30$& $(  23, 245 )$ & $( 208, 246 )$ & $( 264, 469 )$ & $( 274,  75 )$ & $( 281, 373 )$ & $( 328, 429 )$ \\
\hline
$31$& $(  26, 115 )$ & $( 197, 275 )$ & $( 260, 142 )$ & $( 275, 250 )$ & $( 280, 172 )$ & $( 334, 156 )$ \\
\hline
$32$& $(   9, 509 )$ & $( 220,  28 )$ & $( 257, 246 )$ & $( 268, 414 )$ & $( 284, 251 )$ & $( 352, 164 )$ \\
\hline
$33$& $(  53,   9 )$ & $(  68, 451 )$ & $(  94, 230 )$ & $( 289, 323 )$ & $( 307, 367 )$ & $( 345, 381 )$ \\
\hline
$34$& $(  40, 226 )$ & $(  76,  35 )$ & $(  95, 234 )$ & $( 290, 287 )$ & $( 309, 174 )$ & $( 327,  97 )$ \\
\hline
$35$& $(  39,  31 )$ & $(  69, 138 )$ & $(  87, 234 )$ & $( 289, 445 )$ & $( 309,  67 )$ & $( 328,  74 )$ \\
\hline
$36$& $(  35,  79 )$ & $(  71,   1 )$ & $(  84, 264 )$ & $( 292,  29 )$ & $( 320, 133 )$ & $( 347, 426 )$ \\
\hline
$37$& $(  52,   2 )$ & $(  78, 194 )$ & $(  89, 474 )$ & $( 302, 215 )$ & $( 311, 255 )$ & $( 329, 125 )$ \\
\hline
$38$& $(  34,  79 )$ & $(  75, 299 )$ & $(  93, 272 )$ & $( 293, 177 )$ & $( 315,  89 )$ & $( 340, 383 )$ \\
\hline
$39$& $(  37, 254 )$ & $(  67, 254 )$ & $(  86, 127 )$ & $( 291, 228 )$ & $( 306, 132 )$ & $( 349, 278 )$ \\
\hline
$40$& $(  50, 172 )$ & $(  73, 403 )$ & $(  92, 500 )$ & $( 303, 367 )$ & $( 317, 126 )$ & $( 337, 198 )$ \\
\hline
$41$& $(  44, 388 )$ & $(  79, 413 )$ & $(  85,  94 )$ & $( 292,  95 )$ & $( 315, 177 )$ & $( 325, 309 )$ \\
\hline
$42$& $(  49, 206 )$ & $(  72, 233 )$ & $(  87, 489 )$ & $( 301, 220 )$ & $( 313, 174 )$ & $( 344,  86 )$ \\
\hline
$43$& $(  58, 247 )$ & $(  66, 351 )$ & $(  90, 231 )$ & $( 295,   3 )$ & $( 305,  85 )$ & $( 348, 411 )$ \\
\hline
$44$& $(  43, 248 )$ & $(  71, 476 )$ & $(  91, 160 )$ & $( 296, 232 )$ & $( 311, 208 )$ & $( 326,  60 )$ \\
\hline
$45$& $(  54, 151 )$ & $(  66, 218 )$ & $(  84, 403 )$ & $( 298, 160 )$ & $( 313,  72 )$ & $( 336,   6 )$ \\
\hline
$46$& $(  61, 219 )$ & $(  76, 441 )$ & $(  94, 417 )$ & $( 298, 326 )$ & $( 306,  54 )$ & $( 333, 371 )$ \\
\hline
$47$& $(  46, 462 )$ & $(  80, 117 )$ & $(  82,  63 )$ & $( 295, 507 )$ & $( 318, 431 )$ & $( 339, 268 )$ \\
\hline
$48$& $(  60, 443 )$ & $(  72,  83 )$ & $(  81, 508 )$ & $( 291, 164 )$ & $( 307, 354 )$ & $( 343, 413 )$ \\
\hline
$49$& $(  55, 163 )$ & $(  80, 474 )$ & $(  92, 335 )$ & $( 290, 429 )$ & $( 312,  41 )$ & $( 350, 140 )$ \\
\hline
$50$& $(  47, 462 )$ & $(  70, 120 )$ & $(  96, 346 )$ & $( 301, 428 )$ & $( 318, 495 )$ & $( 342, 108 )$ \\
\hline
$51$& $(  59, 312 )$ & $(  67, 324 )$ & $(  95, 277 )$ & $( 299,  16 )$ & $( 314, 270 )$ & $( 322, 189 )$ \\
\hline
$52$& $(  41,  79 )$ & $(  68, 219 )$ & $(  83, 456 )$ & $( 304, 507 )$ & $( 316, 283 )$ & $( 338, 182 )$ \\
\hline
$53$& $(  57, 361 )$ & $(  65, 229 )$ & $(  83, 212 )$ & $( 294, 358 )$ & $( 310, 463 )$ & $( 324,  10 )$ \\
\hline
$54$& $(  56, 203 )$ & $(  77, 264 )$ & $(  82,  57 )$ & $( 303, 488 )$ & $( 304, 201 )$ & $( 341, 419 )$ \\
\hline
$55$& $(  48,  75 )$ & $(  70, 237 )$ & $(  91, 332 )$ & $( 302, 217 )$ & $( 310, 489 )$ & $( 330, 128 )$ \\
\hline
$56$& $(  45, 288 )$ & $(  74, 319 )$ & $(  96, 324 )$ & $( 297, 367 )$ & $( 320, 209 )$ & $( 335,  52 )$ \\
\hline
$57$& $(  63, 468 )$ & $(  73, 266 )$ & $(  85, 286 )$ & $( 300,  67 )$ & $( 308, 276 )$ & $( 334, 159 )$ \\
\hline
$58$& $(  38, 355 )$ & $(  65, 464 )$ & $(  89, 142 )$ & $( 305, 360 )$ & $( 319, 342 )$ & $( 351, 168 )$ \\
\hline
$59$& $(  42, 207 )$ & $(  69,  31 )$ & $(  93,  50 )$ & $( 297, 273 )$ & $( 314, 100 )$ & $( 352,  24 )$ \\
\hline
$60$& $(  36, 286 )$ & $(  77, 454 )$ & $(  88, 442 )$ & $( 296, 154 )$ & $( 308, 457 )$ & $( 323, 164 )$ \\
\hline
$61$& $(  51, 319 )$ & $(  78, 444 )$ & $(  81, 316 )$ & $( 294, 368 )$ & $( 312, 288 )$ & $( 321, 190 )$ \\
\hline
$62$& $(  64, 162 )$ & $(  79, 398 )$ & $(  90, 507 )$ & $( 293,  48 )$ & $( 316, 207 )$ & $( 331, 139 )$ \\
\hline
$63$& $(  62, 493 )$ & $(  74, 119 )$ & $(  86, 314 )$ & $( 300,  15 )$ & $( 317, 186 )$ & $( 346, 303 )$ \\
\hline
$64$& $(  33, 336 )$ & $(  75, 237 )$ & $(  88, 325 )$ & $( 299, 119 )$ & $( 319, 161 )$ & $( 332,   5 )$ \\
\hline
$65$& $(   6,  54 )$ & $(  27, 239 )$ & $(  51, 398 )$ & $( 140, 119 )$ & $( 187, 166 )$ & $( 341, 493 )$ \\
\hline
$66$& $(  10, 336 )$ & $(  22, 303 )$ & $(  47,  82 )$ & $( 147, 332 )$ & $( 177, 119 )$ & $( 328, 252 )$ \\
\hline
$67$& $(  16, 259 )$ & $(  24, 241 )$ & $(  61, 115 )$ & $( 144, 176 )$ & $( 162,  59 )$ & $( 352, 252 )$ \\
\hline
$68$& $(  13, 498 )$ & $(  17,   8 )$ & $(  33, 426 )$ & $( 159, 429 )$ & $( 191, 453 )$ & $( 346, 378 )$ \\
\hline
$69$& $(  12, 460 )$ & $(  22, 283 )$ & $(  55, 283 )$ & $( 158, 334 )$ & $( 184, 430 )$ & $( 330, 367 )$ \\
\hline
$70$& $(   7, 283 )$ & $(  20,  73 )$ & $(  56, 497 )$ & $( 133, 415 )$ & $( 161, 373 )$ & $( 322, 481 )$ \\
\hline
$71$& $(  14, 368 )$ & $(  26, 277 )$ & $(  50, 493 )$ & $( 157, 490 )$ & $( 164, 217 )$ & $( 324, 217 )$ \\
\hline
$72$& $(   3, 485 )$ & $(  27, 283 )$ & $(  54, 426 )$ & $( 154, 499 )$ & $( 186, 261 )$ & $( 321, 332 )$ \\
\hline
$73$& $(  12, 458 )$ & $(  30, 386 )$ & $(  38, 334 )$ & $( 141, 408 )$ & $( 180, 225 )$ & $( 332, 131 )$ \\
\hline
$74$& $(   4, 189 )$ & $(  23, 467 )$ & $(  64, 154 )$ & $( 142,  43 )$ & $( 169, 485 )$ & $( 344,  63 )$ \\
\hline
$75$& $(   7,  31 )$ & $(  25, 262 )$ & $(  44, 173 )$ & $( 150, 474 )$ & $( 165, 183 )$ & $( 351, 171 )$ \\
\hline
$76$& $(  15, 129 )$ & $(  17, 478 )$ & $(  40, 477 )$ & $( 149,  43 )$ & $( 171, 440 )$ & $( 334,  38 )$ \\
\hline
$77$& $(   9,  56 )$ & $(  29, 331 )$ & $(  63, 413 )$ & $( 136, 322 )$ & $( 163, 300 )$ & $( 342, 499 )$ \\
\hline
$78$& $(   9, 198 )$ & $(  20, 194 )$ & $(  43, 467 )$ & $( 160, 501 )$ & $( 189, 242 )$ & $( 349, 457 )$ \\
\hline
$79$& $(   3, 159 )$ & $(  32, 427 )$ & $(  62, 304 )$ & $( 138,   1 )$ & $( 176,  19 )$ & $( 329, 410 )$ \\
\hline
$80$& $(  11,  77 )$ & $(  18, 413 )$ & $(  45, 367 )$ & $( 153, 378 )$ & $( 175, 377 )$ & $( 338, 507 )$ \\
\hline
$81$& $(   8,  30 )$ & $(  19, 166 )$ & $(  59, 204 )$ & $( 139, 151 )$ & $( 178,  62 )$ & $( 336, 343 )$ \\
\hline
$82$& $(  15, 272 )$ & $(  28, 142 )$ & $(  53, 193 )$ & $( 146, 163 )$ & $( 174, 421 )$ & $( 326,  92 )$ \\
\hline
$83$& $(  13, 114 )$ & $(  18, 221 )$ & $(  48, 484 )$ & $( 155, 254 )$ & $( 190, 315 )$ & $( 348, 202 )$ \\
\hline
$84$& $(   1, 315 )$ & $(  28, 204 )$ & $(  46, 467 )$ & $( 143, 485 )$ & $( 172,  13 )$ & $( 345, 271 )$ \\
\hline
$85$& $(   2, 159 )$ & $(  24, 355 )$ & $(  37, 199 )$ & $( 134, 480 )$ & $( 183, 250 )$ & $( 331, 309 )$ \\
\hline
$86$& $(   5,  43 )$ & $(  19,  83 )$ & $(  49, 252 )$ & $( 148, 273 )$ & $( 181, 271 )$ & $( 327, 483 )$ \\
\hline
$87$& $(   4,   2 )$ & $(  31,  15 )$ & $(  36, 157 )$ & $( 156, 417 )$ & $( 179, 438 )$ & $( 335, 419 )$ \\
\hline
$88$& $(   8, 487 )$ & $(  31, 308 )$ & $(  39, 259 )$ & $( 129, 405 )$ & $( 185, 178 )$ & $( 339,  75 )$ \\
\hline
$89$& $(   5, 444 )$ & $(  29, 450 )$ & $(  41, 286 )$ & $( 137, 385 )$ & $( 167, 271 )$ & $( 343,  87 )$ \\
\hline
$90$& $(   2, 222 )$ & $(  32, 341 )$ & $(  52, 191 )$ & $( 130, 201 )$ & $( 170, 270 )$ & $( 350, 143 )$ \\
\hline
$91$& $(  14, 397 )$ & $(  21, 350 )$ & $(  57, 134 )$ & $( 151,  62 )$ & $( 166, 485 )$ & $( 347, 464 )$ \\
\hline
$92$& $(  16, 101 )$ & $(  23, 319 )$ & $(  35, 166 )$ & $( 132, 195 )$ & $( 173, 234 )$ & $( 323, 401 )$ \\
\hline
$93$& $(  11,  41 )$ & $(  30, 256 )$ & $(  34,  61 )$ & $( 131, 193 )$ & $( 192, 403 )$ & $( 333, 206 )$ \\
\hline
$94$& $(   6, 239 )$ & $(  26, 343 )$ & $(  58, 311 )$ & $( 145, 133 )$ & $( 188, 312 )$ & $( 340,  96 )$ \\
\hline
$95$& $(   1,  83 )$ & $(  25, 416 )$ & $(  60, 447 )$ & $( 152, 461 )$ & $( 182, 407 )$ & $( 337, 265 )$ \\
\hline
$96$& $(  10, 411 )$ & $(  21,  36 )$ & $(  42, 278 )$ & $( 135, 442 )$ & $( 168, 179 )$ & $( 325, 467 )$ \\
\hline
$97$& $(  54, 145 )$ & $( 102, 200 )$ & $( 115, 140 )$ & $( 123, 341 )$ & $( 297, 438 )$ & $( 307,  12 )$ \\
\hline
$98$& $(  37, 272 )$ & $( 106, 100 )$ & $( 109, 414 )$ & $( 120, 103 )$ & $( 290,  69 )$ & $( 319,   3 )$ \\
\hline
$99$& $(  57,  68 )$ & $(  97, 491 )$ & $( 113, 466 )$ & $( 125, 207 )$ & $( 291, 314 )$ & $( 305, 151 )$ \\
\hline
$100$& $(  33, 214 )$ & $( 101, 103 )$ & $( 112, 485 )$ & $( 118,  94 )$ & $( 304, 413 )$ & $( 317, 236 )$ \\
\hline
$101$& $(  35, 384 )$ & $( 106, 201 )$ & $( 110,  83 )$ & $( 128, 348 )$ & $( 300, 106 )$ & $( 312, 447 )$ \\
\hline
$102$& $(  53, 260 )$ & $( 103, 218 )$ & $( 117, 285 )$ & $( 126, 416 )$ & $( 294, 274 )$ & $( 311,  85 )$ \\
\hline
$103$& $(  40, 191 )$ & $( 101, 209 )$ & $( 110, 488 )$ & $( 123, 215 )$ & $( 295, 464 )$ & $( 320,  30 )$ \\
\hline
$104$& $(  49,  51 )$ & $( 106, 404 )$ & $( 114, 400 )$ & $( 127, 311 )$ & $( 292, 261 )$ & $( 320, 279 )$ \\
\hline
$105$& $(  47, 447 )$ & $(  99, 468 )$ & $( 117,  17 )$ & $( 119, 337 )$ & $( 303, 213 )$ & $( 314, 352 )$ \\
\hline
$106$& $(  56, 393 )$ & $( 103,  19 )$ & $( 118,  60 )$ & $( 120, 418 )$ & $( 292, 206 )$ & $( 308, 185 )$ \\
\hline
$107$& $(  45,  30 )$ & $( 100,  53 )$ & $( 115, 496 )$ & $( 126, 199 )$ & $( 303, 464 )$ & $( 305, 480 )$ \\
\hline
$108$& $(  41, 423 )$ & $(  98, 443 )$ & $( 115, 280 )$ & $( 124, 129 )$ & $( 289, 100 )$ & $( 310, 179 )$ \\
\hline
$109$& $(  62, 433 )$ & $( 104, 215 )$ & $( 113, 246 )$ & $( 123, 283 )$ & $( 304, 301 )$ & $( 309, 336 )$ \\
\hline
$110$& $(  60, 130 )$ & $(  99, 132 )$ & $( 110,  85 )$ & $( 122,  92 )$ & $( 297,  74 )$ & $( 315, 509 )$ \\
\hline
$111$& $(  51, 414 )$ & $( 100, 326 )$ & $( 113, 230 )$ & $( 121, 375 )$ & $( 302, 283 )$ & $( 318,  48 )$ \\
\hline
$112$& $(  52, 157 )$ & $( 103, 360 )$ & $( 114,  50 )$ & $( 124, 429 )$ & $( 291, 478 )$ & $( 317, 104 )$ \\
\hline
$113$& $(  34, 403 )$ & $(  98, 285 )$ & $( 108, 263 )$ & $( 119, 446 )$ & $( 290, 491 )$ & $( 313, 467 )$ \\
\hline
$114$& $(  43,  93 )$ & $( 105, 308 )$ & $( 111, 505 )$ & $( 121, 109 )$ & $( 301, 435 )$ & $( 312, 282 )$ \\
\hline
$115$& $(  50, 139 )$ & $( 100,  39 )$ & $( 116, 486 )$ & $( 127, 347 )$ & $( 293, 161 )$ & $( 306, 350 )$ \\
\hline
$116$& $(  58, 292 )$ & $( 104, 397 )$ & $( 112, 104 )$ & $( 119, 289 )$ & $( 296, 413 )$ & $( 310,  95 )$ \\
\hline
$117$& $(  59, 390 )$ & $( 107, 201 )$ & $( 111, 298 )$ & $( 118, 259 )$ & $( 295,  46 )$ & $( 307, 178 )$ \\
\hline
$118$& $(  46, 486 )$ & $(  99, 338 )$ & $( 107, 451 )$ & $( 124, 247 )$ & $( 293, 385 )$ & $( 319, 397 )$ \\
\hline
$119$& $(  42, 448 )$ & $(  97, 294 )$ & $( 111, 161 )$ & $( 127, 211 )$ & $( 298, 491 )$ & $( 308, 506 )$ \\
\hline
$120$& $(  44, 126 )$ & $(  98, 459 )$ & $( 109, 331 )$ & $( 122, 136 )$ & $( 299, 307 )$ & $( 311, 481 )$ \\
\hline
$121$& $(  39, 259 )$ & $( 102,  44 )$ & $( 112, 459 )$ & $( 122, 126 )$ & $( 298, 344 )$ & $( 316, 407 )$ \\
\hline
$122$& $(  38, 149 )$ & $( 102, 346 )$ & $( 117, 130 )$ & $( 128, 408 )$ & $( 301, 213 )$ & $( 306, 408 )$ \\
\hline
$123$& $(  55, 387 )$ & $( 104, 312 )$ & $( 116, 333 )$ & $( 120, 172 )$ & $( 302, 192 )$ & $( 316, 438 )$ \\
\hline
$124$& $(  61, 302 )$ & $( 107, 267 )$ & $( 109,  18 )$ & $( 125, 452 )$ & $( 294, 126 )$ & $( 315, 192 )$ \\
\hline
$125$& $(  36, 295 )$ & $( 105, 348 )$ & $( 114, 413 )$ & $( 125, 139 )$ & $( 289, 372 )$ & $( 318, 197 )$ \\
\hline
$126$& $(  64, 178 )$ & $( 101, 340 )$ & $( 108, 270 )$ & $( 121, 464 )$ & $( 300, 272 )$ & $( 309, 271 )$ \\
\hline
$127$& $(  63,   8 )$ & $( 105, 246 )$ & $( 116, 311 )$ & $( 128, 153 )$ & $( 299,  16 )$ & $( 313,  56 )$ \\
\hline
$128$& $(  48, 424 )$ & $(  97,  58 )$ & $( 108, 296 )$ & $( 126,  77 )$ & $( 296, 471 )$ & $( 314,  98 )$ \\
\hline
$129$& $(   3, 182 )$ & $(  21, 262 )$ & $( 242,  13 )$ & $( 291, 188 )$ & $( 304,  43 )$ & $( 310, 270 )$ \\
\hline
$130$& $(   5, 399 )$ & $(  25, 106 )$ & $( 238, 371 )$ & $( 295, 352 )$ & $( 306, 159 )$ & $( 310, 254 )$ \\
\hline
$131$& $(   7, 251 )$ & $(  26,  77 )$ & $( 244, 260 )$ & $( 299, 312 )$ & $( 306, 366 )$ & $( 316, 223 )$ \\
\hline
$132$& $(   3, 383 )$ & $(  22, 484 )$ & $( 240, 132 )$ & $( 292, 429 )$ & $( 307, 322 )$ & $( 318, 197 )$ \\
\hline
$133$& $(   5, 276 )$ & $(  23,  75 )$ & $( 252, 182 )$ & $( 294, 153 )$ & $( 307, 225 )$ & $( 319,  40 )$ \\
\hline
$134$& $(  14,  89 )$ & $(  30, 245 )$ & $( 233, 416 )$ & $( 289,  95 )$ & $( 304, 426 )$ & $( 314, 258 )$ \\
\hline
$135$& $(   9, 437 )$ & $(  21,  38 )$ & $( 232, 196 )$ & $( 297, 480 )$ & $( 302, 454 )$ & $( 317, 169 )$ \\
\hline
$136$& $(   8,  46 )$ & $(  32, 119 )$ & $( 245, 104 )$ & $( 294, 363 )$ & $( 305,  26 )$ & $( 314, 468 )$ \\
\hline
$137$& $(   9, 431 )$ & $(  28, 144 )$ & $( 254, 499 )$ & $( 291, 257 )$ & $( 303,  75 )$ & $( 311, 345 )$ \\
\hline
$138$& $(   6, 460 )$ & $(  19, 123 )$ & $( 227, 151 )$ & $( 298, 426 )$ & $( 310, 128 )$ & $( 320, 393 )$ \\
\hline
$139$& $(  16, 114 )$ & $(  32, 358 )$ & $( 229, 285 )$ & $( 297, 154 )$ & $( 308,  84 )$ & $( 319, 443 )$ \\
\hline
$140$& $(   4, 477 )$ & $(  17, 157 )$ & $( 243, 440 )$ & $( 293,  69 )$ & $( 300, 189 )$ & $( 318, 158 )$ \\
\hline
$141$& $(   4,  43 )$ & $(  27, 228 )$ & $( 248, 296 )$ & $( 298, 234 )$ & $( 303, 226 )$ & $( 315, 449 )$ \\
\hline
$142$& $(  12,  14 )$ & $(  16, 372 )$ & $( 256, 217 )$ & $( 289, 189 )$ & $( 300, 309 )$ & $( 320, 250 )$ \\
\hline
$143$& $(  17,  49 )$ & $(  19, 327 )$ & $( 225, 193 )$ & $( 297, 210 )$ & $( 309,  78 )$ & $( 312, 408 )$ \\
\hline
$144$& $(  13, 240 )$ & $(  29, 292 )$ & $( 236, 481 )$ & $( 292, 461 )$ & $( 301, 387 )$ & $( 311, 360 )$ \\
\hline
$145$& $(  15, 102 )$ & $(  20, 141 )$ & $( 228, 362 )$ & $( 293, 330 )$ & $( 299, 345 )$ & $( 312, 139 )$ \\
\hline
$146$& $(   8, 357 )$ & $(  29, 148 )$ & $( 239, 304 )$ & $( 290, 270 )$ & $( 308, 487 )$ & $( 315, 209 )$ \\
\hline
$147$& $(   2, 304 )$ & $(  26, 135 )$ & $( 249, 398 )$ & $( 292, 113 )$ & $( 305, 450 )$ & $( 317, 463 )$ \\
\hline
$148$& $(   6,  39 )$ & $(  22, 223 )$ & $( 247, 487 )$ & $( 289,  74 )$ & $( 303,   2 )$ & $( 313, 165 )$ \\
\hline
$149$& $(  10, 119 )$ & $(  31, 362 )$ & $( 230, 464 )$ & $( 294, 105 )$ & $( 309,  50 )$ & $( 317, 468 )$ \\
\hline
$150$& $(  12, 171 )$ & $(  20, 284 )$ & $( 231, 119 )$ & $( 295, 306 )$ & $( 308, 210 )$ & $( 313, 507 )$ \\
\hline
$151$& $(  11, 207 )$ & $(  31,  78 )$ & $( 234, 177 )$ & $( 293,  58 )$ & $( 301, 240 )$ & $( 320, 222 )$ \\
\hline
$152$& $(  11, 149 )$ & $(  27, 474 )$ & $( 241, 274 )$ & $( 296,  42 )$ & $( 306,  80 )$ & $( 313, 306 )$ \\
\hline
$153$& $(  15, 116 )$ & $(  24, 318 )$ & $( 255, 306 )$ & $( 290, 100 )$ & $( 304, 298 )$ & $( 311, 259 )$ \\
\hline
$154$& $(  13, 139 )$ & $(  24,  20 )$ & $( 253, 297 )$ & $( 296, 196 )$ & $( 302,  77 )$ & $( 312,  63 )$ \\
\hline
$155$& $(   2, 184 )$ & $(  25,  30 )$ & $( 250, 303 )$ & $( 291, 156 )$ & $( 301, 240 )$ & $( 316, 342 )$ \\
\hline
$156$& $(   1, 235 )$ & $(  30, 377 )$ & $( 226, 430 )$ & $( 296, 337 )$ & $( 307, 511 )$ & $( 316, 489 )$ \\
\hline
$157$& $(  14, 162 )$ & $(  28, 468 )$ & $( 251, 323 )$ & $( 299, 193 )$ & $( 302, 307 )$ & $( 315, 361 )$ \\
\hline
$158$& $(   7, 244 )$ & $(  18, 342 )$ & $( 237, 458 )$ & $( 295,  10 )$ & $( 309, 167 )$ & $( 319, 191 )$ \\
\hline
$159$& $(  10, 252 )$ & $(  23, 112 )$ & $( 246, 305 )$ & $( 290,   4 )$ & $( 305, 265 )$ & $( 318, 351 )$ \\
\hline
$160$& $(   1, 294 )$ & $(  18, 485 )$ & $( 235, 370 )$ & $( 298,   6 )$ & $( 300, 272 )$ & $( 314, 447 )$ \\
\hline
$161$& $(   8, 389 )$ & $(  18, 442 )$ & $(  23, 246 )$ & $( 110,  47 )$ & $( 126,  74 )$ & $( 215, 125 )$ \\
\hline
$162$& $(  12, 482 )$ & $(  21, 195 )$ & $(  31, 398 )$ & $( 100, 241 )$ & $( 119, 141 )$ & $( 199,  85 )$ \\
\hline
$163$& $(   5,  62 )$ & $(  13,  66 )$ & $(  28,  49 )$ & $( 111,   5 )$ & $( 128, 370 )$ & $( 204, 223 )$ \\
\hline
$164$& $(   9, 334 )$ & $(  14, 140 )$ & $(  32, 183 )$ & $( 109, 419 )$ & $( 117, 440 )$ & $( 209, 481 )$ \\
\hline
$165$& $(   2, 390 )$ & $(  11, 299 )$ & $(  29,  18 )$ & $(  99, 494 )$ & $( 113, 508 )$ & $( 201, 360 )$ \\
\hline
$166$& $(   1, 492 )$ & $(  19,  75 )$ & $(  32, 206 )$ & $(  97, 471 )$ & $( 120,  46 )$ & $( 211, 356 )$ \\
\hline
$167$& $(  10, 451 )$ & $(  16, 334 )$ & $(  28,  68 )$ & $( 100, 134 )$ & $( 118,  71 )$ & $( 219, 308 )$ \\
\hline
$168$& $(   7, 256 )$ & $(  16,  74 )$ & $(  29, 469 )$ & $( 112, 360 )$ & $( 128,  52 )$ & $( 213, 202 )$ \\
\hline
$169$& $(   6, 221 )$ & $(  17, 403 )$ & $(  25, 167 )$ & $( 107, 317 )$ & $( 123, 216 )$ & $( 202, 258 )$ \\
\hline
$170$& $(  11, 111 )$ & $(  16, 421 )$ & $(  26, 299 )$ & $( 101, 489 )$ & $( 127, 457 )$ & $( 198, 489 )$ \\
\hline
$171$& $(   3,  96 )$ & $(  15,  30 )$ & $(  30, 418 )$ & $( 108,  86 )$ & $( 116, 178 )$ & $( 218,   7 )$ \\
\hline
$172$& $(   8, 273 )$ & $(  12,  62 )$ & $(  27, 372 )$ & $( 111, 337 )$ & $( 117, 209 )$ & $( 203, 369 )$ \\
\hline
$173$& $(   8, 202 )$ & $(  20, 321 )$ & $(  24, 283 )$ & $( 112,  36 )$ & $( 125, 243 )$ & $( 216, 465 )$ \\
\hline
$174$& $(   3, 495 )$ & $(  18, 292 )$ & $(  28,  80 )$ & $( 102, 150 )$ & $( 114, 189 )$ & $( 195,  11 )$ \\
\hline
$175$& $(   1, 265 )$ & $(  15, 268 )$ & $(  22, 157 )$ & $( 105, 277 )$ & $( 122, 495 )$ & $( 207, 459 )$ \\
\hline
$176$& $(   5,  32 )$ & $(  12, 342 )$ & $(  26, 399 )$ & $( 105, 446 )$ & $( 123, 408 )$ & $( 222, 447 )$ \\
\hline
$177$& $(   9,  93 )$ & $(  11, 421 )$ & $(  32, 440 )$ & $(  98, 266 )$ & $( 118, 430 )$ & $( 196, 397 )$ \\
\hline
$178$& $(   7, 332 )$ & $(  19, 183 )$ & $(  31, 189 )$ & $( 108, 452 )$ & $( 114,  77 )$ & $( 194, 294 )$ \\
\hline
$179$& $(   1,  64 )$ & $(  20, 127 )$ & $(  26, 187 )$ & $( 106, 259 )$ & $( 119, 296 )$ & $( 205, 252 )$ \\
\hline
$180$& $(  10, 196 )$ & $(  13,  51 )$ & $(  27,  81 )$ & $(  98, 325 )$ & $( 127, 490 )$ & $( 224, 454 )$ \\
\hline
$181$& $(   4, 455 )$ & $(  21, 122 )$ & $(  22,  60 )$ & $( 106, 215 )$ & $( 115, 392 )$ & $( 210, 121 )$ \\
\hline
$182$& $(   6, 178 )$ & $(  20, 150 )$ & $(  29, 122 )$ & $( 103, 230 )$ & $( 122, 289 )$ & $( 223, 197 )$ \\
\hline
$183$& $(   7, 260 )$ & $(  17, 421 )$ & $(  30, 171 )$ & $( 104, 268 )$ & $( 126, 179 )$ & $( 212, 171 )$ \\
\hline
$184$& $(   5, 200 )$ & $(  21,  91 )$ & $(  27, 207 )$ & $(  99, 112 )$ & $( 120, 402 )$ & $( 220, 113 )$ \\
\hline
$185$& $(   4, 390 )$ & $(  14,  91 )$ & $(  25,  46 )$ & $( 104,  92 )$ & $( 124, 429 )$ & $( 208, 227 )$ \\
\hline
$186$& $(   3, 199 )$ & $(  13, 312 )$ & $(  23, 216 )$ & $( 103, 440 )$ & $( 113,  88 )$ & $( 221, 351 )$ \\
\hline
$187$& $(  10, 446 )$ & $(  15, 239 )$ & $(  25,   2 )$ & $( 109, 455 )$ & $( 121, 483 )$ & $( 197, 152 )$ \\
\hline
$188$& $(   6,  58 )$ & $(  23, 443 )$ & $(  24,  85 )$ & $(  97, 510 )$ & $( 115, 228 )$ & $( 206, 279 )$ \\
\hline
$189$& $(   4, 347 )$ & $(  19, 359 )$ & $(  24,  43 )$ & $( 102,  71 )$ & $( 116, 202 )$ & $( 217,  65 )$ \\
\hline
$190$& $(   9, 511 )$ & $(  17, 299 )$ & $(  22, 289 )$ & $( 110,  43 )$ & $( 125, 468 )$ & $( 214, 135 )$ \\
\hline
$191$& $(   2,  89 )$ & $(  14, 242 )$ & $(  31,  81 )$ & $( 107, 170 )$ & $( 121,  99 )$ & $( 200, 295 )$ \\
\hline
$192$& $(   2, 248 )$ & $(  18,   4 )$ & $(  30, 163 )$ & $( 101, 352 )$ & $( 124, 128 )$ & $( 193,  73 )$ \\
\hline
$193$& $(  25, 106 )$ & $( 108, 407 )$ & $( 151, 412 )$ & $( 273, 425 )$ & $( 297, 464 )$ & $( 305, 267 )$ \\
\hline
$194$& $(  19, 133 )$ & $( 113,  45 )$ & $( 131,  70 )$ & $( 259, 491 )$ & $( 293, 151 )$ & $( 311, 266 )$ \\
\hline
$195$& $(  11,  47 )$ & $( 117, 254 )$ & $( 154, 194 )$ & $( 265, 215 )$ & $( 300, 208 )$ & $( 316, 431 )$ \\
\hline
$196$& $(  27,  68 )$ & $( 128,   9 )$ & $( 145,  34 )$ & $( 276, 238 )$ & $( 291, 231 )$ & $( 309, 409 )$ \\
\hline
$197$& $(   2, 481 )$ & $( 111, 373 )$ & $( 144, 360 )$ & $( 286, 457 )$ & $( 289, 101 )$ & $( 315, 311 )$ \\
\hline
$198$& $(  31,  83 )$ & $( 104, 185 )$ & $( 150, 405 )$ & $( 266,  21 )$ & $( 291, 510 )$ & $( 308, 261 )$ \\
\hline
$199$& $(  21, 280 )$ & $(  97, 357 )$ & $( 146, 129 )$ & $( 275, 129 )$ & $( 303, 501 )$ & $( 306, 295 )$ \\
\hline
$200$& $(  18,  32 )$ & $( 121,  95 )$ & $( 135,  71 )$ & $( 281, 362 )$ & $( 304, 373 )$ & $( 315, 253 )$ \\
\hline
$201$& $(  16,  46 )$ & $( 116, 246 )$ & $( 129,  48 )$ & $( 263,  50 )$ & $( 295, 295 )$ & $( 312, 185 )$ \\
\hline
$202$& $(  32, 120 )$ & $( 127, 145 )$ & $( 132, 103 )$ & $( 272, 126 )$ & $( 294,  36 )$ & $( 320, 267 )$ \\
\hline
$203$& $(   3, 369 )$ & $( 120, 508 )$ & $( 160, 381 )$ & $( 269, 321 )$ & $( 289, 445 )$ & $( 311, 328 )$ \\
\hline
$204$& $(   6, 509 )$ & $( 106, 431 )$ & $( 159,  45 )$ & $( 279, 122 )$ & $( 302, 463 )$ & $( 308, 359 )$ \\
\hline
$205$& $(  14, 283 )$ & $( 122, 464 )$ & $( 136, 319 )$ & $( 283, 189 )$ & $( 292, 155 )$ & $( 313, 321 )$ \\
\hline
$206$& $(  15, 354 )$ & $( 114, 216 )$ & $( 138,  67 )$ & $( 288, 429 )$ & $( 295, 147 )$ & $( 314, 170 )$ \\
\hline
$207$& $(  30, 178 )$ & $( 102, 324 )$ & $( 140,  53 )$ & $( 262, 376 )$ & $( 290, 510 )$ & $( 310, 192 )$ \\
\hline
$208$& $(   4,  56 )$ & $( 103,  65 )$ & $( 139, 488 )$ & $( 274,  33 )$ & $( 301, 180 )$ & $( 314, 391 )$ \\
\hline
$209$& $(  22, 301 )$ & $( 124, 103 )$ & $( 134, 292 )$ & $( 257,  11 )$ & $( 298,  84 )$ & $( 309,   0 )$ \\
\hline
$210$& $(  28, 180 )$ & $( 107, 108 )$ & $( 149,   7 )$ & $( 258, 197 )$ & $( 296, 278 )$ & $( 320, 179 )$ \\
\hline
$211$& $(   8,  49 )$ & $( 119, 277 )$ & $( 152,  87 )$ & $( 261, 373 )$ & $( 304, 255 )$ & $( 319, 493 )$ \\
\hline
$212$& $(  20, 327 )$ & $( 105, 474 )$ & $( 143, 273 )$ & $( 260,  33 )$ & $( 290, 390 )$ & $( 317, 371 )$ \\
\hline
$213$& $(  29,  51 )$ & $( 125,  31 )$ & $( 153, 200 )$ & $( 284,  60 )$ & $( 303,  91 )$ & $( 319, 471 )$ \\
\hline
$214$& $(  24, 426 )$ & $(  98, 230 )$ & $( 141,  71 )$ & $( 277, 219 )$ & $( 297, 121 )$ & $( 306,  73 )$ \\
\hline
$215$& $(   9,  63 )$ & $( 112,   5 )$ & $( 148, 323 )$ & $( 267, 426 )$ & $( 293, 409 )$ & $( 307, 108 )$ \\
\hline
$216$& $(   1, 294 )$ & $( 115, 470 )$ & $( 147, 357 )$ & $( 280, 464 )$ & $( 301, 221 )$ & $( 317, 373 )$ \\
\hline
$217$& $(   5, 447 )$ & $( 110, 106 )$ & $( 133, 360 )$ & $( 285,  73 )$ & $( 302, 435 )$ & $( 313,  58 )$ \\
\hline
$218$& $(  10, 239 )$ & $( 123,  73 )$ & $( 130, 286 )$ & $( 264,  30 )$ & $( 292, 496 )$ & $( 316, 412 )$ \\
\hline
$219$& $(   7, 408 )$ & $(  99,  40 )$ & $( 155, 114 )$ & $( 270, 454 )$ & $( 296,  49 )$ & $( 318, 327 )$ \\
\hline
$220$& $(  17, 361 )$ & $( 118,   2 )$ & $( 156, 195 )$ & $( 271, 508 )$ & $( 299, 343 )$ & $( 305, 319 )$ \\
\hline
$221$& $(  23, 221 )$ & $( 109, 415 )$ & $( 158, 323 )$ & $( 287, 423 )$ & $( 300, 467 )$ & $( 310, 207 )$ \\
\hline
$222$& $(  26,  64 )$ & $( 100, 470 )$ & $( 137,  39 )$ & $( 278,  87 )$ & $( 298, 490 )$ & $( 312, 123 )$ \\
\hline
$223$& $(  12, 226 )$ & $( 126, 194 )$ & $( 142, 122 )$ & $( 282, 260 )$ & $( 299, 272 )$ & $( 307, 105 )$ \\
\hline
$224$& $(  13, 451 )$ & $( 101, 146 )$ & $( 157,  69 )$ & $( 268, 154 )$ & $( 294, 153 )$ & $( 318, 267 )$ \\
\hline
\end{longtable}}
\end{center}
}



%
\bibliographystyle{ieeetr}

\end{document}